\documentclass{article}

\usepackage{arxiv}

\usepackage[utf8]{inputenc} 
\usepackage[T1]{fontenc}    
\usepackage{hyperref}       
\usepackage{url}            
\usepackage{booktabs}       
\usepackage{amsfonts}       
\usepackage{nicefrac}       
\usepackage{microtype}      
\usepackage{lipsum}
\usepackage{graphicx}
\usepackage{enumitem}
\usepackage{gensymb}
\usepackage{amsmath}
\usepackage{amssymb}
\usepackage{float}
\usepackage{physics}

\usepackage{mathtools}
\DeclarePairedDelimiter{\nint}\lfloor\rceil

\DeclareRobustCommand{\rchi}{{\mathpalette\irchi\relax}}
\newcommand{\irchi}[2]{\raisebox{\depth}{$#1\chi$}}

\usepackage[ruled,linesnumbered]{algorithm2e}
\makeatletter
\newcommand{\nosemic}{\renewcommand{\@endalgocfline}{\relax}}
\newcommand{\dosemic}{\renewcommand{\@endalgocfline}{\algocf@endline}}
\newcommand{\pushline}{\Indp}
\newcommand{\popline}{\Indm\dosemic}
\let\oldnl\nl
\newcommand{\nonl}{\renewcommand{\nl}{\let\nl\oldnl}}
\makeatother

\usepackage[style=numeric-comp, sorting=none]{biblatex}
\title{An image processing pipeline for in-situ dynamic X-ray imaging of directional solidification of metal alloys in thin cells}

\author{
  Mihails Birjukovs\\
  Institute of Numerical Modelling\\
  University of Latvia (UL)\\
  Riga, Latvia, Jelgavas 3, 1004 \\
  \texttt{mihails.birjukovs@lu.lv} \\
   \And
  Natalia Shevchenko\\
  Helmholtz-Zentrum Dresden-Rossendorf (HZDR)\\
  Department of Magnetohydrodynamics\\
  Bautzner Landstraße 400, 01328 Dresden, Germany \\
  \And
    Sven Eckert\\
  Helmholtz-Zentrum Dresden-Rossendorf (HZDR)\\
  Department of Magnetohydrodynamics\\
  Bautzner Landstraße 400, 01328 Dresden, Germany \\
  }

\begin{document}
\maketitle

\begin{abstract}
We present an image processing algorithm developed for quantitative analysis of directional solidification of metal alloys in thin cells using X-ray imaging. Our methodology allows to identify the fluid volume, fluid channels and cavities, and to separate them from the solidified structures. It also allows morphological analysis within the solid fraction, including automatic decomposition into dominant grains by orientation and connectivity. In addition, the interplay between solidification and convection can be studied by characterizing convection plumes in the fluid, and solute concentrations above the developing solidification front. The image filters used enable the developed code (open-source) to work reliably even for single images with low signal-to-noise ratio, low contrast-to-noise ratio, and low image resolution. This is demonstrated by applying the code to several dynamic \textit{in situ} X-ray imaging experiments with a solidifying gallium-indium alloy in a thin cell. Grain (and global) dendrite orientation statistics, convective plume parameterization, etc. can be obtained from the code output. The limitations of the presented approach are also explained.
\end{abstract}


\section{Introduction}
\label{sec:intro}

Solidification is a central aspect of many industrial applications, particularly in metallurgy, e.g. production of nickel-based superalloys, lightweight aluminum and magnesium alloys, etc. \cite{solidification-pattern-growth-modes, solidification-kao-mf-plume-dynamics, book-solidification-fundamentals}. A well-known and common problem is the risk of defect formation during these processes. Segregation of solute species originates at the micro scale, but propagates to and emerges at the macro scale (macrosegregation), leading to non-uniformity of the distribution of the inter-metallic phases in industrial alloys \cite{solidification-beckermann-macrosegregation-modelling}. In addition, during solidification of alloys, partitioning of elements leads to the formation of a solute boundary layer in the vicinity of the liquid–solid interface. In cases where the density of the solute may be lighter than that of the bulk liquid, buoyancy forces in the boundary layer directed back towards the bulk liquid cause the formation of solute plumes that emanate from the solid-liquid interface. Under certain conditions, the escaping solute can form stable channels called \textit{chimneys}. After complete solidification, these disturbances remain as defects in the castings known as \textit{freckles}, which are essentially anisotropic alloy composition inhomogeneities in the form of channels with diameters proportional to few primary dendrite arm spacings and lengths varying from millimeters to centimeters \cite{solidification-alloys-by-design, madison-thesis, solidification-freckles-superalloys, solidification-freckle-prediction, solidification-kao-mf-plume-dynamics, solidification-saad-gain-simulations, solidification-xray-imaging-chimneys-natalia}.

It is therefore desirable to control solidification such that defects do not occur. However, control requires understanding the underlying physics, and solidification processes in liquid metal alloys are very complex, with many possible regimes of pattern formation depending on system parameters (e.g. temperature gradient, cooling rate, alloy component mass fractions) \cite{solidification-pattern-growth-modes, solidification-review-beckermann, solidification-pattern-growth-modes-2, solidification-review-directional-growth, book-solidification-fundamentals}. There exists an interplay of many physical mechanisms on different length scales: primary and secondary dendrite arm growth, liquid–solid interface instabilities, liquid mass flow near the interface and in the bulk (in general, both natural and forced convection), concentration transport, liquid flow in through solidified dendrite structures and remelting, global and local temperature dynamics, etc. \cite{solidification-review-beckermann, solidification-freckle-criterion, solidification-xray-imaging-melt-convection-natalia, solidification-xray-imaging-forced-convection-natalia, book-solidification-fundamentals, solidification-review-annurev-convection, solidification-kao-simulations-solid-mechanics}. One way to control such complex dynamics is by applying magnetic field to the domain where solidification occurs -- however, one then has to consider additional physical mechanisms within the system, such as liquid flow damping or forcing by the Lorentz force, which also includes a thermoelectric contribution. This and other factors introduced by magnetic field application significantly alters solidified microstructures \cite{solidification-kao-mf-plume-dynamics, solidification-kao-simulations-mhd-effects, solidification-xray-imaging-transverse-mf-natalia, solidification-review-mf-effects}.

A very prospective and commonly used method for studying solidification dynamics without or with applied magnetic field is by using downscaled model systems -- Hele-Shaw cells where binary alloy solidification can be observed at the meso-scale (i.e. dendrite grains with spatially resolved individual dendrites) using \textit{in situ} dynamic X-ray transmission contrast radiography. Even though one obtains only the projections of the solidified microstructures, it has proven to be a very effective means of probing systems with solidification processes for physical insights \cite{solidification-xray-imaging-ni-alloy, solidification-saad-gain-simulations, solidification-kao-mf-plume-dynamics, solidification-xray-imaging-alge-alloy, solidification-xray-imaging-alcu-alloy, solidification-xray-imaging-alcu-alloy-bubbles, solidification-xray-imaging-validating-3d-simulations, solidification-xray-imaging-tesa-freckles, solidification-xray-imaging-3d-retrieval, solidification-xray-imaging-melt-convection-natalia, solidification-xray-imaging-forced-convection-natalia, solidification-xray-imaging-chimneys-natalia, solidification-xray-imaging-alcu-equiaxed, solidification-xray-imaging-alcu-periodic-growth, solidification-xray-imaging-transverse-mf-natalia, solidification-xray-imaging-optical-flow-boden}. In addition to the challenges associated with imaging, there is also the matter of extracting valuable information from the acquired images. Ideally, to get the full picture of system dynamics, one has to separate the liquid from the solid, identify the solidification front and any liquid enclosures within the solidified microstructure, and obtain the microstructure skeletons. One could then perform orientation analysis for the skeletons, derive the primary dendrite spacing statistics, determine the local velocity with which the solidification front travels, as well as measure the solute concentration near the solidification front as it moves, since the concentration largely determines the front evolution. In this regard, detecting convective plumes and analyzing their shapes is also of interest, as is velocimetry in the liquid flow regions. In addition, it could also be of interest to detect and separate different grains (if any) within the microstructure seen in the images. Of course, the problem lies in doing all of the above automatically and reliably, which is relevant given the amount of images usually acquired in X-ray radiography experiments and the amount of information captured within each image.

However, while there exist solutions for some of the above problems, most appear to be limited to segmentation/detection of dendritic structures \cite{solidification-denmap-1, solidification-denmap-2, solidification-ml-dendrite-core-detection, solidification-dendrite-tip-tracking, solidification-equiaxed-dendrite-segmentation-neural, solidification-equiaxed-dendrite-segmentation-neural-2}. In \cite{solidification-denmap-1, solidification-denmap-2, solidification-ml-dendrite-core-detection}, the focus is the detection of the dendrite cores from images of planes normal to the solidification direction -- this, and the fact the algorithm presented in \cite{solidification-denmap-1, solidification-denmap-2} uses template matching as one of its stages, makes it hardly applicable to studying X-ray images with directionally solidifying dendrite "forests" (e.g., as in \cite{solidification-kao-mf-plume-dynamics, solidification-saad-gain-simulations, solidification-xray-imaging-chimneys-natalia, solidification-xray-imaging-melt-convection-natalia, solidification-xray-imaging-optical-flow-boden}) where planes parallel to the growth direction are imaged. In addition, at least observing the demonstrated application examples, it seems that these methods should be reliably applicable in the cases when the images are fully filled with dendrites, i.e., in instances where there are both liquid and solid regions, one must first be separated from the other using different methods. Dendrite tip tracking is performed in \cite{solidification-dendrite-tip-tracking} by segmenting the upper part of the solidified structures growing upwards. The key aspect of the segmentation procedure is to use the difference between two consecutive frames to highlight the newly formed solid, segment the relevant region and then derive the tip coordinates. The utilized approach also enables tracking the solidification front. However, in cases where the differences between the frames are smaller and significant noise is present, the algorithm could be expected to run into performance issues and a more general approach is desired. A versatile approach using neural networks for automated detection of equiaxed dendrite detection was reported in \cite{solidification-equiaxed-dendrite-segmentation-neural}. Another important example of automated dendrite segmentation using neural networks was shown in \cite{solidification-equiaxed-dendrite-segmentation-neural-2} where, unlike in \cite{solidification-equiaxed-dendrite-segmentation-neural}, a binary mask for the solid structure was predicted instead of detecting separate dendrites. It is, however, worth pointing out that the example images/cases presented in \cite{solidification-equiaxed-dendrite-segmentation-neural, solidification-equiaxed-dendrite-segmentation-neural-2} do not exhibit significant noise, which is often present even after some temporal averaging in dynamic X-ray radiography experiments where exposure times are relatively low -- it is therefore not clear how well these methods will perform under such conditions.

The lack of a systematic approach to image processing beyond methods for segmentation presents a problem, since it has been clearly demonstrated that the microstructure evolution must be analyzed in conjunction with the other processes in solidifying systems. Currently the most common tool used for image analysis in the field is \textit{ImageJ} with its many custom plugins developed by the community \cite{imagej-1, imagej-2}. While \textit{ImageJ} is open-source, with an impressive arsenal of methods, many of them are not automated, robust, or publically available. In contrast, it would be very convenient to have an open-source all-in-one solution for X-ray image analysis. It should also be noted that such code could be applied to the output of numerical simulations as well, the difference being that the latter do not have the image noise associated with experimental measurements. Thus, more direct comparisons between simulations and experiments, which seem to be largely lacking, could be possible.

This was the motivation for us to present our solution -- the first version of the open-source code developed for automatic analysis of dynamic X-ray radiography images of directional solidification processes studied using Hele-Shaw cells. The current version does not yet have an integrated optical flow component (e.g., like the code used in \cite{solidification-xray-imaging-optical-flow-boden}), but otherwise it meets the above mentioned analysis functionality requirements. Moreover, it was designed for robustness and is quite resilient to image noise and low image contrast. Utilized approaches to image and data processing combine both well-known state of art and our original methods, particularly for solid structure segmentation and dendrite grain analysis. The performance of the methodology implemented in the code is demonstrated on data from \textit{in situ} X-ray radiography experiments \cite{solidification-xray-imaging-melt-convection-natalia,solidification-xray-imaging-chimneys-natalia}.

\section{Image characterization}

\subsection{Image acquisition}

All images used in this paper were acquired at the X-ray lab at Helmholtz-Zentrum Dresden-Rossendorf (HZDR). The Hele-Shaw solidification cell with dimensions $35 ~ mm \times 25 ~ mm \times 0.15 ~ mm$ was imaged at 1 frame per second with a $1~s$ exposure time. The imaging system utilized the \textit{Phoenix X-ray XS225D-OEM} X-ray tube and is described in more detail in \cite{solidification-xray-imaging-transverse-mf-natalia, solidification-saad-gain-simulations, solidification-kao-mf-plume-dynamics, solidification-xray-imaging-chimneys-natalia, solidification-xray-imaging-melt-convection-natalia}. For each image sequence recording, dark current signals and X-ray beam profile signals were recorded for subsequent image correction and normalization during pre-processing. For every set of system parameters, repeated recordings were made to ensure the results are reproducible, as well as for redundancy.

\subsection{Image properties}
\label{sec:image-props}

\begin{figure}[htbp]
\centering
\includegraphics[width=0.9\textwidth]{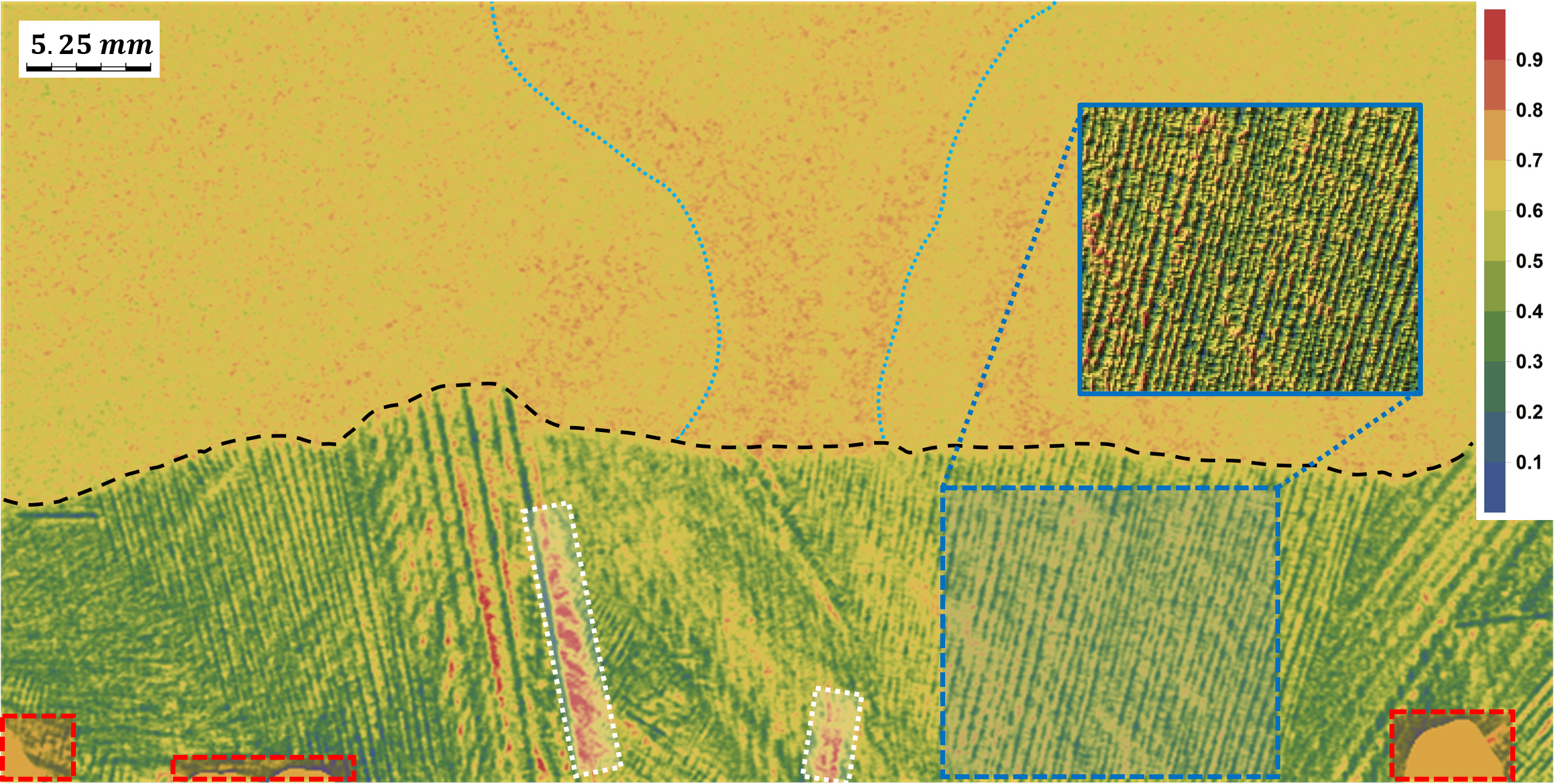}
\caption{A typical (false color) FOV in one of the imaging experiments after pre-processing (Algorithm \ref{alg:pre-processing}). The black dashed curve is a rough outline of the solidification front. The light-blue dashed line indicates one of the convective plumes within the FOV. Interior liquid metal pools are highlighted with white dashed frames. Note also the image artifacts highlighted with red dashed frames at the bottom of the FOV. The dark blue dashed frame indicates a region that belong to one of the formed dendrite grains -- the inset (dark blue frame) shows its relief plot for visual clarity. Both the false-color FOV image and the relief plot share the same color scale (color bar to the right) and scale (top-left corner).}
\label{fig:example-image}
\end{figure}

Images are 16-bit gray-scale TIFFs and the field of view (FOV) typically has a $\sim 760 \times 576~ px$ (pixel) image size with a pixel size $[13.7;37.6]~\mu m$ (actual image size varies between different image sequences due to boundary cropping). Figure \ref{fig:example-image} is an example of an acquired image of a dendritic network in the solidifying Ga-In alloy \cite{solidification-xray-imaging-melt-convection-natalia,solidification-xray-imaging-chimneys-natalia}.

To simplify further description and analysis, it is important to at least informally define key image features. In Figure \ref{fig:example-image}, one can see the \textit{solidification front} (SF) outline with a black dashed curve. Here we define the SF as an the envelope of the FOV region with the solidified structures. A more precise \textit{operational} definition in the image processing context will be given in Section \ref{sec:liquid-solid-separation}. The region (or multiple) containing solidified structures delimited by the solidification front is the \textit{solid zone} (SZ). This zone may also include liquid \textit{cavities} isolated from the bulk liquid and \textit{channels} that are connected to the bulk liquid (above the SF), e.g. one such channel will later form from the larger of the closed liquid pools highlighted in Figure \ref{fig:example-image}. Thus, the \textit{liquid zone} (LZ) is the difference between the FOV and the SZ, minus the cavities and channels. These definitions will be used throughout the rest of this article.

The images exhibit Poisson (multiplicative) noise, as well as salt-and-pepper noise due to momentarily overexposed or unresponsive ("dead") camera pixels. The X-ray beam flux over the FOV is non-uniform with a fall-off near the edges of the acquired images. The contrast-to-noise ratio (CNR) is different for the LZ and the SZ. The convective plume CNR in the LZ is initially rather good , but typically degrades over time as the \textit{solid fill factor} (SFF, the ratio of the SZ area to the FOV area) of the cell increases -- this is because the solute is ejected above the SF, the LZ is saturated, and the contrast between the liquid alloy components diminishes. In addition, the CNR in the SZ may also vary over the image since there is solute flow across the solidified structures, potentially occluding them. The signal-to-noise ratio (SNR) is usually adequate for structures in the SZ, but it is rather low for the convective plumes in the LZ. In addition, some of the images may exhibit larger-scale artifacts -- for instance, as shown in Figure \ref{fig:example-image} with red dashed frames. In this case the artifacts are the spots where the two parallel walls of the Hele-Shaw cells were fused together.

\section{Image processing}

\subsection{Assumptions and considerations}

The developed image processing code must enable in-depth analysis of both the LZ and SZ over time, as well as the dynamics associated with the SF evolution. Therefore, the objectives are as follows:

\begin{enumerate}[noitemsep,topsep=7.5pt,leftmargin=0.1cm]
\setlength{\itemindent}{1.0cm}
    \item Segment the LZ and SZ.
    \item Derive the SF.
    \item Identify channels connected to the LZ, and also liquid cavities within the SZ.
    \item Segment the convective plumes within the LZ for shape analysis.
    \item Extract skeletons of the structures (in this case dendrites) within the SZ.
    \item Perform orientation analysis for the structures identified within the LZ.
    \item Decompose the SZ structures into sub-domains (grains) by orientation and connectivity.
    \item Measure the solute concentration near the SF.
\end{enumerate}

The following assumptions are made regarding the images and the physical system:

\begin{itemize}[noitemsep,topsep=7.5pt,leftmargin=0.1cm]
\setlength{\itemindent}{1.0cm}
    \item When treating the SZ, the noise is considered white Gaussian. This is because after pre-processing (Algorithm \ref{alg:pre-processing}) the image luminance does not vary too much at length scales much greater then the dendrite thickness and inter-dendrite spacing (Figure \ref{fig:example-image}) -- this is in contrast with the LZ (compare the luminance distribution within the dark-blue dashed frame against that above the SF).
    \item Persistent larger-scale artifacts (i.e. not pixels with outlying luminance values) in the images, if present, are considered stationary over time.
    \item Dendrites in the SZ have linear or only slightly curved shapes.
    \item Dendrites may overlap in the imaging plane and thus their X-ray radiography projections may cross.
\end{itemize}

Given the above, the following considerations determine the methods of choice:

\begin{itemize}[noitemsep,topsep=7.5pt,leftmargin=0.1cm]
\setlength{\itemindent}{1.0cm}
    \item Low image resolution means that care must be taken when attempting to remove noise from dendrites. This is because the pixel size is a significant fraction of the dendrite width. At the same time, the textures in the solid domain are rather fine (inter-dendrite spacings are roughly of the same order of magnitude as the dendrite width). The SNR is such that the dendrites are corrupted by noise enough that methods which are not texture-/morphology-aware cannot achieve satisfactory non-destructive denoising.
    \item Liquid flow across the dendrites acts as correlated noise when attempting to derive dendrite morphology. This further complicates the solid structure analysis -- treating this issue jointly with the Gaussian noise stemming from under-exposure does not produce good enough results and a separate approach is needed.
    \item While the CNR at the SF is rather high in the example seen in Figure \ref{fig:example-image}, in other cases the SF is not as smooth and is less contrast. Note that segmenting liquid cavities and channels can be even more difficult. Therefore, dedicated filters are required to significantly increase the CNR of the liquid/solid phase boundary before SZ/LZ segmentation.
    \item The segmentation method for SZ/LZ separation must reliably work under potentially varying image quality: the ray produced by the X-ray tube may flicker and has non-uniform intensity; solid fraction increase and solute ejection into the bulk liquid strongly change the image luminance distribution both locally and globally; these effects should be modelled by the utilized segmentation method.
    \item Methods used for LZ denoising must be such that the shapes of convective plumes are not overly deformed or smeared out, but denoising here is much less constrained than in the SZ.
    \item Image quality varies greatly across different experiments and image sequences, both our own and those performed by other researchers -- it is therefore worthwhile to develop a code that is resilient and can operate under adverse conditions potentially much worse that what is seen in Figure \ref{fig:example-image}.
    \item Such a code with many components and methods will inevitably have a rather large number of parameters -- these should either be mostly fixed/general or should be quickly optimizeable.
\end{itemize}

Image processing is organized in stages outlined in Algorithm \ref{alg:overall-pipeline-structure}. Throughout the paper we will provide the default parameters for the various procedures involved.

\begin{algorithm}

    \nonl \textbf{Input}: Raw image sequence
    
    Pre-process images (Algorithm \ref{alg:pre-processing})
    
    Remove image artifacts (Algorithm \ref{alg:artifact-removal})
    
    Segment the SZ (Section \ref{sec:filtering-before-phase-segmentation} \& Algorithm \ref{alg:mil-based-segmentation})
    
    Identify channels connected to the LZ, and also liquid cavities within the SZ (Algorithm \ref{alg:lz-cavity-channel-front-derivation})
    
    Derive the SF and segment the LZ (Algorithm \ref{alg:lz-cavity-channel-front-derivation})
    
    Extract skeletons of the structures (in this case dendrites) within the SZ (Algorithms \ref{alg:image-partitioning}, \ref{alg:solid-skeleton-segmentation} \& \ref{alg:resolving-unoriented-structures})
    
    Perform orientation analysis for the structures identified within the LZ (Algorithms \ref{alg:detecting-dominant-grains} \& \ref{alg:resolve-grain-ambiguities})
    
    Decompose the SZ structures into sub-domains (grains) by orientation and connectivity (Algorithms \ref{alg:detecting-dominant-grains} \& \ref{alg:resolve-grain-ambiguities})
    
    Measure the solute concentration near the SF (Algorithm \ref{alg:concentration-measurement-above-sf})
    
    Segment the convective plumes within the LZ for shape analysis (Algorithm \ref{alg:convective-plume-segmentation})
    
    \nonl \textbf{Output:} 
    \begin{itemize}[noitemsep,topsep=0pt]
    \popline    \item SF shape, height map and growth rate over time
        \item Solute concentration dynamics near the SF
        \item Shape dynamics for convective plumes in the LZ
        \item Dendrite structure maps with highlighted features
        \item Dendrite orientation spectra for the SZ
        \item Dendrite orientation spectra and relative areas for grains identified within the SZ
    \end{itemize} 

\caption{Overall structure of the image processing pipeline}
\label{alg:overall-pipeline-structure}
\end{algorithm}

\subsection{Pre-processing}

Image pre-processing is performed in \textit{ImageJ} as shown in Algorithm \ref{alg:pre-processing}. For dark current correction, the mean projection of recorded dark current noise images is subtracted from all images in the raw image sequence. Then the dark current-compensated flat field is computed from the mean projection of X-ray beam flux distribution images. Afterwards the dark current-compensated FOV images are normalized with respect to the flat field compensated for the dark current. All of the above can be formulated as follows:

\begin{algorithm}

    \nonl \textbf{Input}: Raw image sequence: $16$-bit 1-channel TIFFs
    
    Crop images to the FOV of interest, in our case typically yielding image resolutions of $\sim 760 \times 380$ pixels
    
    (Optional) Temporal filtering with a Gaussian kernel
    
    Compensate the images for camera dark current (mean)
    
    Compensate the flat field (reference) images for dark current (mean)
    
    Perform flat field correction (FFC, 32-bit precision)
    
    Convert to 16-bit
    
    Remove outlying bright luminance values (median thresholding)
    
    Normalize the resulting images (pixel luminance re-scaled to $[0;1]$)
    
    \nonl \textbf{Output:} Temporally averaged, dark current and flat-field corrected normalized images

\caption{Pre-processing for raw images}
\label{alg:pre-processing}
\end{algorithm}

\begin{equation}
    I = \frac{I_0 - \left< I_\text{dark} \right>}{ \left< I_\text{beam} \right> - \left< I_\text{dark} \right>}
\label{eq:image-correction-dark-beam}    
\end{equation}
where $I$ are the luminance maps of the corrected images, $I_0$ are the cropped raw images, and $I_\text{dark}$ and $I_\text{beam}$ are the dark current and X-ray beam flux images. This transformation, specifically the flat-field correction (FFC) results in the spatial dependence of the SNR.

Afterwards, bright outliers are removed using median thresholding with a 1- to 2-$px$ radius. The threshold is chosen such that outlier removal modifies only the pixels where luminance by far exceeds the local median (within the designated radius). Pixels with luminance above the threshold are then assigned the local median values. Finally, the images are normalized and saved, then passed to \textit{Wolfram Mathematica} for further processing.

\subsection{Liquid/solid zone separation}
\label{sec:liquid-solid-separation}

\subsubsection{Artifact removal}

Sometimes the images will still contain artifacts even after cropping and pre-processing (for example as in Figure \ref{fig:example-image}). While FFC ensures that such artifacts are no longer strong outliers, these image areas still significantly affect image luminance histograms and may interfere with image filtering (especially BM3D, significantly disrupting patch matching) and segmentation. We therefore use a procedure that identifies and inpaints these defects, i.e. makes them seamless with respect to the surrounding image textures. The artifact mask for an image sequence is obtained as outlined in Algorithm \ref{alg:artifact-removal}:

\begin{algorithm}

\nonl \textbf{Input:} \\
    \begin{itemize}[noitemsep,topsep=0pt]
    \popline  \item Averaged reference images (Algorithm \ref{alg:pre-processing})
        \item Pre-processed image sequence (Algorithm \ref{alg:pre-processing})
    \end{itemize}
    
    \nl Artifact mask: binarization (possibly multi-pass) with a user-defined threshold or another \\ appropriate method (e.g. Otsu \cite{otsu-thresholding})
    
    \nl Morphological dilation
    
    \nl Artifact inpainting using texture synthesis
    
    \nonl \textbf{Output:} Images without artifacts

\caption{Artifact removal for pre-processed images}
\label{alg:artifact-removal}
\end{algorithm}

Depending on the nature of artifacts, it may be necessary to perform segmentation in multiple stages. In our cases, a single pass with a user-defined threshold (usually $>0.975$ for bright artifacts as in our case) was sufficient. Note that dark artifacts can be detected using an identical procedure applied to an inverted image. Morphological dilation using disk structural elements \cite{images-mathematical-morphology} is performed so that the artifact mask has a safety buffer. The latter is necessary for texture synthesis-based inpainting to properly fill the artifact zones using samples from adjacent textures sufficiently far from the artifacts. A 5-$px$ element radius is used for structural elements for dilation, and inpainting is performed with a maximum $N_\text{neigh}=150$ neighboring pixels used for texture comparison and a maximum of $N_\text{samp}=300$ sampling instances for texture fitting \cite{wolfram-mathematica-inpaint}.

\subsubsection{Image filtering}
\label{sec:filtering-before-phase-segmentation}

Prior to segmentation, image filtering is performed to increase the CNR for the LZ/SZ boundaries, including liquid cavities (CNR tends to be especially low) and channels. Here the filters were applied such that they also eliminate dendrite structures while preserving larger-scale liquid zones and larger spaces between dendrites that are filled with liquid. It was decided to use block-matching 3D (BM3D) filtering \cite{bm3d-first-og, bm3d-ieee-og, bm3d-ipol-lebrun, bm3d-exact}, since, unlike other tested solutions, it consistently preserved the SZ shape well and also strongly increased the CNR of channels and cavities within the SZ.

We perform image filtering in two stages. First, BM3D is applied. BM3D works by exploiting structural similarity between different patches within an image and filtering the sufficiently similar patches collectively as 3D blocks, and then aggregating the filtered patches at respective positions to construct a filtered image \cite{bm3d-first-og, bm3d-ieee-og, bm3d-ipol-lebrun, bm3d-exact}. Similarity between patches is measured using the $L^2$-norms of the differences between hard-thresholded (based on the assumed noise variance $\sigma_\text{BM3D}$) spectra of their discrete cosine transforms (DCTs) or discrete wavelet transforms (DWTs) with biorthogonal spline wavelets (in general, depends on the implementation and settings) \cite{bm3d-ieee-og}. 3D block collaborative filtering is performed via combined hard-thresholded 2D transform (DCT/DWT) and 1D transform (Haar wavelet DWT) \cite{bm3d-ieee-og}. Importantly, the denoised image estimate produced this way can be used for a second stage of collaborative patch filtering, but now using the Wiener filter -- this significantly improves the output image in terms of feature preservation and noise reduction \cite{bm3d-first-og, bm3d-ieee-og, bm3d-ipol-lebrun}.

\clearpage

We use the latest \href{https://webpages.tuni.fi/foi/GCF-BM3D/index.html#ref_software}{\textit{MATLAB} implementation} by Tampere University of Technology (Python version also available) which is based on \cite{bm3d-exact}. It is integrated into our \textit{Wolfram Mathematica} code using \href{http://matlink.org/documentation/}{\textit{MATLink}} for seamless image processing. This version of BM3D introduces exact transform-domain noise variance calculation, improving patch matching, denoising (including handling different types of correlated noise) and filtered image reconstruction from aggregated filtered patches. In addition, a re-filtering procedure is introduced that prevents the loss of finer details within an image by performing secondary collaborative hard-thresholded DCT and Wiener filtering on intermediate noisy image estimates where initially smoothed out fine features separated (in the image frequency domain) from the filtered noise have been added back in \cite{bm3d-exact}.

For filtering prior to the SF segmentation, we configure BM3D as follows: \textit{"normal profile"} (developer-optimized presets for patch size, patch search neighborhood size, maximum patch count in 3D blocks, etc., for both DWT hard-thresholding and Wiener filtering stages), assuming Gaussian white noise with $\sigma_\text{BM3D} = 0.02$ (re-scaled internally according to the image bit depth) and with re-filtering enabled \cite{bm3d-exact}.

The second filtering stage is the non-local means (NM) filter \cite{non-local-means-filter} which we use to mitigate any artifacts left over and/or produced by BM3D and further increase the CNR for the SF and cavities and channels within the SZ. NM can be viewed as a simplified/precursor version of BM3D where similar patches within search neighborhoods are averaged with weights $w_{ij}$ computed via

\begin{equation}
    w_{ij} =
    \exp{ \left[
    - \text{max} \left( 0, ~
    \frac{1}{k_\sigma^2} \cdot
    \left(
    \frac{E_{ij}^2}{p_\text{n}} - 2  
    \right)
    \right)
    \right]
    }; ~~ \tilde{w}_{ij} = \frac{w_{ij}}{\text{max} (w_{ij})}; ~~ \tilde{w}_{00} = 1
    \label{eq:nonlocal-means-weight}
\end{equation}
where $i$ and $j$ are the patch indices, $E_{ij}$ is the Euclidean distance between patches, $p_\text{n}$ is the noise power factor and $k_\sigma$ is the filtering parameter \cite{non-local-means-filter-new-weight-function}. Here $k_\sigma=0.75$ and $p_\text{n}$ is computed based on the internally estimated noise variance \cite{wolfram-mathematica-nonlocal-means}. The patch size is chosen to be $r_\text{l} = 3~ px$ with the patch search neighborhood size $r_\text{p} = 5 r_\text{l}$. Note that, because of the fast exponential decay of weights in (\ref{eq:nonlocal-means-weight}), large Euclidean distances lead to nearly zero weights, acting as an automatic (smooth) patch similarity threshold (unlike the hard thresholding in BM3D) \cite{non-local-means-filter}. Once the filtering is done, one can proceed with the SZ segmentation.

\subsubsection{Solid zone segmentation}

Despite significant improvements to SNR and CNR due to filtering (Section \ref{sec:filtering-before-phase-segmentation}), segmentation is still a challenge due to generally imperfect reference-based FFC, spatially varying SNR and CNR, X-ray beam fluctuations and strongly varying SFF and solute concentration in the LZ over time, as well as formation, growth/shrinking and disappearance of cavities and channels. The tests show that (at least in our cases) global segmentation methods, even advanced ones, are not able to detect the SZ stably and accurately over an entire image sequence that usually starts with no solid zone and possibly ends with $\text{SFF} \approx 1$. We have therefore opted for an empirical "physics-aware" model that computes an adaptive binarization threshold for the filtered images. The segmentation steps are summarized in Algorithm \ref{alg:mil-based-segmentation}.

\begin{algorithm}

\nonl \textbf{Input:} \\
    \begin{itemize}[noitemsep,topsep=0pt]
    \popline  \item Raw image sequence (Algorithm \ref{alg:pre-processing})
        \item Filtered image sequence (Section \ref{sec:filtering-before-phase-segmentation})
    \end{itemize}
    
    \nl (Optional) Apply Gaussian/median filtering to the raw images
    
    \nl Compute the mean inverse luminance (MIL) for the raw images
    
    \nl (Optional) Filter the MIL time series
    
    \nl Compute the MIL-based adaptive threshold time series (\ref{eq:mil-based-adaptive-threshold})
    
    \nl Segment the SZ from the filtered (i.e. post BM3D and NM) images  using the adaptive threshold
  
    \nonl \textbf{Output:} SZ masks for the entire image sequence

\caption{SZ segmentation}
\label{alg:mil-based-segmentation}
\end{algorithm}

The initial sequence of \textit{raw} images is used, because it is desirable to capture the beam fluctuations as well. Optionally, small-radius median/Gaussian filtering may be applied first to mitigate outliers in the images. Afterwards the mean inverse luminance (MIL) $\left< 1 - I(t) \right>$ is computed for all (normalized) images in a sequence where $t$ is the time/frame index. To avoid over-fitting the adaptive threshold to the MIL time series, it is (optionally) filtered using the Gaussian total variation (TV) filter \cite{total-variation-rof-model}. The adaptive threshold $\tau (t)$ for images is computed from MIL as follows:

\begin{equation}
    \tau (t) = 1 - C_1 \cdot f_1 (t) \cdot f_2^p (t); ~~~~ 
    \underbrace{ 
    f_1 (t) = L_{\text{TV}}  \langle 1 - I(t) \rangle 
    }_{\text{SFF correction}
    }; 
    ~ \underbrace{
    f_2 (t) = (g_1 \circ g_2) ( C_2 + 1 - f_1 (t) )
    }_{\text{LZ saturation correction}}
\label{eq:mil-based-adaptive-threshold}    
\end{equation}
where $g_1 (X) = X/ \min (X)$, $g_2 (X) = X/ \max (X)$, $X(t)$ is time series, $C_1 > 0$, $C_2 \geq 0$, $p \in \mathbb{R}$ and $L_{\text{TV}}$ is the TV filtering applied (if necessary) to the MIL time series.

The adaptive part of the threshold consists of two contributions: $f_1(t)$ and $f_2(t)$, where $f_1(t)$ is tied to the SFF of the FOV -- the greater the SFF, the greater the MIL is because the SZ attenuates the X-ray significantly more intensely then the LZ; $f_2(t)$ is the correction based on $f_1(t)$ that accounts for the fact that, as the SFF increases, the solute is pushed out of the solid zone and the LZ becomes significantly more saturated. Both $f_1(t)$ and $f_2(t)$ also capture the instantaneous global illumination changes that could result from X-ray beam flickering. Moreover, $f_2(t)$ plays a very important role in cases where flow in the LZ rapidly transitions between natural and forced convection regimes (e.g. under the influence of applied magnetic field), since these transitions are accompanied by significant changes in the LZ luminance.

The SZ is segmented by binarizing the inverted filtered images with respective thresholds $\tau (t)$. In the cases considered in this paper, we opted not to perform Gaussian/median pre-filtering for the raw image before MIL computation. TV filtering is performed for the MIL time series using a regularization parameter equal to $0.05$ and restricting the number of TV iterations to $500$ \cite{total-variation-rof-model, wolfram-total-variation}. Parameters $C_1$, $C_2$ and $p$ for $\tau (t)$ can vary rather significantly between cases, so they will be provided for each considered example in Section \ref{sec:results}. However, we observe that $C_1 \in [0; 10] \cdot 10^{-2}$ in all cases. $f_2(t)$ in (\ref{eq:mil-based-adaptive-threshold}) is designed such that the threshold $\tau (t)$ is stricter for greater SFF -- this is because the difference in luminance between the SZ and LZ is much lower in the initial stages of solidification than it is when the cell with the alloy is almost entirely filled with the solid phase. This is due to the fact that much of the SZ is initially permeated by the solute some which is later expelled into the LZ. Corrections $f_1(t)$ and $f_2(t)$ can be completely different for various image sequences, therefore examples will be shown for specific cases in Section \ref{sec:results}, in addition to the strategy for quickly optimizing $C_1$, $C_2$ and $p$. Generally, increasing $C_1$ and $p$, and lowering $C_2$ imposes a stricter threshold on the SZ mask.

\subsubsection{Liquid zone, solidification front, cavity and channel segmentation}
\label{sec:front-cavity-channel-identification}

Next, liquid cavities and channels are identified, the SF is derived and the LZ is segmented -- the steps involved are outlined in Algorithm \ref{alg:lz-cavity-channel-front-derivation}. Steps 1 generates the SZ mask does not contain the areas occupied by artifacts -- this is the mask used for the dendrite structure analysis in Section \ref{sec:solid-domain-analysis}. Then the liquid phase mask is obtained in Step 2. Image padding in Step 3 is done because the cavities can also have a boundary at the bottom and the sides of the FOV. The top boundary is not padded because once the SF has passed the top of the FOV it is impossible to distinguish cavities at the FOV top from channels. This way the boundary component removal leaves only the cavities in the processed masks.

Steps 4 to 9 fill and exclude the detected cavities from the liquid phase mask (Step 4) and, if necessary, one can also remove leftover artifacts (if any) from SZ segmentation (Step 5); the closing transform \cite{images-mathematical-morphology} (Step 6) is performed using disk structuring elements to fill the asperities in the 0- and 1-valued level sets in the mask -- the asperities with length scales below a user-defined disk element size are filled conformally to the nearby level set boundary shape; the filling transform \cite{book-digital-image-processing} (Step 7) completes the channel filling wherever it was imperfect after closing; Steps 8 and 9 detect channels as the differences between the output of Step 7 and the masks after cavity filling while removing small-scale segments that physically do not correspond to channels, i.e. are either too small to be classified as such or simply are artifacts.

To segment the LZ, one adds the artifact segments outside of the SZ to the masks with filled channels and cavities (Steps 10 to 13). The SF is obtained by applying the Canny edge detection \cite{canny-edge-detection} to the LZ mask. The SF is then smoothed with a small-radius Gaussian filter to eliminate noise introduced by the preceding operations, after which Otsu binarization \cite{otsu-thresholding} is applied and the thinning transform \cite{book-digital-image-processing} is performed so that the SF is exactly $1~px$ thick. The physics that can be derived from the sequence of the SF states over time are shown further in Section \ref{sec:results}.

Other chosen parameters are as follows: a 25- to 35-$px$ radius (varies by case) for the structural elements for the closing transform (Step 6), 2-$px$ radius for the Canny kernel (Step 14) and a 3-$px$ radius for the Gaussian kernel (Step 15). Segment size thresholds are different for various image sequences, but mostly vary slightly about $300~ px^2$ for artifact elimination in the liquid phase and $500~ {px}^2$ for channels.

\clearpage

\begin{algorithm}

    \nonl \textbf{Input}: SZ masks for the entire image sequence (Algorithm \ref{alg:mil-based-segmentation})
    
    \nonl \underline{\textit{Get the SZ \& liquid phase masks with artifact areas excluded}}
    
    \pushline SZ: multiply the SZ masks by the inverted artifact mask
    
    Liquid phase: invert the result and multiply by the inverted artifact mask
    
    \nonl \popline \underline{\textit{Segment cavities \& channels}}
    
    \pushline Cavities: apply 1-$px$ image padding (pixel value 0, all image boundaries except for the top) to the output of Step 2, remove border components, then revert the image padding
    
    Subtract the cavity masks from the corresponding liquid phase masks (Step 2) and invert the resulting images
    
    (Optional) Perform segment size thresholding for the resulting masks
    
    Invert the masks, then apply the closing transform
    
    Apply the filling transform and invert the result
    
    From the output of Step 7, subtract the difference between \textit{it} and its segment size-thresholded version
    
    Channels: find the image difference between the output of Steps 4(5) \& 8 and perform segment size thresholding for the result
    
    \nonl \popline \underline{\textit{Derive the SF \& segment the LZ}}
    
    \pushline Apply 1-$px$ image padding (pixel value 0, all image boundaries except for the bottom) to the artifact mask, remove border components, then reverse the image padding
    
    Apply the closing transform
    
    Add the outputs of Steps 7 \& 11
    
    LZ: invert the output of Step 12
    
    SF: perform edge detection for the output of Step 12
    
    \nonl \popline \underline{\textit{SF correction \& smoothing}}
    
    \pushline Apply small-radius Gaussian filtering to the SF masks
    
    Normalize the images and apply Otsu thresholding
    
    Perform morphological thinning

    \nonl \popline \textbf{Output:} 
    \begin{itemize}[noitemsep,topsep=0pt]
    \popline    \item SZ with artifacts excluded
        \item Segmented LZ with artifacts excluded
        \item Cavity masks
        \item Channel masks
        \item SF states
    \end{itemize}
    
\caption{LZ, cavity and channel separation \& SF derivation}
\label{alg:lz-cavity-channel-front-derivation}
\end{algorithm}

\subsection{Solid domain analysis}
\label{sec:solid-domain-analysis}

\subsubsection{Image partitioning \& scan region identification}

Once the SZ has been segmented, one can proceed with the analysis of the solidified structures within the SZ. Since the local textures generally vary strongly throughout the SZ, as does their CNR, it was decided to split the FOV image into a number of partitions, determine which ones contain enough of the SZ for significant analysis (\textit{scan regions}), and perform local filtering and feature extraction. This procedure is detailed in Algorithm \ref{alg:image-partitioning}.

\begin{algorithm}

\nonl \textbf{Input:} \\
    \begin{itemize}[noitemsep,topsep=0pt]
    \popline  \item Pre-processed images (Algorithm \ref{alg:pre-processing})
        \item SZ masks with artifact areas excluded (Algorithm \ref{alg:lz-cavity-channel-front-derivation})
    \end{itemize}
    
    \nl Partition the pre-processed images into grids square patches with side lengths based on image dimensions (\ref{eq:partition-size})
    
    \nl Partition the SZ masks into corresponding square patches; \textit{memoize output}
    
    \nl Assign all patches their position indices; \textit{memoize}
    
    \nl Compute the SFF for the SZ mask patches and assign the values to the respective image patches
    
    \nl Designate the image patches with $\text{SFF} > \varepsilon_\textit{SFF} $ ($\varepsilon_\text{SFF} > 0$, user-defined) as \textit{scan regions}
  
    \nonl \textbf{Output:} Scan regions for further analysis (Algorithm \ref{alg:solid-skeleton-segmentation})

\caption{Image partitioning \& scan region identification}
\label{alg:image-partitioning}
\end{algorithm}

The pre-processed images (Algorithm \ref{alg:pre-processing}) are partitioned into a regular grid of square patches with side lengths given by

\begin{equation}
L =  s \cdot \min{\text{dim}(z)}
\label{eq:partition-size}
\end{equation}
where $z$ is the input image and $s$ is the scaling factor. Partitioning is performed such that the image area coverage is maximized. In general, before proceeding further, one should also check the partition quality by re-assembling the obtained patches into a recovered image $z'$ and comparing $\text{dim}(z')$ versus $\text{dim}(z)$ to ensure, if the partitioning was not pixel-perfect (very rarely), that no more than a few pixel rows/columns were lost. Once the images are partitioned, the SZ masks are partitioned in the same way, i.e. image and mask patches with identical image position indices (assigned to all patches) correspond pixel-to-pixel.

Some patches, especially for the images near the beginning of the sequence, are going to be mostly filled with liquid, and thus are not eligible for solid structure analysis. Identifying which regions to scan for solid structures saves a significant amount of computation time. To determine the patches where significant amounts of the solid phase are present, the SFF is computed for every patch using the SZ mask patches. The patches with the SFF greater than a user-defined threshold $\varepsilon_\text{SFF}$ are designated as \textit{scan regions} and are passed to Alrorithm \ref{alg:solid-skeleton-segmentation} for further analysis. The memoized image position indices for the patches will be used later for the FOV reconstruction.

Tests indicated that the optimal settings for the considered image sequences are $s = 0.1$ and $\varepsilon_\text{SFF} = 0.3$ for all the tested images sequences. This also accounts for the methods discussed further in Sections \ref{sec:dendrite-skeleton-extraction}-\ref{sec:dgd-final-step}.

\subsubsection{Scan region filtering \& dendrite skeleton extraction}
\label{sec:dendrite-skeleton-extraction}

The identified scan regions are processed as indicated in Algorithm \ref{alg:solid-skeleton-segmentation}.

\begin{algorithm}

\nonl \textbf{Input:} scan region images (Algorithm \ref{alg:image-partitioning})

    \nonl \underline{\textit{Prepare filter input}}
    
    \pushline Re-scale the images and perform color tone mapping (CTM)
    
    Re-scale the images again and perform FFC
    
    \nonl \popline \underline{\textit{Filter the images}}
    
    \pushline Apply BM3D filtering
    
    Perform two iterations of non-local means masking (NMM)
    
    Apply soft color tone map masking (SCTMM)
    
    Apply FFC
  
    \nonl \popline \underline{\textit{Extract dendrite skeletons}}
    
    \pushline Segment dendrites using 2-threshold hysteresis binarization
    
    Invert the mask
    
    Apply the thinning transform
    
    Perform size thresholding
  
    Multiply the resulting masks with their corresponding SZ mask patches (Algorithm \ref{alg:image-partitioning}, Step 2)
  
    Perform morphological pruning (optionally in multiple passes)
    
    Remove border pixels
    
    Perform size thresholding
  
    \nonl \popline \textbf{Output:} Dendrite skeleton masks

\caption{Scan region filtering \& dendrite skeleton extraction}
\label{alg:solid-skeleton-segmentation}
\end{algorithm}

Here the strategy is to use BM3D to denoise the dendrite textures as non-destructively as possible. However, we have found that in general it can be difficult to obtain good results without preparing the images first. This is why the first stage is image normalization, color tone mapping (CTM) and FFC. The $\text{CTM}(x,c)$ operation maps the colors (in this case the gray-scale values) of the input image $x$ image using gamma compression with a global compression factor $c$ \cite{reproduction-of-color-chapter-6} and thus compresses the dynamic range -- this has the effect of dramatically increasing the CNR of the dendrites. Meanwhile FFC uses a coarse polynomial fit of the image luminance map to perform background correction (flattening) without reference \cite{wolfram-brightness-equalize}. In some cases this helps to reduce the large-wavelength correlated noise due to liquid metal flow across the dendrites. We use $c = 0.5$ for CTM and second-order FFC polynomials.

Next the scan regions are filtered. One first applies BM3D to restore the dendrite textures in images, then two iterations of non-local means masking (NMM) are applied to mitigate any leftover correlated noise and increase dendrite CNR. NMM could be viewed as a generalized, locally adaptive version of unsharp masking -- it takes the input image $x$ and transforms it into an output image $y$ as follows:

\begin{equation}
y = 2*x - w_\text{nm} * \text{NM}(x,r_\text{l},r_\text{p})
\label{eq:nm-correction}
\end{equation}
where $\text{NM}(x,r_\text{l},r_\text{p})$ is the NM filter, $w_\text{nm}$ is the NM mask weight, and $r_\text{l}$ and $r_\text{p}$ are explained in Section \ref{sec:filtering-before-phase-segmentation} where the NM filter weights are given by (\ref{eq:nonlocal-means-weight}). We have previously applied NMM to particle detection in neutron radiography images of particle-laden liquid metal flow for a similar purpose, with good results \cite{birjukovs-particle-EXIF, birjukovs-particle-track-curvature-stats}. Afterwards the soft color tone map masking (SCTMM) is applied for further background reduction and CNR enhancement. SCTMM transforms an original normalized image $x$ to output $y$ in the following way:

\begin{equation}
y = x * 
\left( 
x - \left( 1 - \text{CTM} (x,c) \right) 
\right)
\label{eq:ctm-correction}
\end{equation}
The motivation and principles behind SCTMM are explained in detail in \cite{birjukovs-neutrons-bubbles-latest} and the applications in neutron imaging of bubble and particle flow in liquid metal are demonstrated in \cite{birjukovs-neutrons-bubbles-latest, birjukovs-particle-EXIF}. Finally, another FFC iteration is performed. For this stage, the BM3D filter is configured as in Section \ref{sec:filtering-before-phase-segmentation} except that $\sigma_\text{BM3D} = 0.035$. For NMM, we use $r_\text{l} = 1~px$, $r_\text{p} = 5 r_l$, $w_\text{nm} = 1.25$ and $p_\text{n} = 0.05$. SCTMM is assigned $c = 0.75$.

Dendrite skeleton extraction is done in eight steps. One starts with double-Otsu hysteresis binarization \cite{book-digital-image-processing}, followed by mask inversion, morphological thinning and size thresholding (8-connectivity). Size threshold for the dendrites is set to $10~{px}^2$. Then the resulting skeleton masks are multiplied with their respective SZ mask patches which crops the skeleton parts that are within the liquid phase areas and thus cannot actually be dendrites. Afterwards the remaining skeleton asperities are removed with multi-pass morphological pruning (in our case 3 passes, each pruning branches of at most $1~px$ length -- a balance between non-destructiveness and effectiveness) \cite{book-digital-image-processing}, then border pixels are removed and size thresholding is done again (this time with a $5~{px}^2$ threshold, also using 8-connectivity).

\subsubsection{Resolving unoriented \& overlapping structures}

After dendrite detection, the skeleton segment orientations ($\varphi$, with respect to the image $X$ axis) must be measured. However, it may be the case that the image filters do not properly resolve primary/secondary dendrites or cases with dendrite overlaps. In addition, some segments may not have clearly defined orientations, or could simply be leftover artifacts. To identify and correct such \textit{unresolved} (in the sense of orientation) structures within all IWs, a procedure outlined in Algorithm \ref{alg:resolving-unoriented-structures} is performed for all IWs.

\begin{algorithm}

\nonl \textbf{Input:} IWs with dendrite skeletons (Algorithm \ref{alg:solid-skeleton-segmentation})

    \nonl \underline{\textit{Identify unresolved skeletons}}
    
    \pushline Compute orientation angles ($\varphi$, with respect to the image $X$ axis) and aspect ratio ($\rchi$) for dendrite segments
    
    Colorize resolved (oriented) segments ($\rchi > \rchi_c$, $\rchi_c \geq 1$ is user-defined) by their $\varphi$
    
    Separate resolved and unresolved segments into different masks
    
    \nonl \popline \underline{\textit{Resolve the skeletons (for masks with unresolved segments)}}
    
    \pushline Detect skeleton corner points
    
    Detect skeleton branch points
    
    Threshold and filter corner and branch points
    
    Remove the remaining corner and branch points from skeletons
  
    \nonl \popline \underline{\textit{Reassemble IW skeletons}}
    
    \pushline Re-evaluate segment orientations for masks with (formerly) unresolved segments (Step 7)
    
    Colorize resolved (oriented) segments ($\rchi > \rchi_c$) by their $\varphi$
    
    Add the output of Step 9 to the previously set aside masks with initially resolved segments (Step 3)
  
    \nonl \popline \textbf{Output:} IWs with resolved dendrite skeletons colorized by their orientations

\caption{Resolve unoriented skeletons}
\label{alg:resolving-unoriented-structures}
\end{algorithm}

Algorithm \ref{alg:resolving-unoriented-structures} uses the aspect ratio $\rchi$ as a criterion to determine if the skeleton segments have a resolved (i.e. well-defined) orientation. Since morphological thinning is one of the steps in Algorithm \ref{alg:solid-skeleton-segmentation} and Algorithm \ref{alg:resolving-unoriented-structures} does not add new pixels to masks, most of the skeleton lines should have a $1~px$ thickness and, provided they are long enough, therefore also high $\rchi$. We found that $\rchi_c = 5$ yields good results. With this setting both very small segments, as well as large skeletons with unresolved overlapping branches, will have low $\rchi$ and will be passed for further processing. Colorizing the segments by $\varphi$ will play a key role later during dendrite grain decomposition (Section \ref{sec:grain-decomposition}), but in Step 3 of algorithm \ref{alg:resolving-unoriented-structures} it is used to separate the masks with initially resolved and unresolved segments. The choice of the color scale does not matter as long as it is normalized (see Section \ref{sec:grain-decomposition}) and the palette is visually convenient for the user. Note that to determine segment $\varphi$ and $\rchi$ we use best-fit ellipses, since we found this procedure to be more robust and accurate than line detection with the Hough transform or the RANSAC method, even for isolated and clearly-oriented $1~px$-thick segments.

Once the masks with unresolved skeletons are separated, corner point detection is performed. We use the Harris-Stevens method \cite{harris-stephens-og, harris-stephens-ipol} with first-order Gaussian derivatives and set the corner detection range to $0.5~ px$. The corner detection threshold is set to $4.5 \cdot 10^{-6}$ with a minimum corner distance of $0~px$. Then morphological branch points are detected \cite{wolfram-mathematica-branch-points}. Afterwards the detected corner points are filtered by selecting corner pixel clusters with pixel count $< 3$ (using 8-connectivity), combined with the detected branch points, and then pixel clusters with $\leq 2$ pixels (8-connectivity) are removed from the resulting mask. Finally, the size-thresholded combined mask is subtracted from the input mask with unresolved skeletons.

With this done, the resolved skeletons are now assigned colors based on $\varphi$ subject to the $\rchi_c = 5$ criterion, and the resulting masks are recombined with those containing initially resolved skeletons.

\subsubsection{Assembling the global dendrite skeleton}
\label{sec:global-skeleton-reassembly}

At this point all the remaining unresolved (white-colored) skeletons within IWs are considered unoriented dendrites and/or artifacts. These are excluded from any subsequent analysis. One can now reassemble the global (FOV) dendrite skeleton image by tiling the IWs according to their position indices from Step 3 of Algorithm \ref{alg:image-partitioning}. It is also now easy to generate maps with color-coded dendrite orientations with highlighted cavities and liquid/solid boundaries.

\subsection{Dendrite grain decomposition}
\label{sec:grain-decomposition}

Before proceeding with decomposition of the resulting global dendrite skeleton into grains, an \textit{global} orientation ($\varphi$) spectrum must computed for the assembled skeleton. It is not only of physical interest, but will be used in Algorithm \ref{alg:detecting-dominant-grains} as well. While a global $\varphi$ spectrum is certainly relevant, it is often desirable to distinguish areas of "coherent" dendrite growth, i.e. \textit{dendrite grains} with their areas and mean dendrite $\varphi$. The presented \textit{dendrite grain decomposition} (GDG) method does this by considering $\varphi$ similarity and proximity of the dendrites detected within the FOV, and it does so by exploiting the color-space representation of $\varphi$ generated by Algorithm \ref{alg:resolving-unoriented-structures}.

\subsubsection{Detecting dominant dendrite grains}

GDG is performed in three stages: a primary scan which detects dominant dendrite grains; a refined scan which checks if the larger grains should be subdivided further and if the smaller grains are eligible; a filtering step which resolves ambiguities and overlaps between the detected dendrite grains. The first step of the GDG procedure is outlined in Algorithm \ref{alg:detecting-dominant-grains}.

\begin{algorithm}

\nonl \textbf{Input:} Assembled global dendrite skeleton (Section \ref{sec:global-skeleton-reassembly})

    \pushline Compute the global $\varphi$ spectrum from the global skeleton
    
    Filter the $\varphi$ spectrum \& detect dominant peaks
    
    Find dendrite segments near the $\varphi$ peaks in the color-space
    
    Build grain masks that cover the dendrite segments
  
    \nonl \popline \textbf{Output:} Separate masks for dominant dendrite grains

\caption{Detecting dominant dendrite grains}
\label{alg:detecting-dominant-grains}
\end{algorithm}

The $\varphi$ spectrum is computed by measuring dendrite segment $\varphi$ (this time without the $\rchi > \rchi_c$ constraint) and lengths, and then constructing a histogram with uniformly-sized bins with values $\rho \in [0;1]$ weighed by the dendrite lengths. Length weights are used to account for boundary pixel removal in IWs which may break up longer dendrites into fragments. The $\varphi$ bin values are then filtered using the total variation (TV) filter followed by the mean filter -- this is to remove outliers and insignificant peaks from the spectrum. Then the peaks are detected and the ones with $\rho > \rho_\text{c}$ are kept. Here the regularization parameter for the TV filter is $0.15$, the mean filter radius is $2$, the $\varphi$ spectrum resolution is $200$ bins and $\rho_\text{c} = 0.12$.

Since there is a mapping between $\varphi \in (-\pi/2;\pi/2]$ and the dendrite skeleton color values (normalized) due to Algorithm \ref{alg:resolving-unoriented-structures}, one can now find the segments in the global skeleton that correspond to the selected $\varphi$ peaks. To do this, peak $\varphi$ values are converted to coordinates $r_\varphi$ in the \textit{CIELAB} (\textit{CIE76}) color-space and the segments with Euclidean distance within $\delta r_\text{LAB}$ from $r_\varphi$ are selected. The selected segments are further filtered by assigning them weights $w_\text{LAB} \in [0;1]$ based on their distance from $r_\varphi$ (farthest to closest) and keeping segments with $w_\text{LAB} > w_\text{c}$. This makes the process more resilient to noise in the $\varphi$ spectrum and helps to avoid grain mask overlaps later. Sometimes, however, there may be groups of two or more very close peaks $r_\varphi$ that survive the thresholding by $\rho > \rho_\text{c}$. In these cases we replace such groups of peaks with mutual Euclidean \textit{CIELAB} distances  $< \delta r_\varphi$ by their mean $r_\varphi$ values. We found that it is optimal for most cases to set $\delta r_\text{LAB} = 0.05$, $w_\text{c} = 0.5$ and $\delta r_\varphi = 2 \cdot \delta r_\text{LAB}$. Another issue that might come up is that $\varphi \in (-\pi/2;\pi/2]$ and the $\varphi$ spectrum does not have periodic boundary conditions. This means that if there were an \textit{actual} peak of some width for dendrite orientations near $\varphi = \pm \pi/2$, it would be treated as two $\varphi$ peaks by Algorithm \ref{alg:detecting-dominant-grains}. This problem can be solved by checking if the peaks closest to the $(-\pi/2;\pi/2]$ boundaries (\textit{edge peaks}) are close enough to these boundaries and to one another \textit{across} the $\varphi = \pm \pi/2$ boundary. In cases where two or more peaks are detected, the edge peaks $\varphi_1$ and $\varphi_2$ are subjected to constraints

\begin{equation}
    \min{ \left( \pi/2 - \abs{ \varphi_{1,2} } \right) } \leq \delta_1; ~~~  
    \abs{ \abs{\varphi_1} - \abs{\varphi_2} } \leq \delta_2
\label{eq:edge-peak-constraints}
\end{equation}
and, if both are satisfied, the respective masks with dendrites within the color-space peak ranges are added before color-space proximity thresholding, effectively treating the edge peaks as one. By default we use $\delta_1 = \delta _2 = 5$ degrees.

When the dendrite clusters corresponding to each $r_\varphi$ peak are found, a mask must be created for them that will delimit and separate them as one grain. This is done by applying the closing transform with disk structuring elements to the skeleton clusters, which fills the spaces between the dendrite skeletons while not affecting the outlying parts of the skeletons, i.e. preserving the shapes of the dendrites that protrude from the bulk of the cluster. Note that this may generate more than one grain mask per $r_\varphi$ peak since dendrite clusters with very similar orientations may be sufficiently far apart. Thus, grains are identified accounting for both dendrite orientations and spatial distribution. Afterwards the resulting grain-covering binary masks are thresholded by their area, and the remaining masks are separated for further analysis with a refined scan. Here we use structuring elements with a $5~px$ radius for closing and the minimum grain area is set to $S_\text{min} = 2000~{px}^2$.

\subsubsection{Refined dendrite grain scan}
\label{sec:refined-grain-scan}

Once dominant grains are identified for each $r_\varphi$ peak, they are subjected to a secondary scan that is designed to check whether the originally recognized grains need to be further subdivided. This is done to both (implicitly) ensure grain uniqueness, minimize overlaps, and resolve smaller areas withing the larger grains that have distinct enough orientations. The scan follows steps similar to those of Algorithm \ref{alg:detecting-dominant-grains}, but with the following modifications:

\begin{enumerate}[noitemsep,topsep=7.5pt,leftmargin=0.75cm]
    \item Steps 1 and 2 are now applied to the dendrites within the grain masks, not the global skeleton.
    \item Area-adaptive $\varphi$ spectrum resolution is used.
\end{enumerate}

Prior to $\varphi$ spectra calculation, the dendrite skeletons belonging to the grains are isolated by multiplying the grain masks by the global dendrite skeleton. The area-adaptive resolution is set up such that, on the one hand, the algorithm can resolve finer differences in orientations within the initial grains and detect the underlying $\varphi$ peaks, while on the other hand not using exceedingly large resolution for smaller grains with relatively few dendrites. In the latter case the algorithm would otherwise treat the noise in the spectrum as significant peaks despite the filtering. The adaptive spectrum bin count $N_\varphi$ (integer) is given by

\begin{equation}
\label{eq:adaptive-angle-bin-count}
    N_\varphi = \min \left( N_\text{min}, \nint{ N_0 \cdot \frac{S}{S_0} } \right)
\end{equation}
where $N_\text{min}$ is the bin count lower bound, $N_0$ is the baseline bin count, $S$ is the grain area and $S_0$ is the area of the entire SZ. Note that before computing $S$ the filling transform is applied to the grain masks -- this is necessary because the grain masks conform to the dendrite skeleton and may have holes. Here we alter $\rho_\text{c} = 0.2$ and $w_\text{c} = 0.75$ and set $N_0 = 200$, $N_\text{min}=3$. The other relevant parameters as in Algorithm \ref{alg:detecting-dominant-grains}.

Note that in some cases the refined scan may eliminate a grain instead of simply keeping or subdividing it. The former can happen if the areas of the resulting resolved grains are below the threshold, although such cases should be quite rare.

\subsubsection{Resolving ambiguities \& performing cleanup}
\label{sec:dgd-final-step}

When the initial grains have been scanned again and kept or decomposed further and/or eliminated, the final DGD step is performed -- ambiguities and overlaps between the grains are resolved. Two cases must be treated here: the previous stages of DGD have, in separate instances, generated two grains with almost identical (i.e. overlapping) masks and this is indeed one and the same grain; there is partial overlap between the grains, but it is physical since the grains are adjacent and the dendrite orientation changes very slowly from one grain to the other, i.e. there exists a transition zone instead of a sharp boundary. In the former case one of the masks is redundant, and in the latter the overlap zone must be identified and designated as such, since no clear distinction between the two grains can be made in the transition zone. Finally, there is also the matter of potentially leftover dendrites with $\varphi$ values that are significant outliers with respect to the mean $\varphi$ for the grains. These issues are addressed by Algorithm \ref{alg:resolve-grain-ambiguities}.

\begin{algorithm}

\nonl \textbf{Input:} Dendrite grains output by the refined scan (Section \ref{sec:refined-grain-scan})

    \pushline Compute the overlap masks for dendrite grain pairs
    
    Compute the \textit{uniqueness factors} $u \in [0;1]$ for the grain pairs
    
    Discard redundant grains ($u < u_\text{c}$, $u_\text{c}$ is user-defined)
    
    Designate the overlap masks for the pairs of remaining unique grains as \textit{overlap zones} and subtract them from the grain masks
    
    Perform total area thresholding for the resulting masks, then secondary size thresholding for the underlying segments -- the output gives the final grain masks
    
    Multiply the global skeleton by the final grain masks to isolate the respective dendrites
    
    Remove dendrites with outlying $\varphi$ values and perform dendrite size thresholding
  
    \nonl \popline \textbf{Output:} final dendrite grain masks \& skeletons

\caption{Resolving ambiguities \& performing grain cleanup}
\label{alg:resolve-grain-ambiguities}
\end{algorithm}

The first step is performed by multiplying all possible pairs of dendrite grain masks. Then the uniqueness factors defined as $u = 1 - S_\cap/\langle S \rangle$ are computed for all dendrite grain pairs, where $S_\cap$ is the overlap mask area and $\langle S \rangle$ is the mean grain area for the pair. Pairs with $u < u_c$ are considered redundant and only one grain mask from such pairs is kept. The overlap masks for all the other grains are kept as overlap zones and subtracted from the unique masks. The remaining segments are then thresholded both by total and individual areas. Afterwards, the resulting masks are multiplied by the local skeleton to isolate the grain dendrites. Finally, the dendrite skeletons are filtered by orientation and length: $\langle \varphi \rangle$ is measured for the grain dendrites and segments outside of the $\langle \varphi \rangle \pm 5 \sigma$ interval (by default) are eliminated, followed by length thresholding. This concludes the DGD process.

Here we use $S_\text{min} = 2000~{px}^2$ for the total grain fragment area and $S_\text{min}' = \nint{ S_\text{min}/5 }$ for grain segment area thresholding, a $2~px$ length threshold for dendrites, and set $u_c = 0.05$.

\subsection{Liquid domain analysis}

\subsubsection{Measuring solute concentration above the solidification front}

Measuring the solute concentration above the SF involves the following considerations:

\begin{itemize}[noitemsep,topsep=7.5pt,leftmargin=0.75cm]
    \item Filtering the noise in the liquid region above the SF as non-destructively as possible, i.e. not to alter the luminance field too much, as it can later be used with the Beer-Lambert law to assess the concentration of the solute.
    \item LZ/SZ segmentation will never be perfect and dendrite tips could be slightly above the SF, yielding errors in the luminance/concentration measurements -- this must be mitigated.
\end{itemize}
Both are addressed by Algorithm \ref{alg:concentration-measurement-above-sf}.

\begin{algorithm}

\nonl \textbf{Input:} \\
    \begin{itemize}[noitemsep,topsep=0pt]
    \popline  \item Pre-processed FOV images without artifacts (Algorithm \ref{alg:artifact-removal})
        \item Solidification front masks (Algorithm \ref{alg:lz-cavity-channel-front-derivation})
    \end{itemize}
    
    \nl Apply median filtering to the FOV images
    
    \nl Apply the bilateral filter
    
    \nl Perform NM filtering
    
    \nl Define the SF-conformal buffer zone by shifting the SF contour mask upwards
    
    \nl Define the concentration sampling zone by extruding the SF-shaped boundary upwards from the buffer zone upper boundary
    
    \nl Compute the mean luminance for vertical pixel bands within the sampling zone over the FOV width
    
    \nl (Optional) Use the Beer-Lambert law to convert the luminance to the solute concentration
  
    \nonl \textbf{Output:} Mean luminance/concentration above the SF over the FOV width for all time stamps

\caption{Measuring solute concentration above the solidification front}
\label{alg:concentration-measurement-above-sf}
\end{algorithm}

We find that the combination of median, bilateral \cite{bilateral-filter-og} and NM (in this order) filters in Steps 1-3, after some parameter tuning, yields a sufficiently well-filtered luminance/concentration field without offsetting the values too much or significantly affecting the larger-wavelength features. The bilateral filter is chosen in particular because of its luminance value range filter component, since with the right settings it should preserve luminance level sets within the images that are fairly close. Once the images are processed, sampling zones are defined for every frame. To address the above mentioned issue with dendrite tips possibly being above the SF, a buffer zone is created where no sampling occurs -- this is done by shifting the SF mask (curve) upwards by a distance $d_\text{buf}$. Starting from this level, the SF curve is then extruded over a distance $d_\text{samp}$ to create the SF-conformal sampling zone for the concentration measurements from the filtered images. Here the mean values are computed for $1~px$-wide vertical strips of $d_\text{samp}$ pixels above the buffer zone over the width of the FOV.

In our cases the default settings which work well for all examples herein are a $1$-$px$ radius for the median filter kernel, $\mu_\text{b} = 3$ pixel value range factor and Gaussian kernel scale $\sigma_\text{b} = 21$ for the bilateral filter \cite{wolfram-mathematica-bilateral-filter}, and for the NM filter we set $r_\text{l} = 1~ px$ with $r_\text{p} = 5 r_\text{l}$. For the sampling zone generation, we set $d_\text{buf} = 15$ and $d_\text{samp} = 30$.

\subsubsection{Convective plume segmentation}

With convective plume segmentation, the luminance value preservation is not as serious a concern as long as the shapes are preserved. Here the idea is to filter the images and then decompose the resulting filtered LZ luminance map into $N_\text{level}$ level sets (clusters) ranked by luminance. One can then, depending on the case, re-assemble $N_\text{plume}$ level sets with the highest luminance back together to obtain the masks for the convective plumes above the SF. This is done via Algorithm \ref{alg:convective-plume-segmentation}.

\begin{algorithm}

\nonl \textbf{Input:} \\
    \begin{itemize}[noitemsep,topsep=0pt]
    \popline  \item Pre-processed FOV images without artifacts (Algorithm \ref{alg:artifact-removal})
        \item LZ masks with filled channels and cavities (Algorithm \ref{alg:lz-cavity-channel-front-derivation})
    \end{itemize}
    
    \nl Perform Steps 1-3 from Algorithm \ref{alg:concentration-measurement-above-sf} (with different parameters)
    
    \nl Gaussian filtering
    
    \nl NM filtering
    
    \nl Create a buffer zone above the LZ boundary (SF) in the LZ mask with filled cavities and channels using morphological dilation
    
    \nl Multiply the filtered image by the resulting mask
    
    \nl Decompose the resulting image into level sets
    
    \nl Re-assemble selected level sets to get convective plume masks 
  
    \nonl \textbf{Output:} Mean luminance/concentration above the SF over the FOV width for all time stamps

\caption{Convective plume segmentation}
\label{alg:convective-plume-segmentation}
\end{algorithm}

Here the median filter radius is $1~px$ and $\mu_\text{b} = 3$ with $\sigma_\text{b} = 21$ as before, but the NM filter parameters are now $r_\text{l} = 2~ px$ with $r_\text{p} = 5 r_\text{l}$, respectively. We have noticed that segmentation yields better results when, after the previous three steps, Gaussian filtering with a $\sigma_\text{g} = 3$ kernel and then NM filtering with $r_\text{l} = 1$ or $2~ px$ and $r_\text{p} = 15$ are applied.

To isolate the liquid region within the FOV, the image is multiplied by the LZ mask. However, before multiplication, a buffer zone is created above the LZ mask by morphological dilation using disk structuring elements with a radius of $15~px$. The rationale here is the same as in the case with the concentration measurements above the SF -- to eliminate the dendrite tips that may have been imperfectly segmented. If not removed, these would form "parasitic" level sets and reduce the actual level set count within the LZ.

The luminance (concentration) level sets for the convective plumes are obtained using the K-medoids method \cite{k-medoids-clustering} with $N_\text{level}$ medoids -- the level sets are then ranked by their mean luminance (with the excluded solid and buffer region having the lowest, zero luminance after mask multiplication) and the top $N_\text{plume}$ are re-assembled into the output mask for the convective plumes. The two involved parameters may vary notably for different image sequences, but we found that setting $N_\text{level} \in [5;9]$ and $N_\text{plume} \in [1;3]$ usually yields good results.

\clearpage

\section{Application examples \& performance}
\label{sec:results}

To demonstrate the resilience of the proposed solution, it was decided to test the developed methodology and code for conditions that are purposefully made worse than one would expect. Specifically, he have opted to use only a \textit{single} frame for FFC in Algorithm \ref{alg:pre-processing} for an image sequence. This way the background features, i.e. setup elements caught within the FOV, artifacts, and the X-ray beam profile are still compensated for, but the resulting SNR is significantly lower.

\subsection{Image filtering \& segmentation}

\begin{figure}[htbp]
\centering
\includegraphics[width=0.85\textwidth]{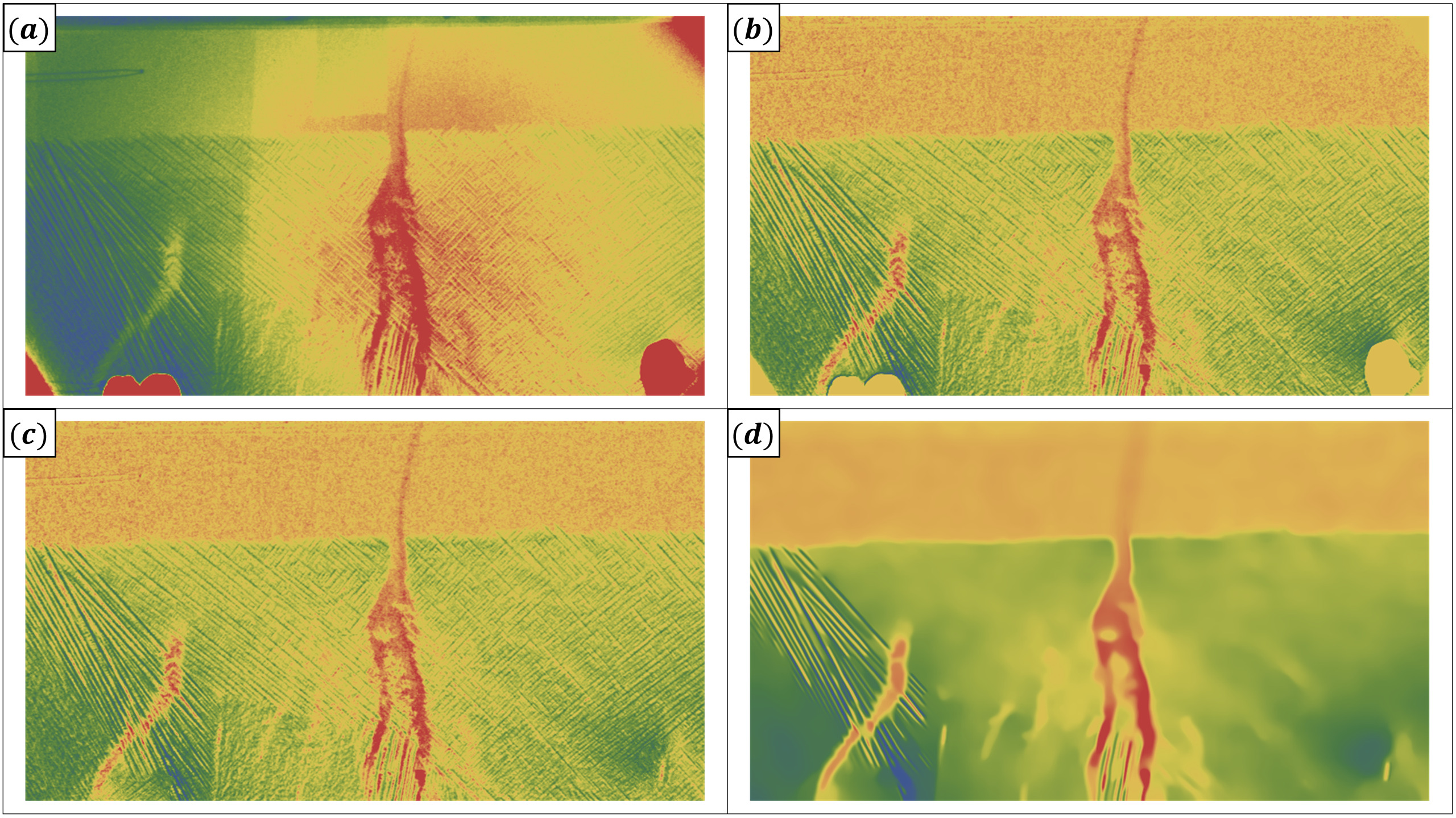}
\caption{FOV image processing: (a) raw image, (b) image pre-processed as in Algorithm \ref{alg:pre-processing}, (c) image with artifacts (note the image corners in (a)) inpainted via Algorithm \ref{alg:artifact-removal} and (d) final image after BM3D and NM filtering (Section \ref{sec:filtering-before-phase-segmentation}). The color scheme here and further, unless stated otherwise, is identical to Figure \ref{fig:example-image}. Note the enhanced contrast of the SF, liquid channels and cavities.}
\label{fig:fov-image-processing-1}
\end{figure}

\begin{figure}[H]
\centering
\includegraphics[width=0.85\textwidth]{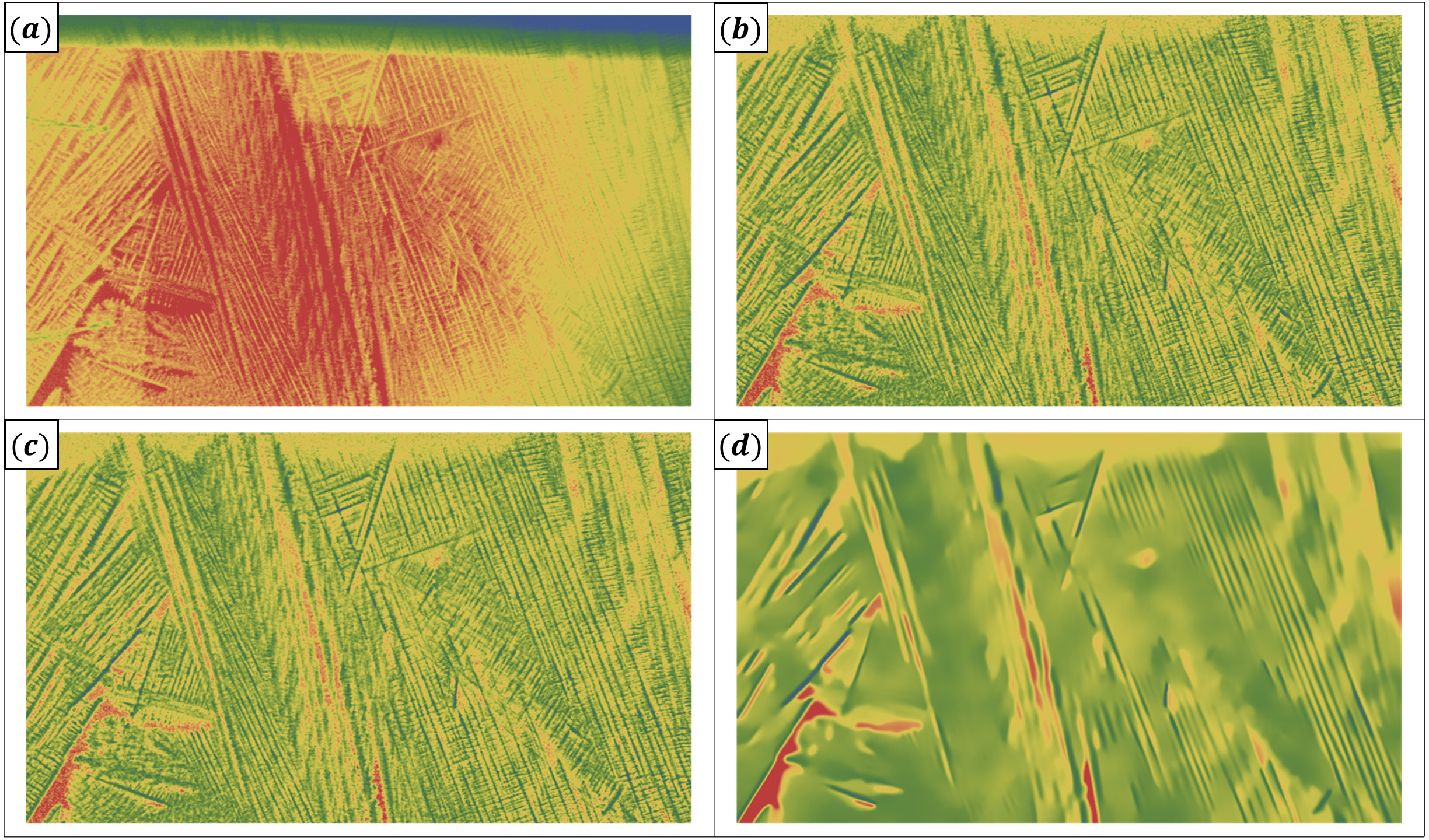}
\caption{Another example of FOV image processing with stages as outlined for Figure \ref{fig:fov-image-processing-1}, here with more complex diverse dendrite orientations and greater solid fill factor.}
\label{fig:fov-image-processing-2}
\end{figure}

\begin{figure}[H]
\centering
\includegraphics[width=0.85\textwidth]{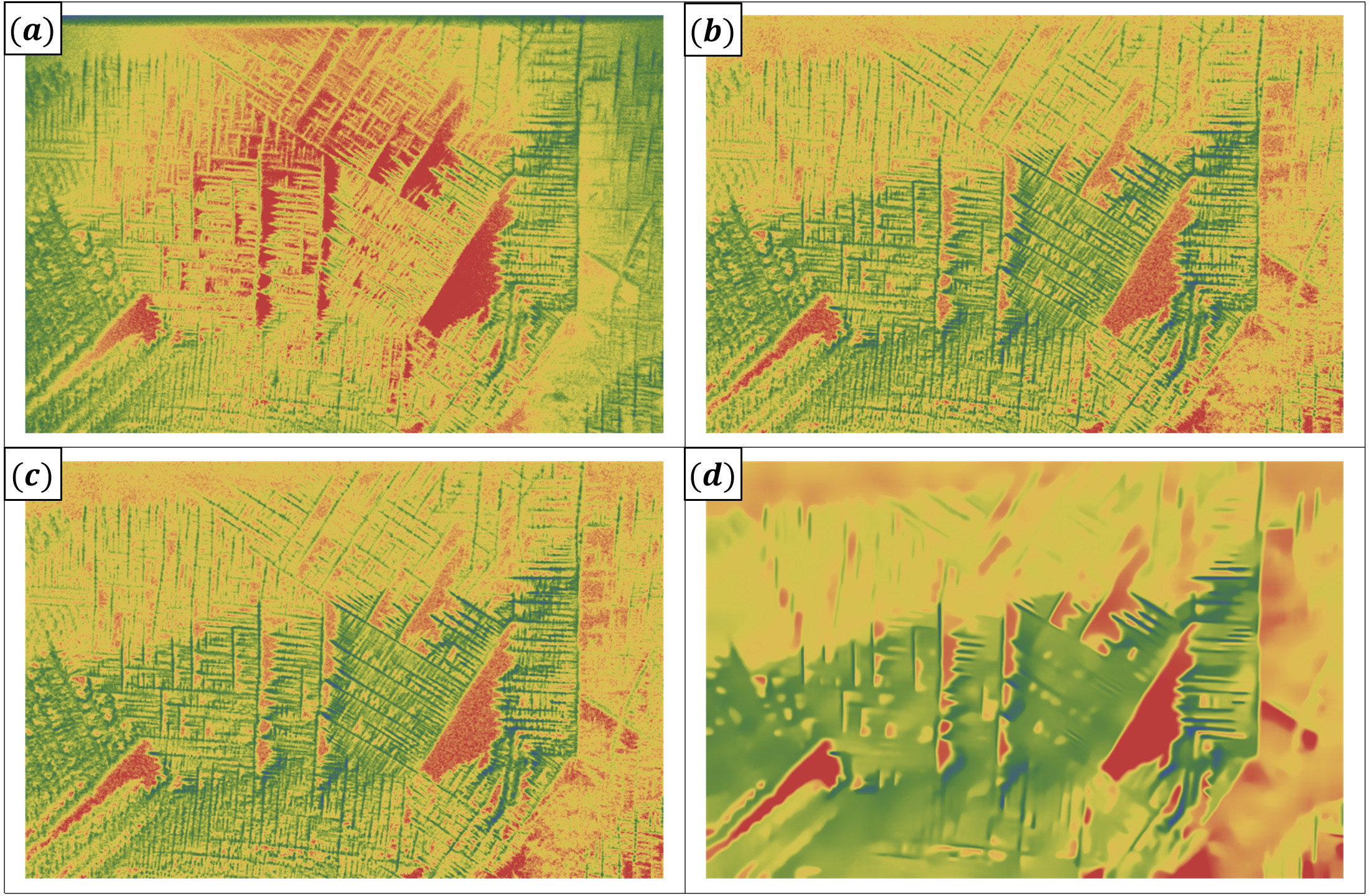}
\caption{Another example of FOV image processing (stages referenced in Figure \ref{fig:fov-image-processing-1}). The less transparent (greener) structures seen in (b) in the middle of the FOV were formed under externally induced forced convection, and the upper, more transparent and less contrasting part of the dendrite structures developed after abruptly switching forced convection off and leading back to the regime of natural convection (more details in \cite{solidification-xray-imaging-melt-convection-natalia}).}
\label{fig:fov-image-processing-3}
\end{figure}

First consider the stages of FOV processing prior to SZ and LZ segmentation. Figures \ref{fig:fov-image-processing-1}-\ref{fig:fov-image-processing-3} show characteristic examples of input images versus the pre-processed and then filtered output. Figure \ref{fig:fov-image-processing-1} is a good example where many of the image features that can realistically be expected are present. Artifacts appear as overexposed corners and areas at the bottom image boundary in (a). A thermocouple is located in the upper-left corner of (a), visibly protruding from the left image boundary). Note also the tape for affixing the thermocouples, visible as rectangular areas with greater opacity at the left and upper boundaries of (a). Liquid channels and cavities are present in the lower part of the FOV, better visible in (b). One can observe that even with single frame FFC, the luminance distribution in (a) about the X-ray beam axis is almost entirely negated in (b), and so are the elements attached to the imaged liquid metal cell. However, as opposed to (a), noise is amplified in (b) and becomes visibly coarser-grained. Afterwards, the artifacts are removed and the result is shown in (c) -- the formed artifact areas should no longer affect the NM and BM3D filter patch matching procedures. Finally, BM3D and NM are applied, in that order, resulting in an output seen in (d) with much more contrast LZ/SZ boundaries, as well as cleaner liquid metal cavities and channels. While the liquid metal plume in the upper part of (b) has been dramatically diffused, it is of no concern here as the result of this image filtering routine is used only for LZ/SZ separation (filters in Algorithm \ref{alg:concentration-measurement-above-sf} are used for concentration measurements in the LZ instead).

Figure \ref{fig:fov-image-processing-2} is an example where the SFF is $\sim 1$ -- again, significant liquid metal cavities are clearly resolved, even the narrow ones forming between dendrites. Note that the finer solid structure features are smoothed out with the given filter settings. A more complicated case is shown in Figure \ref{fig:fov-image-processing-3} where the boundary between liquid and solid appears to be much less clear in the upper part of the image, which is remedied by the filtering. The goodness of filtering, however, is only clear in the context of the objective, which is to segment the SZ/LZ.

\begin{figure}[H]
\centering
\includegraphics[width=0.85\textwidth]{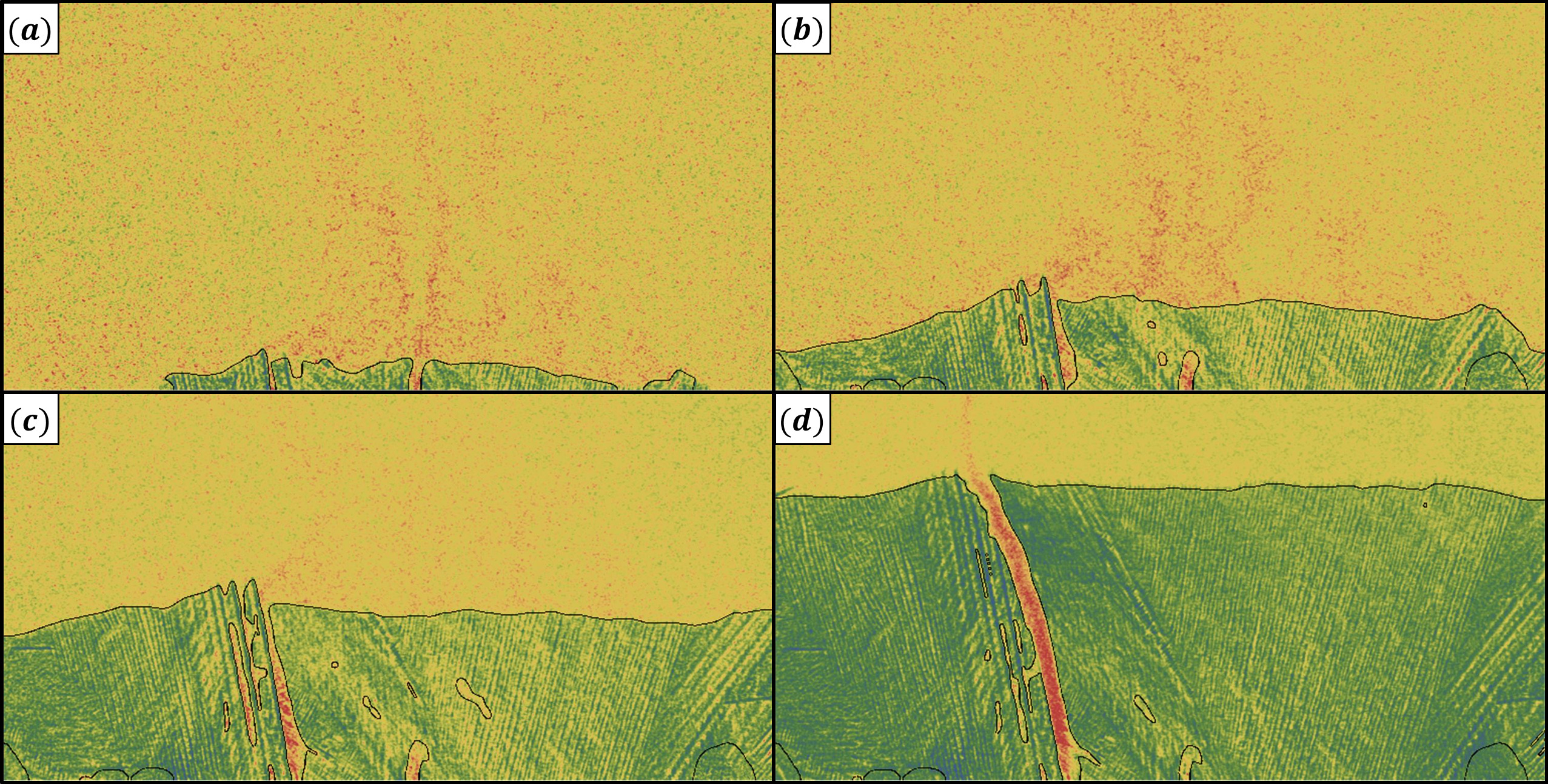}
\caption{Boundaries between segmented SZ/LZ (black contours), accounting for areas with removed persistent artifacts, for different time stamps (a-d) in ascending order obtained by segmentation using Algorithm \ref{alg:mil-based-segmentation} and \ref{eq:mil-based-adaptive-threshold}.}
\label{fig:fov-segmentation-1}
\end{figure}

\begin{figure}[H]
\centering
\includegraphics[width=0.85\textwidth]{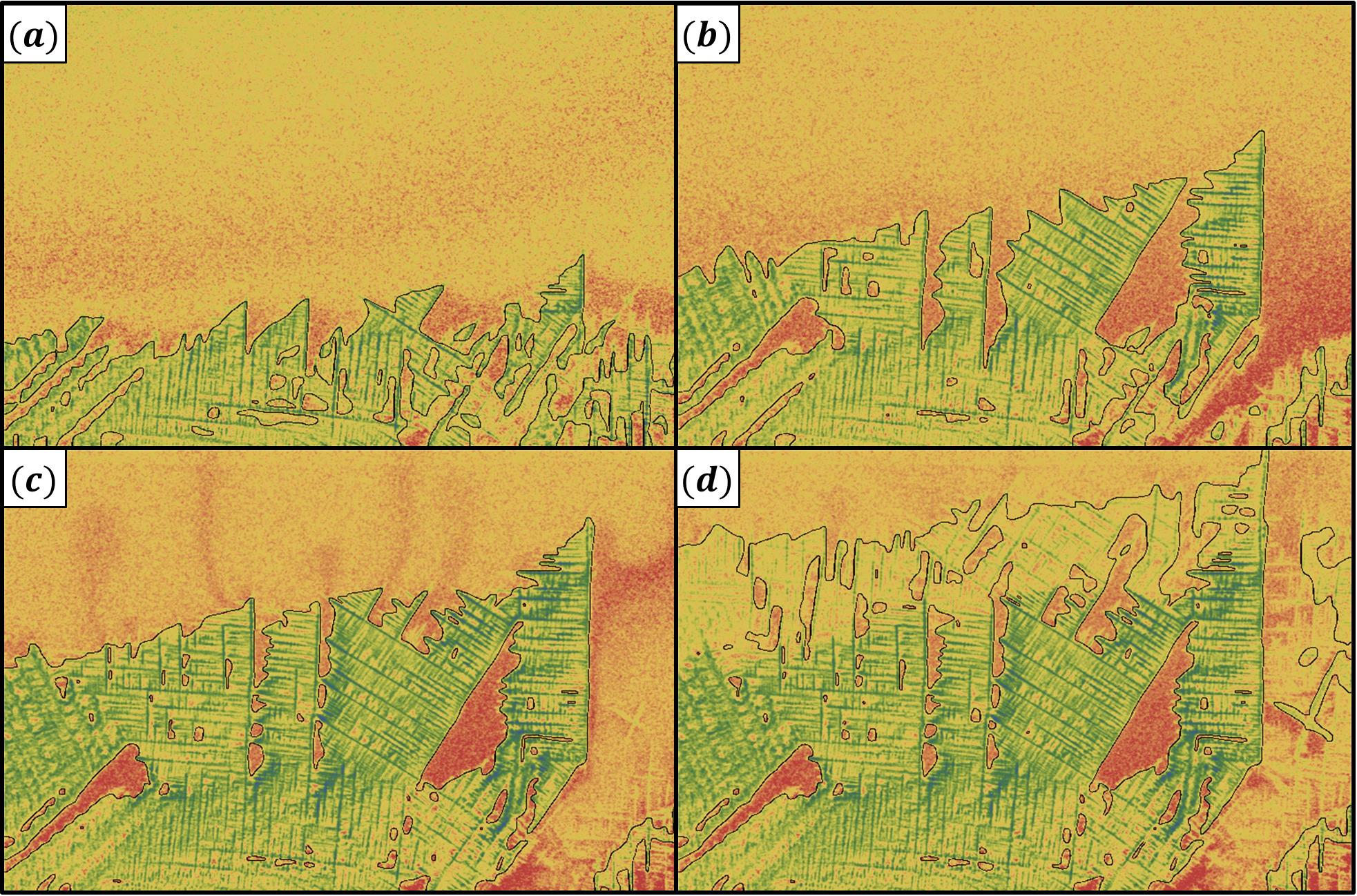}
\caption{Another example demonstrating the performance of the developed SZ/LZ segmentation method. Here a more challenging case is shown compared to Figure \ref{fig:fov-segmentation-1}.}
\label{fig:fov-segmentation-2}
\end{figure}

The results of segmentation (Algorithm \ref{alg:mil-based-segmentation}, using (\ref{eq:mil-based-adaptive-threshold})) after FOV filtering are presented in Figures \ref{fig:fov-segmentation-1} and \ref{fig:fov-segmentation-2}. Figure \ref{fig:fov-segmentation-1} represents an easier case where the front has a rather simple shape, with cavities forming and disappearing, and a channel forming over time to the left of the cell center line. A more challenging case is shown in Figure \ref{fig:fov-segmentation-2}, which corresponds to image filtering results seen in Figure \ref{fig:fov-image-processing-3}. Here it is demonstrated that the segmentation algorithm has no issues detecting the earlier formations in (a) and (b) where forced convection is observed, is not hindered by the rapid onset of convective plumes in (c), and also captures the much less contrast structures seen in the upper part of (d) where the SFF is close to $1$.

\begin{figure}[htbp]
\centering
\includegraphics[width=0.55\textwidth]{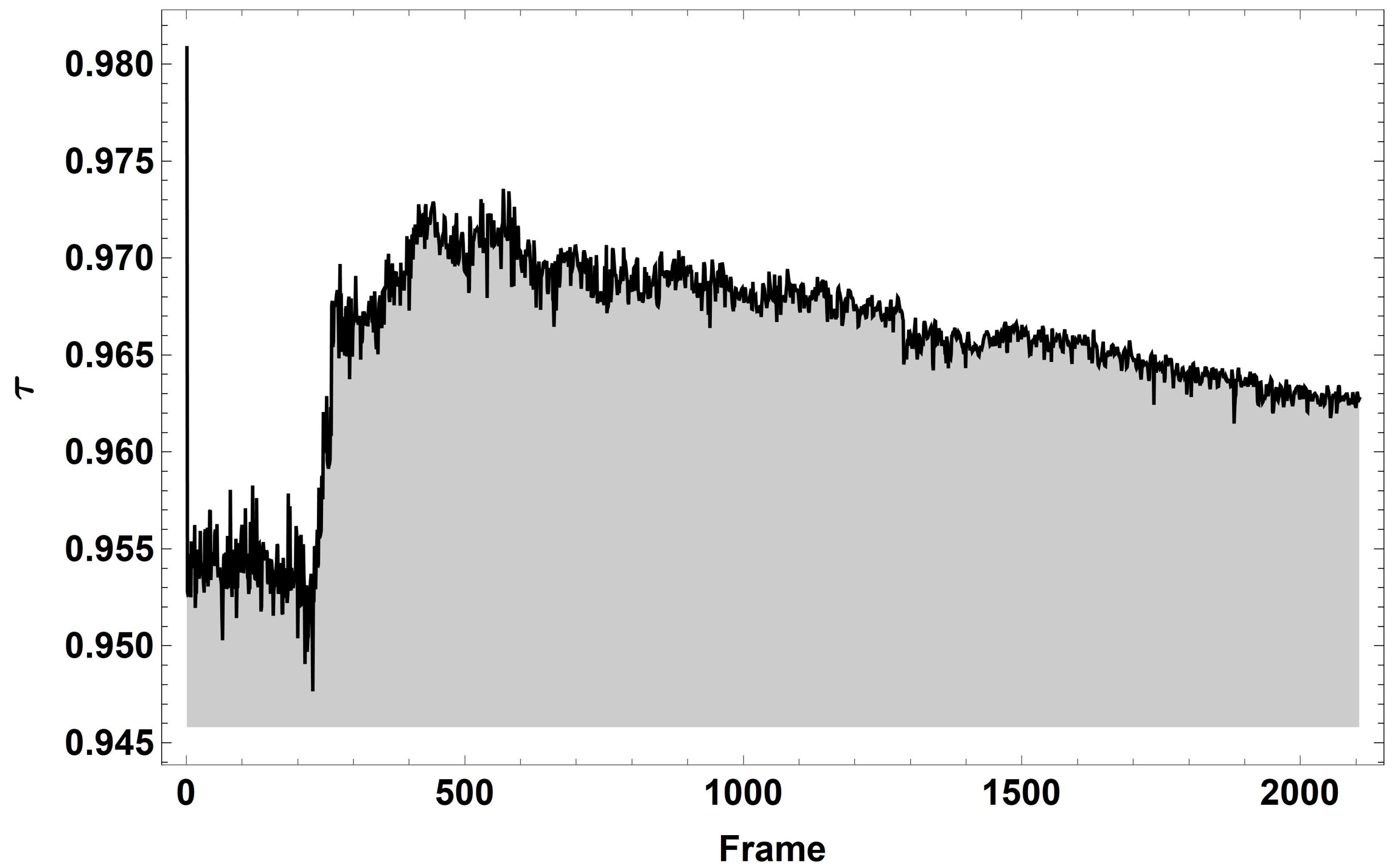}
\caption{An example of a MIL-based adaptive segmentation threshold $\tau$ (\ref{eq:mil-based-adaptive-threshold}) time series for an image sequence represented here by examples in Figures \ref{fig:fov-image-processing-3} and \ref{fig:fov-segmentation-2}.}
\label{fig:segmentation-threshold-series}
\end{figure}

The adaptive MIL-based threshold $\tau$ for the case in Figure \ref{fig:fov-segmentation-2} is shown in Figure \ref{fig:segmentation-threshold-series}. Note that prior to frame $\sim 250$ the FOV is devoid of solidified structures and only convective plumes are present. Once such structures appear in within the FOV, $\tau$ changes significantly to account for the SFF dynamics. In this case the parameters for (\ref{eq:mil-based-adaptive-threshold}) were set to $C_1 = 9 \cdot 10^{-2}$, $C_2 = 15$, $p = -0.5$. A general rule of thumb for setting up the parameters is to adjust $C_1$ such that the SZ is correctly segmented in earlier frames of the sequence (i.e. in the initial phases of growth with smaller SFF values), then tune $C_2$ and $p$ until the later frames in the image sequences are also treated correctly. Initial $C_1$ values prior to fine-tuning can also be inferred from later frames, since there is more information to work with. In our experience, parameter tuning can be done fairly quickly, i.e. $\sim 5$ minutes per image sequence (segmentation iterations are fast since threshold values are given by (\ref{eq:mil-based-adaptive-threshold}) instead of more complex algorithms) was enough to achieve good and stable results. Note that for the case considered in Figure \ref{fig:fov-segmentation-1}, the parameter set was $C_1=8.5\cdot10^{-3}$, $C_2=3$ and $p=15$. While different $C_2$ and $p$ combinations could, in general, yield rather similar results and in some cases tweaking $C_2$ alone can yield desirable results quickly, $p$ adjustments together with $C_2$ often allow one to get the same results faster.

\begin{figure}[htbp]
\centering
\includegraphics[width=1\textwidth]{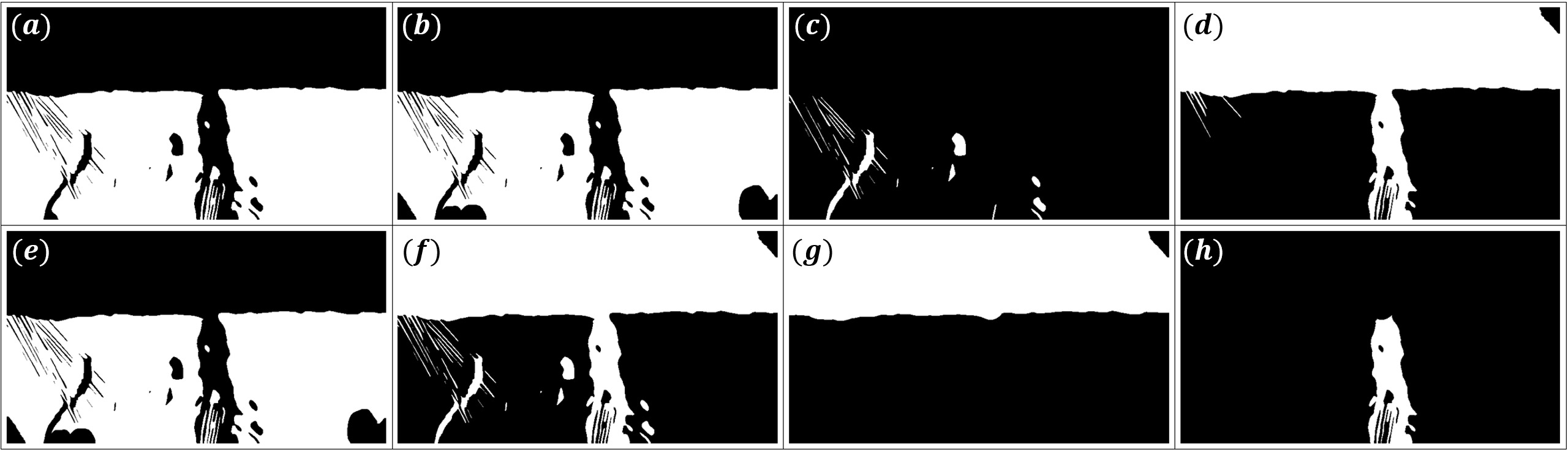}
\caption{Stages of segment classification for separated SZ/LZ: (a) segmented SZ, (b) mask (a) with non-border artifacts removed (case none present here), (c) liquid cavity mask, (d) LZ mask without cavities and border or persistent artifacts, (e) mask (b) but without border artifacts (none were present), (f) combined masks (c) and (d), (g) masks separated by the SF, and (h) liquid channel mask. This case corresponds to the image sequence considered in Figure \ref{fig:fov-image-processing-1}.}
\label{fig:segment-classification-1}
\end{figure}

\begin{figure}[H]
\centering
\includegraphics[width=1\textwidth]{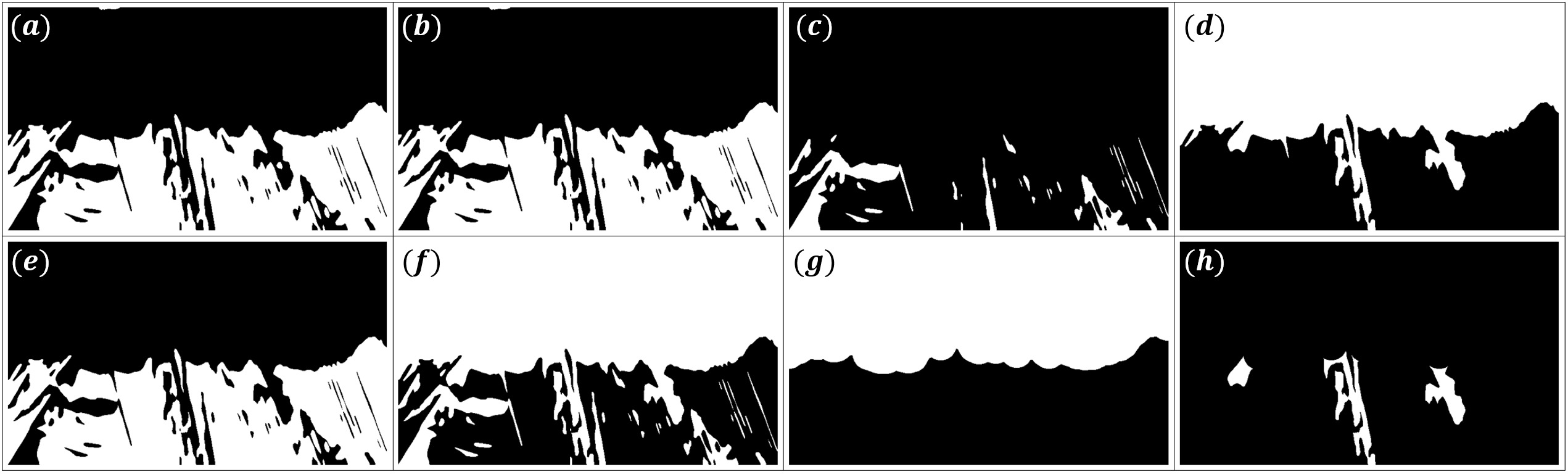}
\caption{Another example of segment classification, showing results for an earlier frame in the same image sequence as the frame shown in Figure \ref{fig:fov-image-processing-2}. Note that the border artifact seen at the top of (a) and (b) has been removed from and is no longer present in (d-h). This case corresponds to Figure \ref{fig:fov-image-processing-3}.}
\label{fig:segment-classification-2}
\end{figure}

After SZ/LZ segmentation is complete, these segments are further differentiated to separate liquid cavities and channels from the SZ, and to determine the shape of the SF. There are also safeguards against artifacts that can potentially be left over after LZ/SZ segmentation, as it is hardly possible to \textit{always} find optimal $C_1$, $C_2$ and $p$ immediately. Therefore, it is good if the code has backup options. Figures \ref{fig:segment-classification-1} and \ref{fig:segment-classification-2} illustrate the steps involved in separating segment classes.

\begin{figure}[H]
\centering
\includegraphics[width=0.5\textwidth]{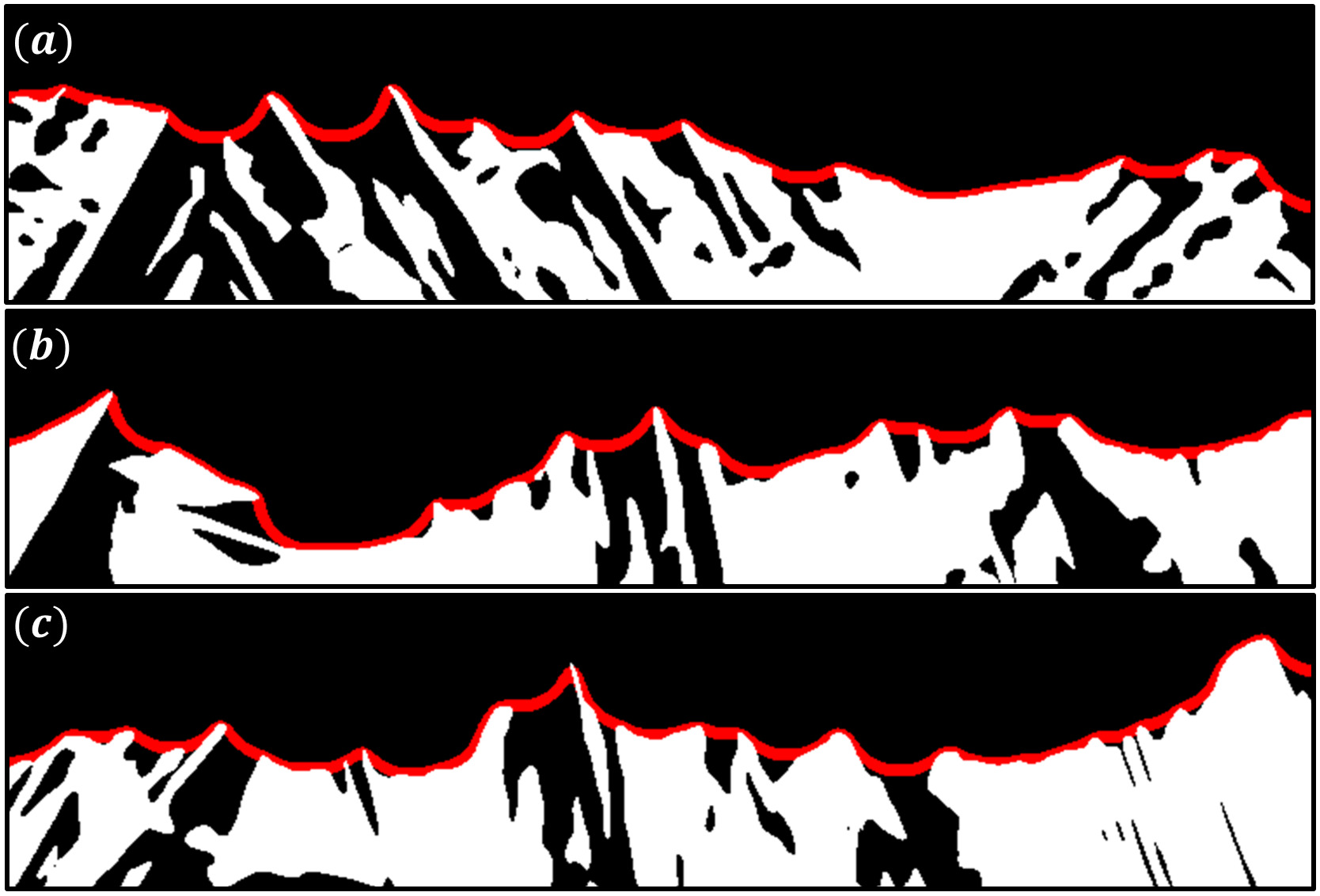}
\caption{Examples showing how the SF (red curves) is traced based on the final SZ mask, i.e. how Figure \ref{fig:segment-classification-2}g results from Figure \ref{fig:segment-classification-2}e/f.}
\label{fig:example-solidification-front-tracing}
\end{figure}

Subfigures (a) and (b) in Figures \ref{fig:segment-classification-1} and \ref{fig:segment-classification-2} are identical, since the SZ/LZ segmentation was performed well, but notice the boundary artifact present in both (a) and (b) in Figure \ref{fig:segment-classification-2}, which is addressed later in step (d). In both cases, once cavities are identified (Algorithm \ref{alg:lz-cavity-channel-front-derivation}), one readily obtains the \textit{bulk} LZ segment, and can then derive the SF and find the channels extending below the SF. Examples of how the SF is traced along the SZ segment boundaries is shown in Figure \ref{fig:example-solidification-front-tracing}. The resulting SF edge masks will be used later for concentration measurements above the SF and for convective plume segmentation in the LZ.

\subsection{Analysis of solidified structures}

\begin{figure}[htbp]
\centering
\includegraphics[width=0.75\textwidth]{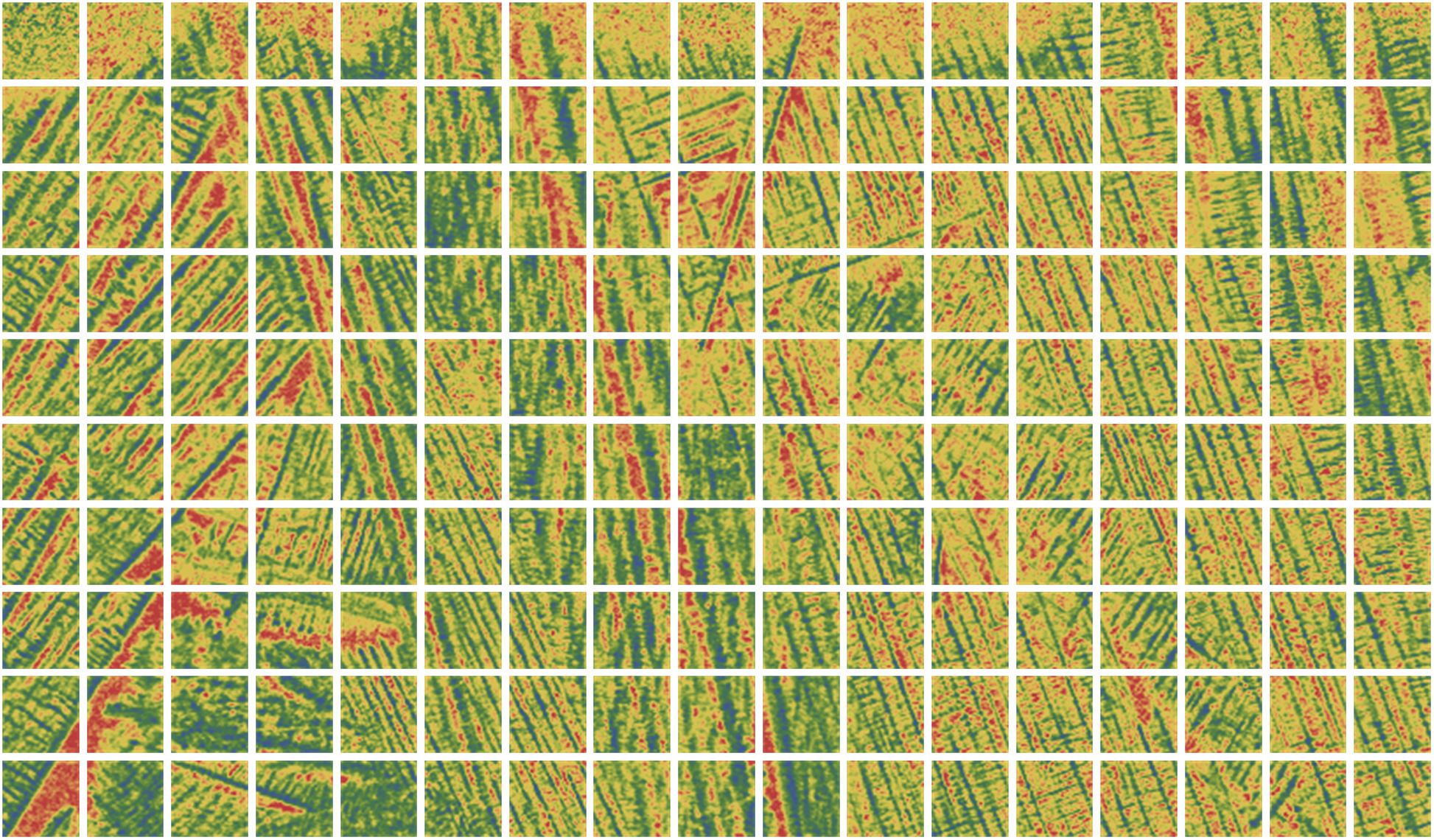}
\caption{An example of how an image is partitioned into IWs prior to solidified structure analysis.}
\label{fig:example-image-partitioning}
\end{figure}

\begin{figure}[htbp]
\centering
\includegraphics[width=0.5\textwidth]{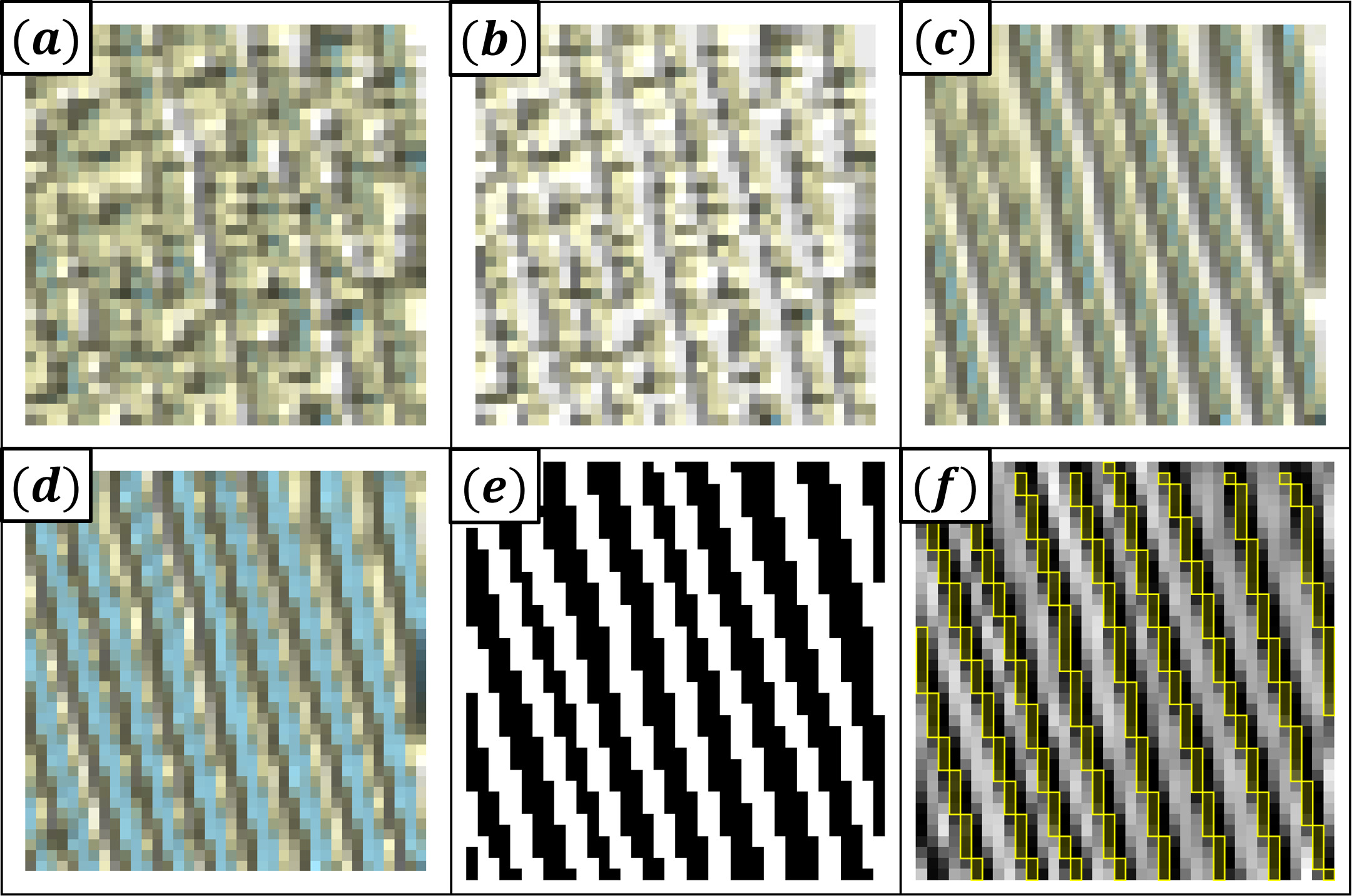}
\caption{IW processing via Algorithm \ref{alg:solid-skeleton-segmentation}: (a) a colorized relief plot of an IW (projection of a pre-processed image), (b) IW after re-scaling, CTM and FFC, (c) BM3D output, (d) results after 2 iterations of NMM, one SCTMM iteration and FFC, (e) output after 2-Otsu (hysteresis) binarization and image inversion, and (f) final result after size thresholding and thinning, overlaid on top of the grayscale version of (d). Image luminance increases from color light blue to white.}
\label{fig:example-iw-processing-1}
\end{figure}

\begin{figure}[htbp]
\centering
\includegraphics[width=0.5\textwidth]{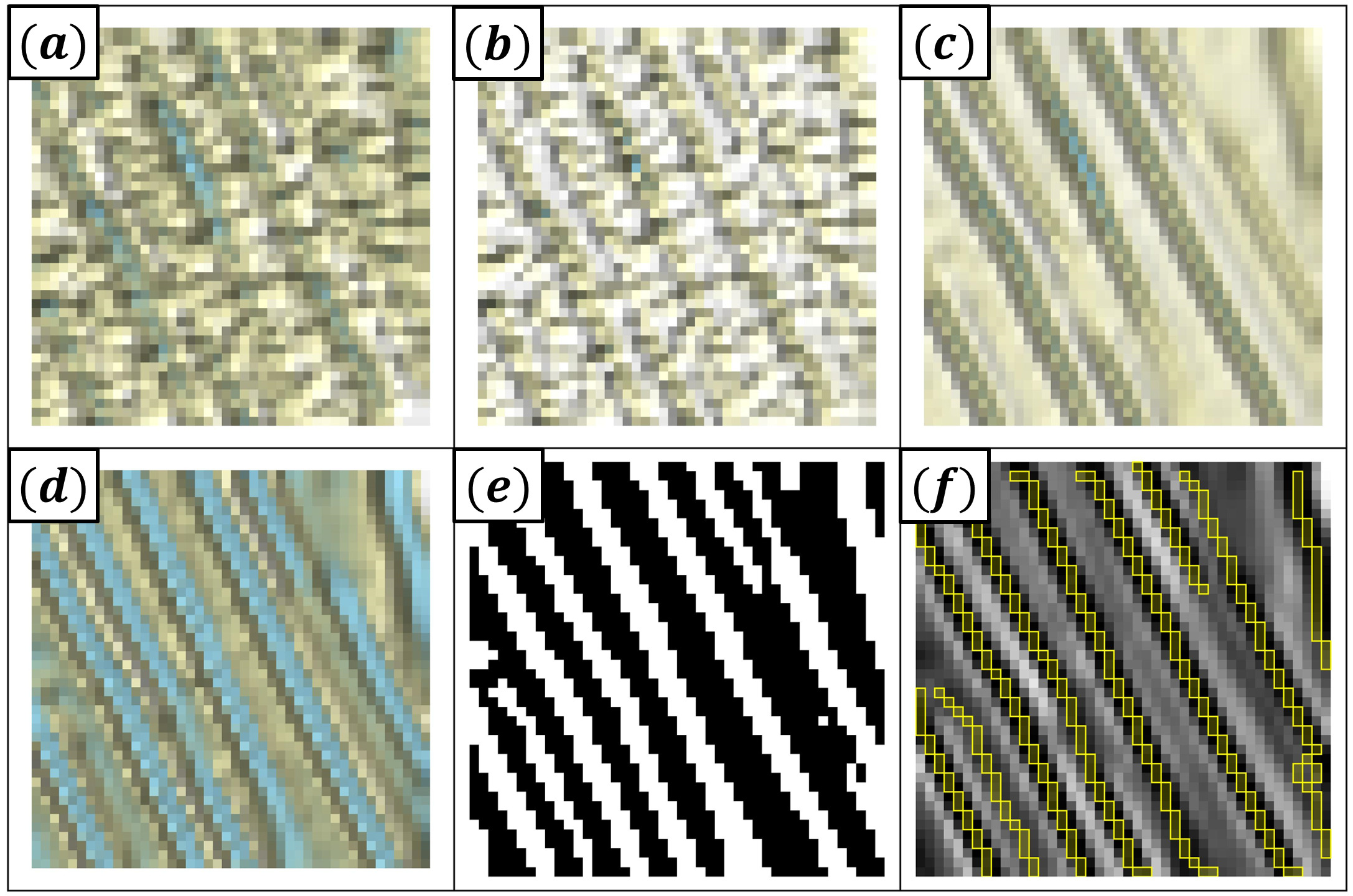}
\caption{Another example of steps taken during IW processing.}
\label{fig:example-iw-processing-2}
\end{figure}

With the SZ identified along with cavities and channels, one can now turn to the analysis of the solidified structures within the SZ. As outlined in Algorithm \ref{alg:image-partitioning}, first the FOV image is partitioned into IWs, as seen in Figure \ref{fig:example-image-partitioning}, and then each partition is processed using Algorithm \ref{alg:solid-skeleton-segmentation}. Examples of this are provided in Figures \ref{fig:example-iw-processing-1} and \ref{fig:example-iw-processing-2}. Notice that the noise makes the identification of dendrites in the initial images (a) quite difficult, and Figure \ref{fig:example-iw-processing-2}a additionally exhibits very low CNR for dendrites, mainly due to larger-wavelength correlated noise stemming from liquid flow. Note also that the dendrites have lower X-ray transparency that the surrounding liquid, and therefore it is the liquid that is colored white in (a).

Subfigures (b) show that CTM and reference-less FFC make the dendrites and the spacings in between much clearer, but the SNR and CNR are unchanged. However, we found that this stage dramatically boosts the performance of BM3D, which is the next stage, the output of which is seen in (c). While the structures in Figure \ref{fig:example-iw-processing-1}c are already clearly discernible to the human eye, the CNR is still lower that desired for reliable segmentation. Figure \ref{fig:example-iw-processing-2}c, on the other hand, is problematic, since BM3D does not mitigate the correlated noise (uncorrelated Gaussian noise model is used for BM3D). This is where the next stage comes in with two iterations of NMM correction and SCTMM, after which reference-less FFC is applied again. Here NMM takes care of much of the correlated noise, SCTMM boosts CNR, and FFC acts as post-NMM large-wavelength background cleanup.

\begin{figure}[H]
\centering
\includegraphics[width=0.65\textwidth]{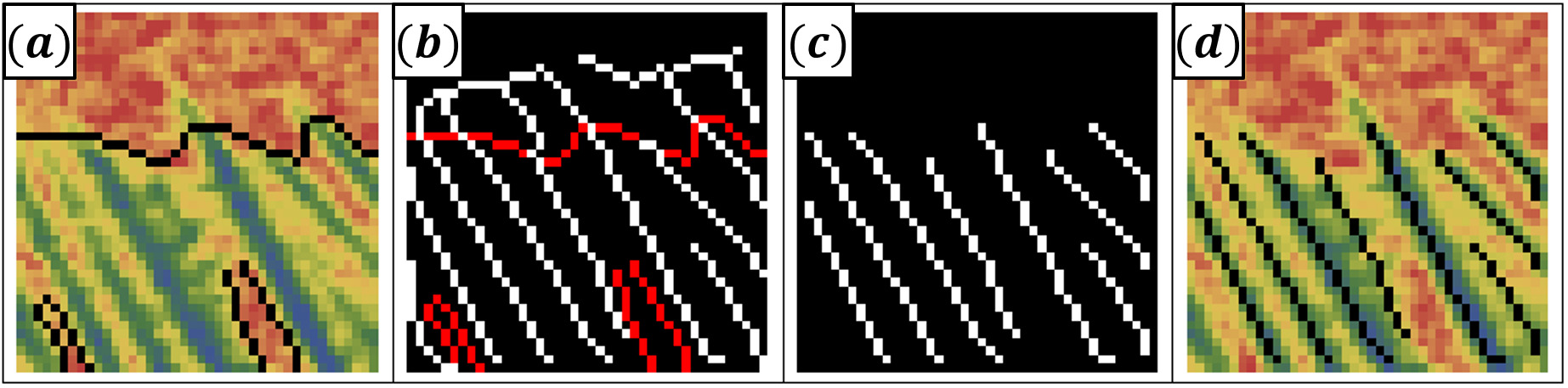}
\caption{An example of IW skeleton cropping using the LZ mask: (a) colorized IW image with a SZ boundary (black contours), (b) IW skeleton with the SZ boundary overlay (red contours), (c) cropped IW skeletons and (b) cropped IW skeletons overlaid on top of the colorized IW image (black lines).}
\label{fig:iw-skeleton-crop-example-1}
\end{figure}

\begin{figure}[H]
\centering
\includegraphics[width=0.65\textwidth]{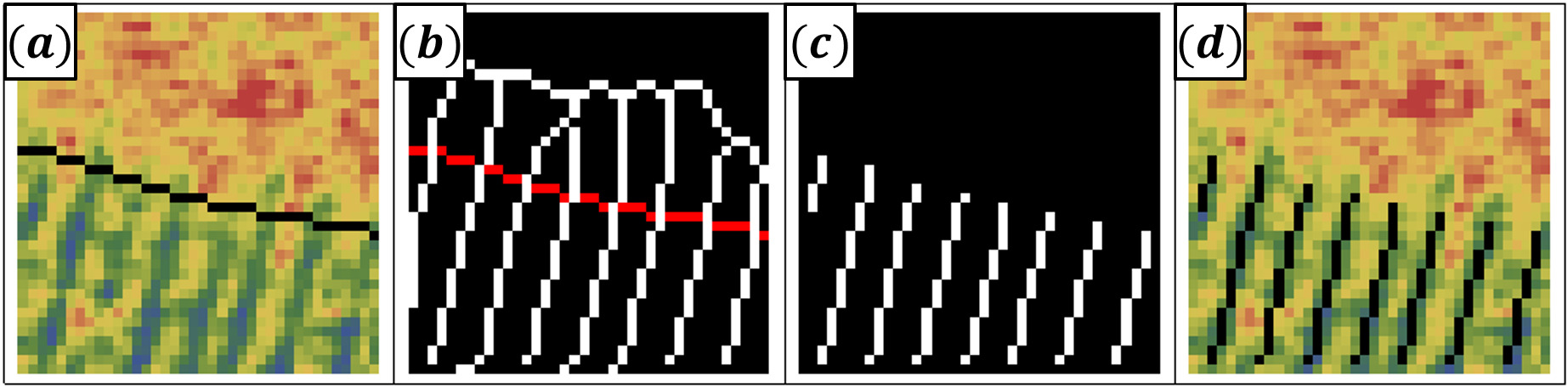}
\caption{Another example of IW skeleton cropping using the LZ mask.}
\label{fig:iw-skeleton-crop-example-2}
\end{figure}

\begin{figure}[H]
\centering
\includegraphics[width=0.75\textwidth]{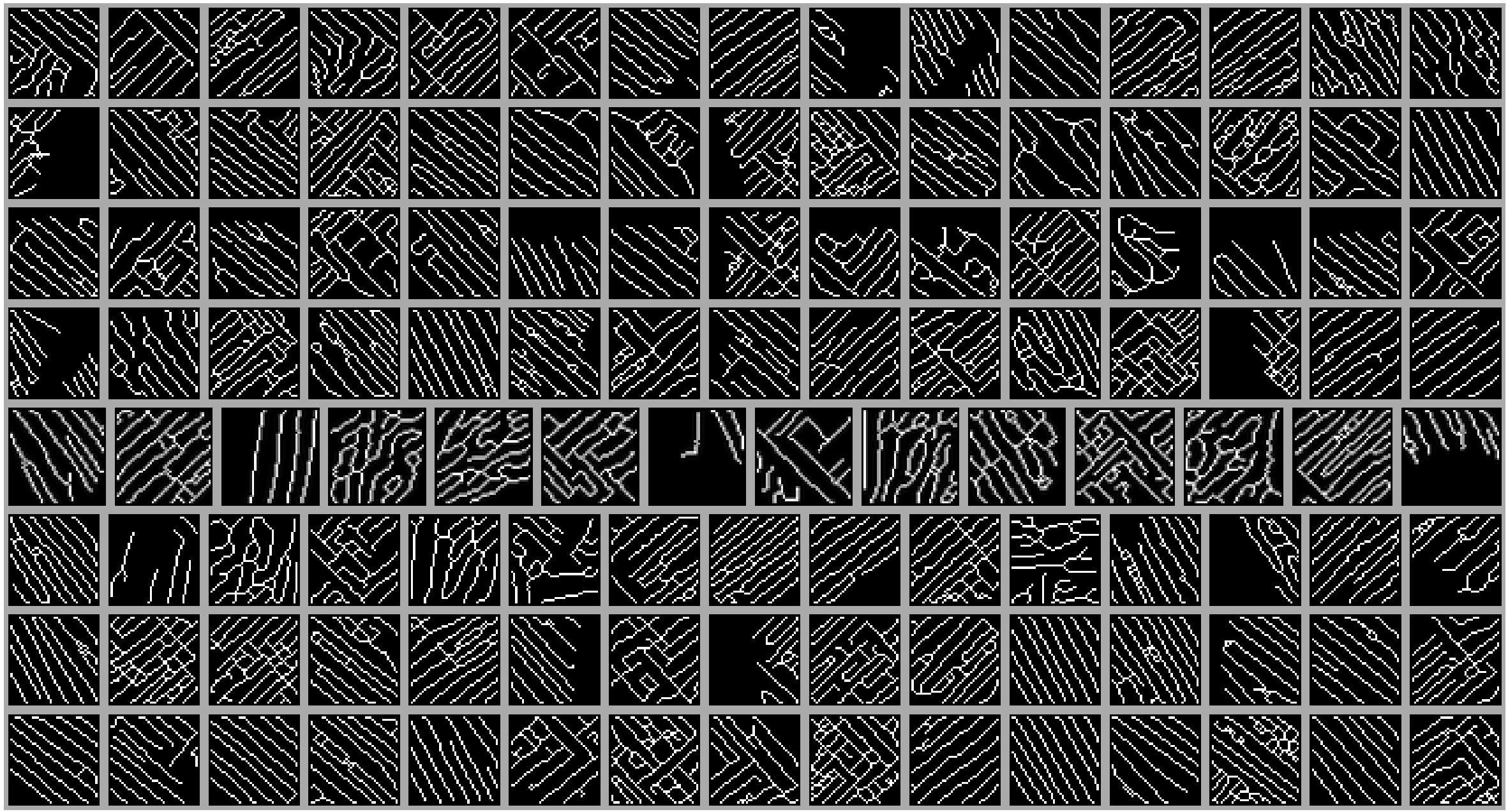}
\caption{An example of detected IW skeletons for an image after cropping to the SZ mask, e.g. Figure \ref{fig:segment-classification-1}(e). This case corresponds to Figures \ref{fig:fov-image-processing-1} and \ref{fig:segment-classification-1}.}
\label{fig:iw-skeletons-after-cropping}
\end{figure}

Afterwards, double-Otsu hysteresis segmentation and image inversion yields the dendrite mask (e), and morphological thinning with size thresholding generates the IW dendrite skeleton (f). Now all that remains is to crop the dendrite skeletons, multiplying their masks by the SZ mask projected onto IWs -- this process is demonstrated in Figures \ref{fig:iw-skeleton-crop-example-1} and \ref{fig:iw-skeleton-crop-example-2}. After cropping is done for every IW, one has solid structure skeletons for the entire FOV -- an example of IW skeletons output by Algorithm \ref{alg:solid-skeleton-segmentation} for the case considered in Figures \ref{fig:fov-image-processing-1} and \ref{fig:segment-classification-1} is shown in Figure \ref{fig:iw-skeletons-after-cropping}.

\begin{figure}[H]
\centering
\includegraphics[width=0.75\textwidth]{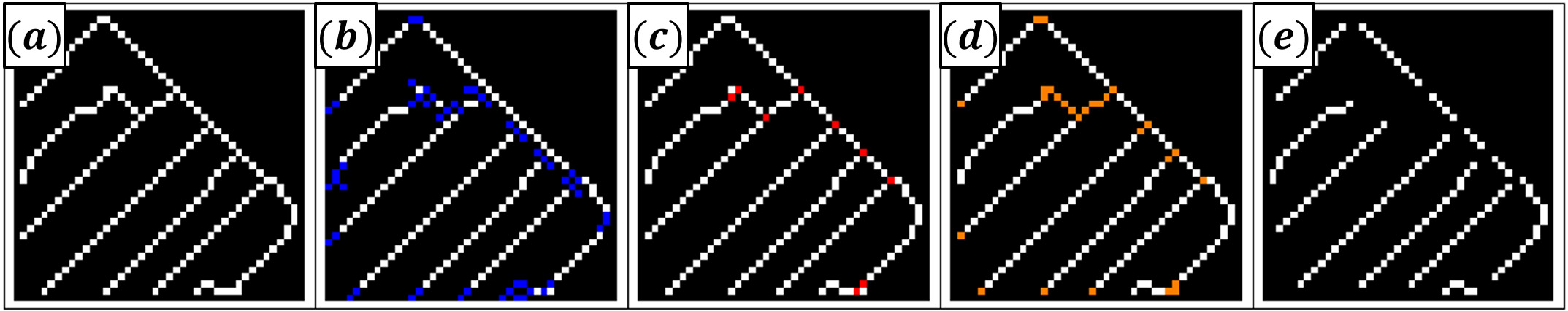}
\caption{The procedure for resolving dendrite crossings (Algorithm \ref{alg:resolving-unoriented-structures}): (a) detected corner pixels, (b) identified morphological branch points, (c) combined and thresholded mask for to be removed from (a), (d) dendrite skeleton mask with resolved crossings.}
\label{fig:example-resolving-crossings-1}
\end{figure}

\begin{figure}[H]
\centering
\includegraphics[width=0.75\textwidth]{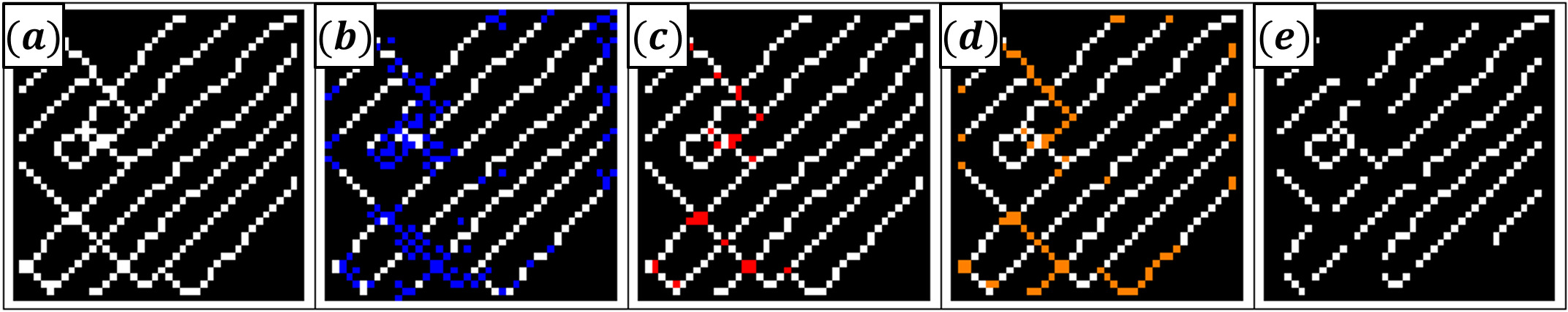}
\caption{Another example of the steps for resolving skeleton crossings.}
\label{fig:example-resolving-crossings-2}
\end{figure}

\begin{figure}[H]
\centering
\includegraphics[width=0.75\textwidth]{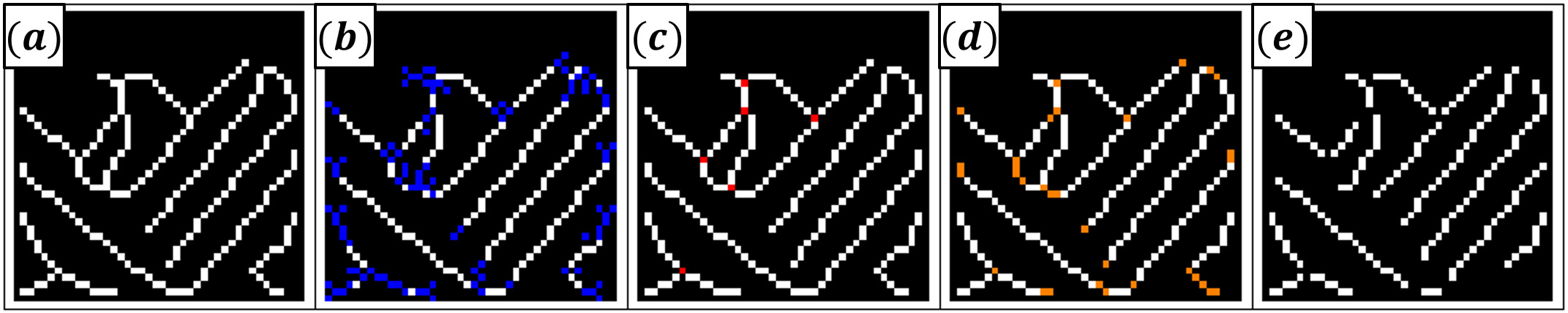}
\caption{Another case of Algorithm \ref{alg:resolving-unoriented-structures} applied to IW skeletons.}
\label{fig:example-resolving-crossings-3}
\end{figure}

\begin{figure}[H]
\centering
\includegraphics[width=0.8\textwidth]{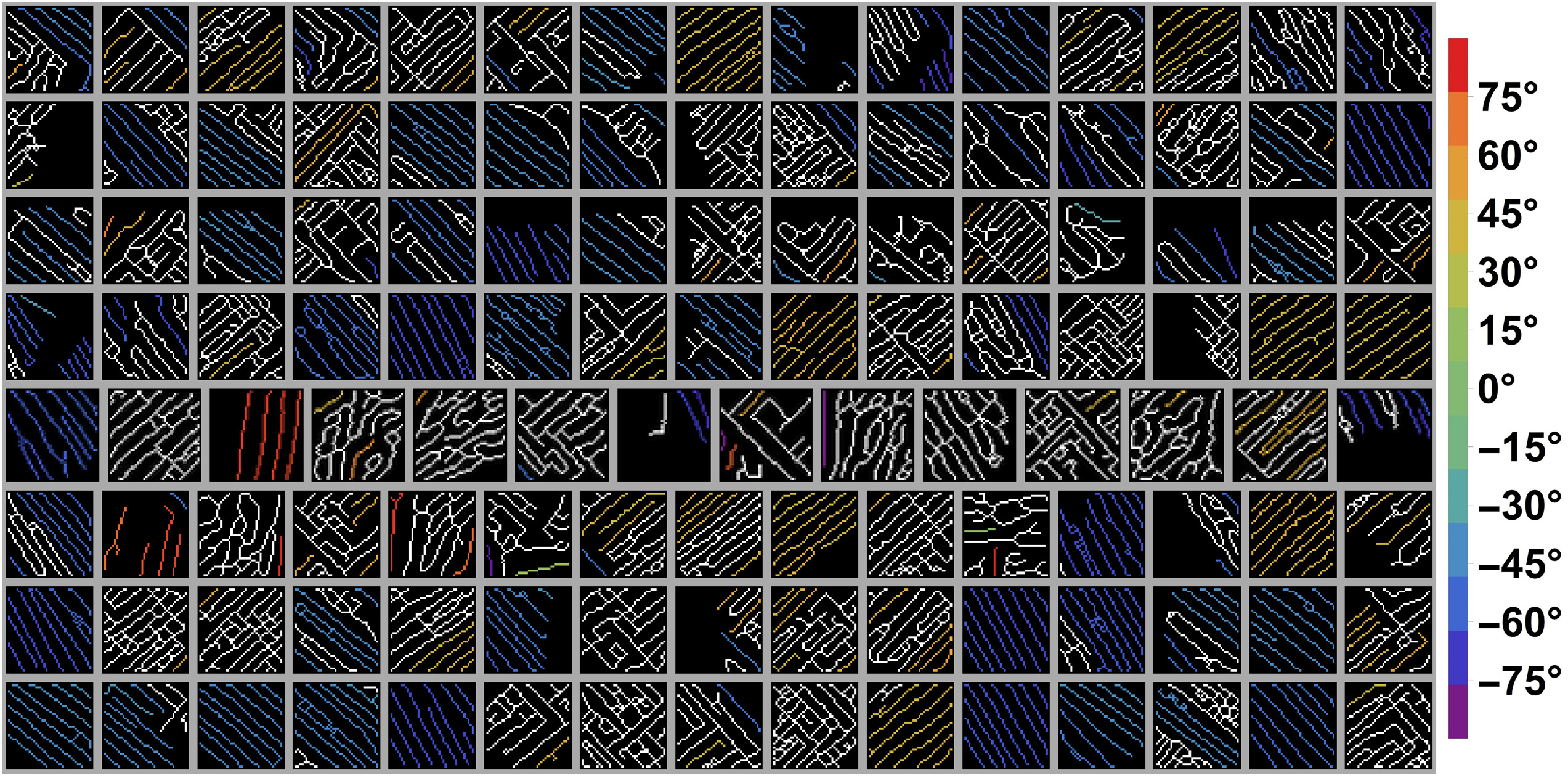}
\caption{IW skeletons shown in Figure \ref{fig:iw-skeletons-after-cropping} colorized by their orientations with respect to the horizontal image axis. The white-colored segments are unoriented according to the criteria established in Algorithm \ref{alg:resolving-unoriented-structures}.}
\label{fig:unresolved-iw-skeletons}
\end{figure}

\begin{figure}[H]
\centering
\includegraphics[width=0.8\textwidth]{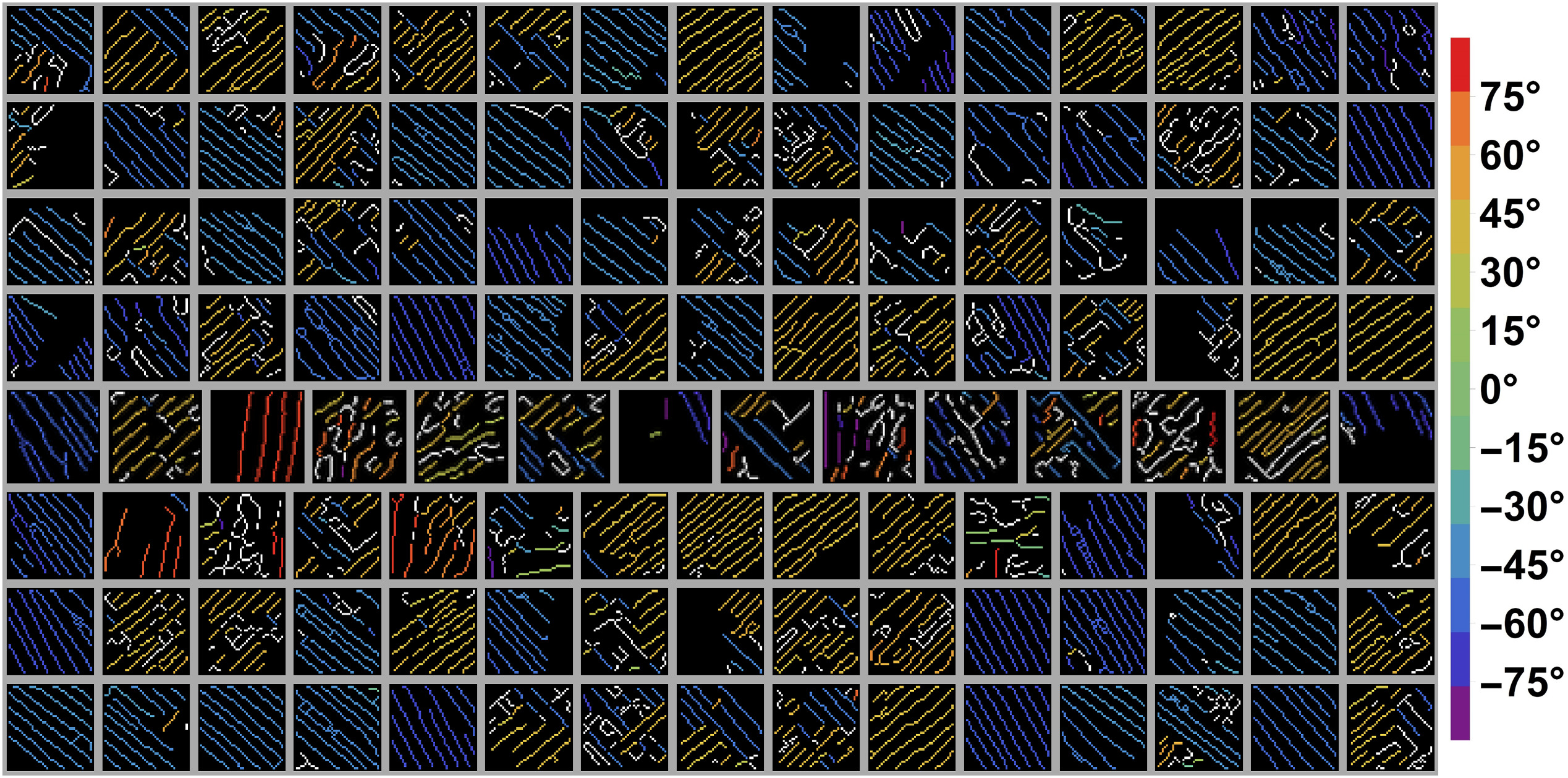}
\caption{The same IW skeletons as in Figure \ref{fig:unresolved-iw-skeletons} after Algorithm \ref{alg:resolving-unoriented-structures} was applied to every IW. The color convention matches Figure \ref{fig:unresolved-iw-skeletons}.}
\label{fig:unoriented-structures-resolved}
\end{figure}

Observe, however, that many of the IWs in Figure \ref{fig:iw-skeletons-after-cropping} exhibit multiple dendrite crossings and branching, making the orientation analysis problematic. To mitigate this issue, Algorithm \ref{alg:resolving-unoriented-structures} is applied to each of the IWs. Figures \ref{fig:example-resolving-crossings-1}-\ref{fig:example-resolving-crossings-3} show the steps involved in this procedure. Although the results have certain imperfections and a small amount of information may be lost in IW skeletons in general, we find that the set of parameters that we have selected for Algorithm \ref{alg:resolving-unoriented-structures} allows to recover much more information that would otherwise be lost. This is very clearly illustrated in Figures \ref{fig:unresolved-iw-skeletons} and \ref{fig:unoriented-structures-resolved} where one can see the significant difference between before and after Algorithm \ref{alg:resolving-unoriented-structures} is applied.

\begin{figure}[htbp]
\centering
\includegraphics[width=1\textwidth]{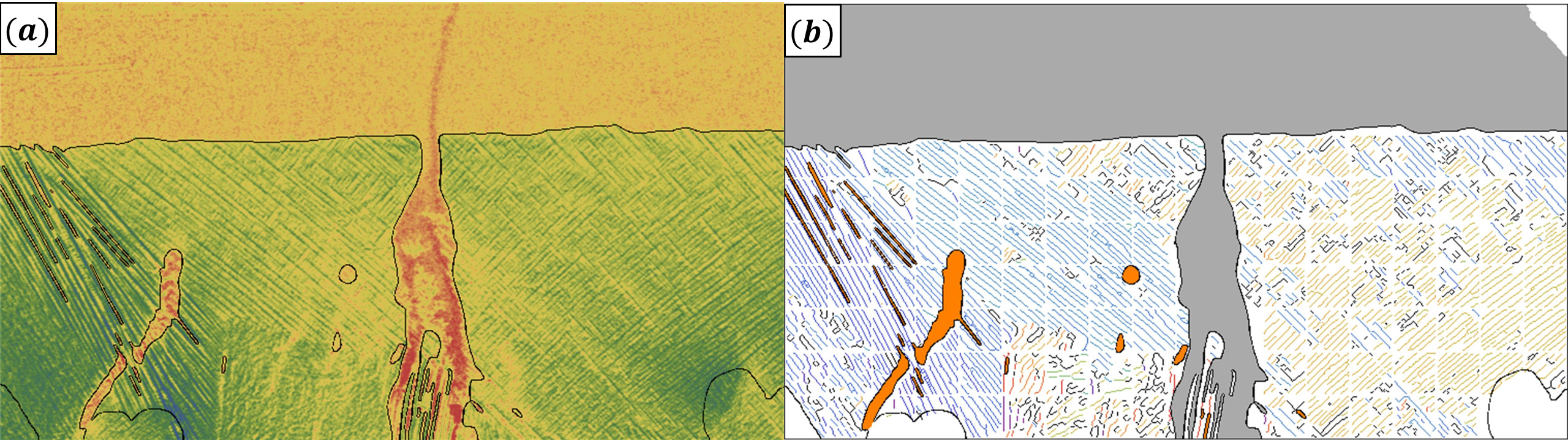}
\caption{SZ analysis results for an image from the case considered in \ref{fig:unoriented-structures-resolved}: (a) pre-processed image with overlaid SZ/LZ boundaries (black contours, image artifact areas included) and (b) the same SZ/SZ boundaries with dendrite skeletons (colorized by orientatiom otherwise black if unoriented) and liquid metal cavity areas (orange) within. The gray area in (b) is the bulk liquid above the SF plus the liquid metal channel connected to it. The orientation color map is as in Figure \ref{fig:unoriented-structures-resolved}.}
\label{fig:assembled-global-skeleton-1}
\end{figure}

\begin{figure}[H]
\centering
\includegraphics[width=1\textwidth]{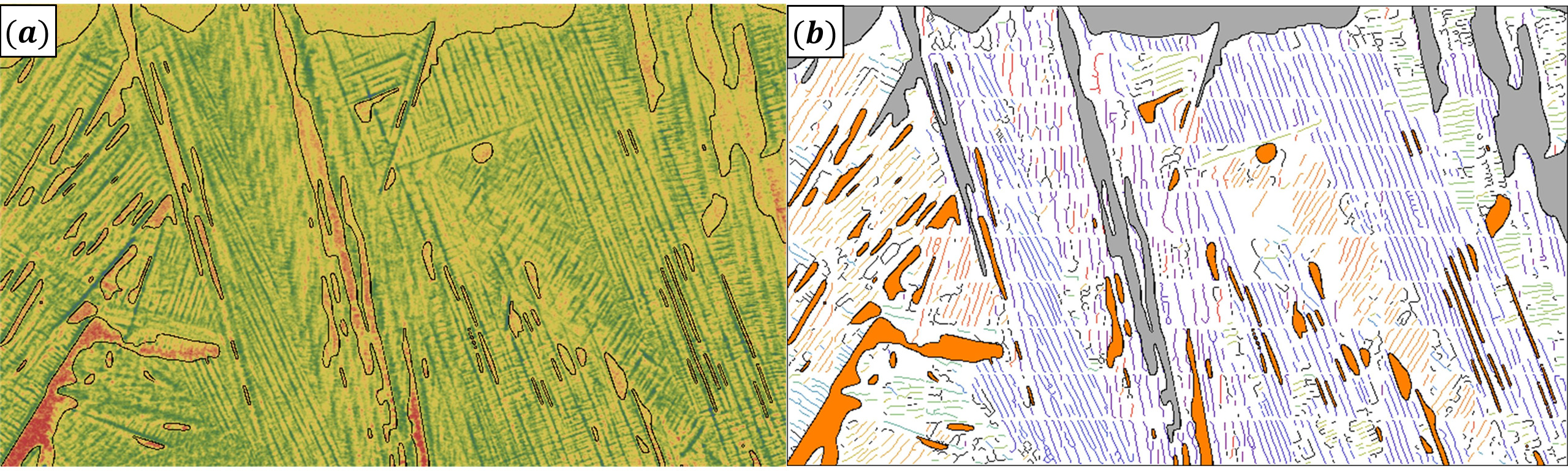}
\caption{Another example of the SZ analysis, this time for the case considered in Figures \ref{fig:fov-image-processing-2}, \ref{fig:segment-classification-2} and \ref{fig:example-image-partitioning}.}
\label{fig:assembled-global-skeleton-2}
\end{figure}

As noticeable in Figure \ref{fig:unoriented-structures-resolved}, some of the new segments still do not exhibit clear orientations or are otherwise ambiguous, but most of the information otherwise inaccessible from Figure \ref{fig:unresolved-iw-skeletons} has been recovered with minimal losses. With this, one can now re-assemble the processed IWs into a global skeleton, which yields results showcased in Figures \ref{fig:assembled-global-skeleton-1} (corresponding to Figure \ref{fig:unoriented-structures-resolved}) and \ref{fig:assembled-global-skeleton-2} (the case shown in Figures \ref{fig:fov-image-processing-2}, \ref{fig:segment-classification-2} and \ref{fig:example-image-partitioning}).

\begin{figure}[htbp]
\centering
\includegraphics[width=0.95\textwidth]{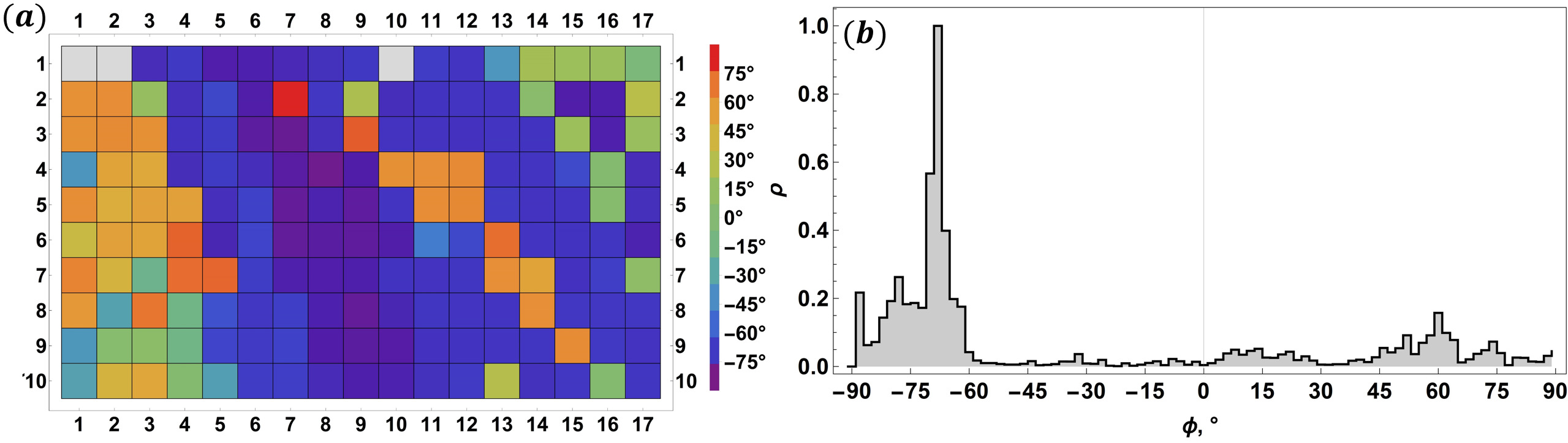}
\caption{(a) Median dendrite orientation $\varphi$ for the IWs (Figure \ref{fig:example-image-partitioning}) and (b) orientation spectrum for the assembled dendrite skeleton (Figure \ref{fig:assembled-global-skeleton-2}), excluding unoriented segments.}
\label{fig:dendrite-orientation-spectrum}
\end{figure}

While Figures \ref{fig:assembled-global-skeleton-1} and \ref{fig:assembled-global-skeleton-2} are already quite informative, there is more to be extracted from the assembled skeletons. One can measure the orientation ($\varphi$) spectrum, computing the relative orientation frequency by weighing over dendrite segment lengths to account for gaps due to IW boundaries and dendrite interruptions due to other reasons. An example of this is shown in Figure \ref{fig:dendrite-orientation-spectrum}b where the $\varphi$ spectrum is computed for the image seen in Figure \ref{fig:assembled-global-skeleton-2}. Note that the unoriented dendrite skeletons (colored black in Figure \ref{fig:assembled-global-skeleton-2}b) do not count towards the spectrum. One can also compute median $\varphi$ for every IW to have a coarser but simplified overview of how dendrite $\varphi$ are distributed over the FOV -- this is seen in Figure \ref{fig:dendrite-orientation-spectrum}a. In addition, by computing $\varphi$ spectra for an entire image sequence, one can observe the dynamics over time. Statistics for liquid cavity (orange areas in Figure \ref{fig:assembled-global-skeleton-2}b) areas, aspect ratio, orientations, etc. can also be computed per frame and their dynamics visualized. The same is true for channels extending into the SZ (e.g., Figure \ref{fig:assembled-global-skeleton-1}b).

\clearpage

\begin{figure}[htbp]
\centering
\includegraphics[width=0.85\textwidth]{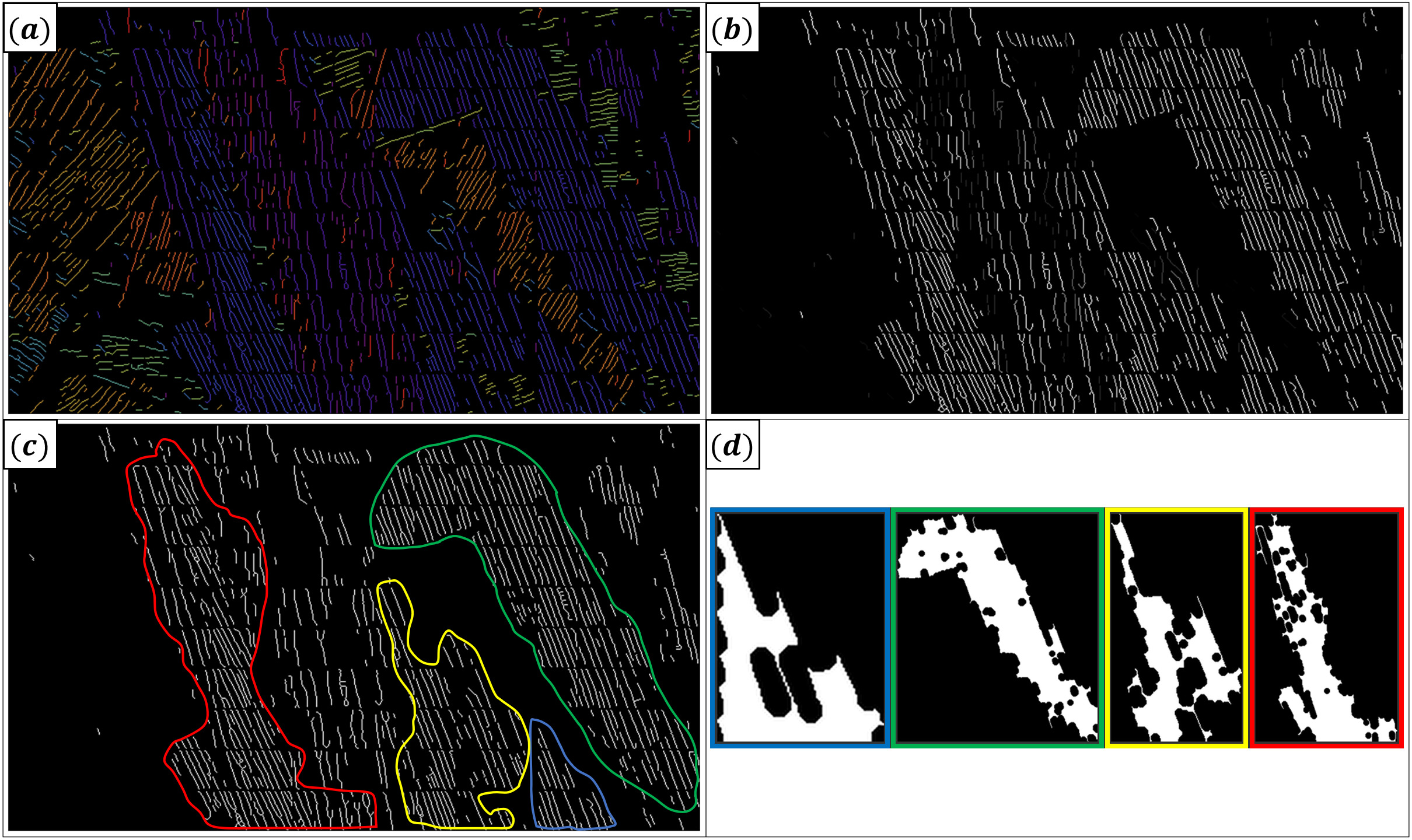}
\caption{The GDG process (Algorithm \ref{alg:detecting-dominant-grains}): (a) global dendrite skeleton with unoriented segments removed, colorized by orientation as in Figure \ref{fig:assembled-global-skeleton-2}b, (b) dendrite skeleton segments detected about the main orientation spectrum peak seen in Figure \ref{fig:dendrite-orientation-spectrum}b, with grayscale color map encoding the distance to the peak in color-space for each segment, (c) segments from (b) after color-space distance thresholding, with rough outlines for the four detected grains, (d) masks for the detected dendrite grains corresponding to the outlined in (c).}
\label{fig:dgd-example-1}
\end{figure}

\begin{figure}[H]
\centering
\includegraphics[width=0.85\textwidth]{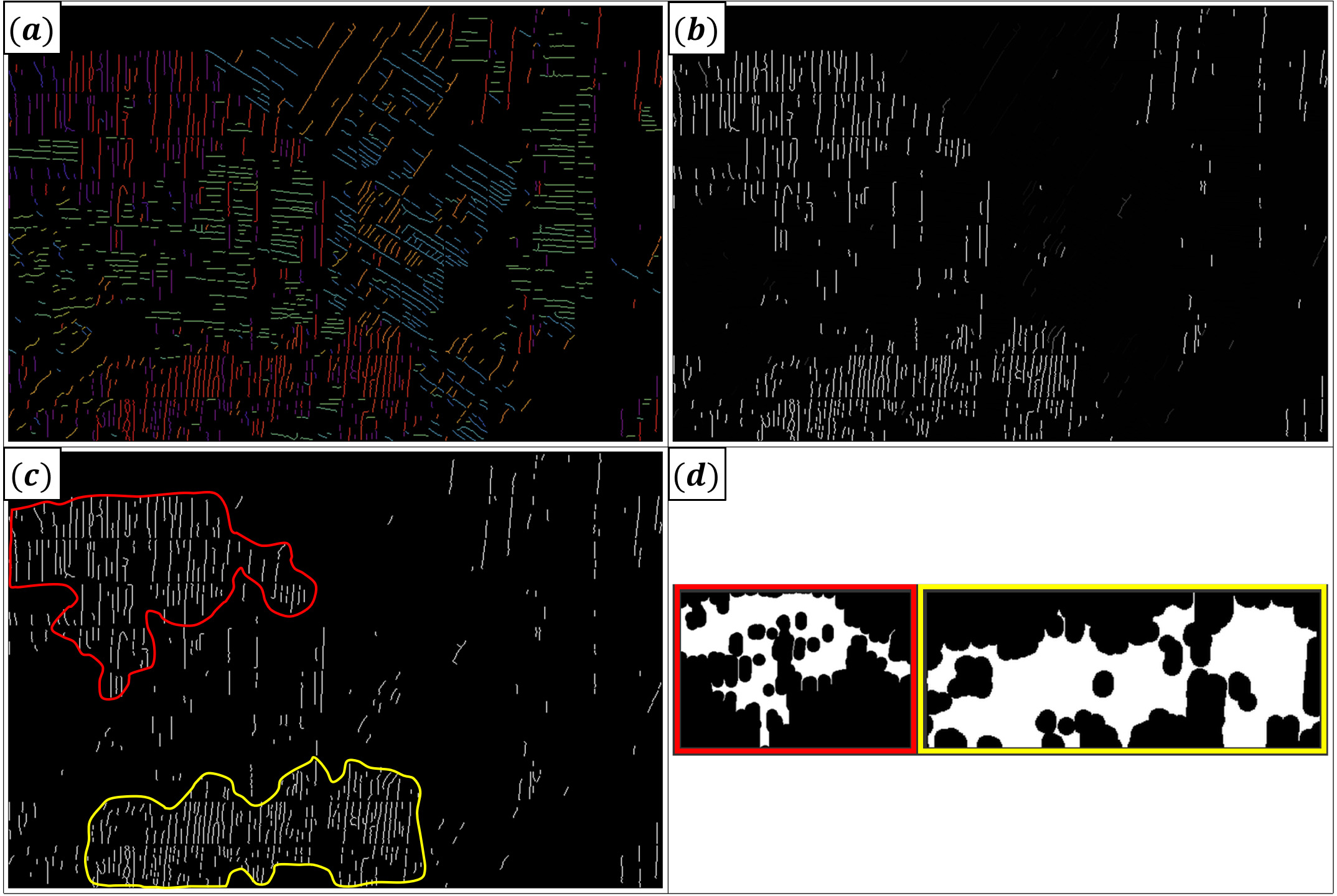}
\caption{Another example of the DGD process steps, this time for the case corresponding to Figure \ref{fig:fov-segmentation-2}. Note the segments with orientations near $\varphi \sim \pm \pi/2$ correctly recognized by DGD as adjacent in the color-space due to the edge constraints from (\ref{eq:edge-peak-constraints}).}
\label{fig:dgd-example-2}
\end{figure}

\begin{figure}[H]
\centering
\includegraphics[width=0.85\textwidth]{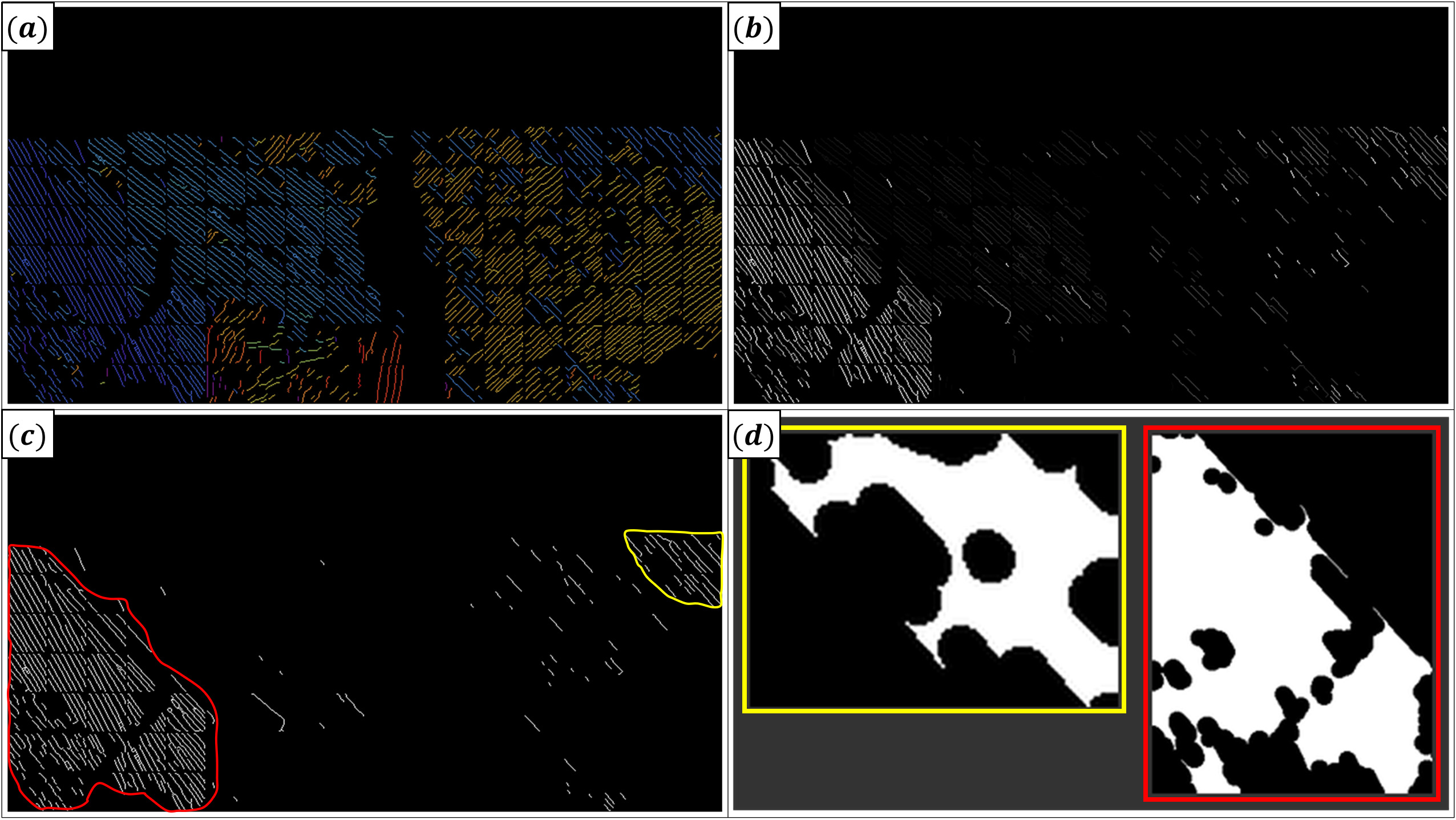}
\caption{Another example, in this case DGD steps are shown for the case in Figure \ref{fig:assembled-global-skeleton-2}. Observe that the thresholding step (b-c) here clearly separates the grains linked by a gradual orientation transition zone (a).}
\label{fig:dgd-example-3}
\end{figure}

Importantly, the $\varphi$ spectrum shown in \ref{fig:dendrite-orientation-spectrum}b is later used for the DGD procedure (Algorithm \ref{alg:detecting-dominant-grains}), which is used to decompose the global dendrite skeleton seen in Figure \ref{fig:assembled-global-skeleton-2} into grains. Examples of this process are shown in Figures \ref{fig:dgd-example-1}-\ref{fig:dgd-example-3}. Figure \ref{fig:dgd-example-1}a is Figure \ref{fig:assembled-global-skeleton-2}b stripped of all background with only dendrites remaining -- here the most intense $\varphi$ spectrum peak shown in Figure \ref{fig:dendrite-orientation-spectrum}b is considered. In (b), one can see the segments with $\varphi$ within an interval of the selected $\varphi$ peak (Algorithm \ref{alg:detecting-dominant-grains}) and (c) shows the result of thresholding by color-space distance. Afterwards, the grain masks are constructed using the closing transform, then size-thresholded, which results in grain masks as seen in (d). These operations are performed for every peak in the global $\varphi$ spectrum that survives filtering and thresholding (Algorithm \ref{alg:detecting-dominant-grains}).

More examples of this are shown in Figures \ref{fig:dgd-example-2} and \ref{fig:dgd-example-3}, which showcase characteristic cases that can be encountered. Specifically, Figure \ref{fig:dgd-example-2} is the case where peaks with $\sim \pm \pi/2$ are detected, recognized as edge peaks and unified. Therefore, the dendrites seen in Figure \ref{fig:dgd-example-2}a are correctly grouped together by color-space distance (which encodes $\varphi$) and then unified into grains by proximity. Figure \ref{fig:dgd-example-3}a, in turn, shows the case where there are two large grains that have a rather gradual transition into one another, akin to a buffer zone between the two. This is where thresholding after color-space distance measurements (b-c) is key.

\begin{figure}[H]
\centering
\includegraphics[width=1\textwidth]{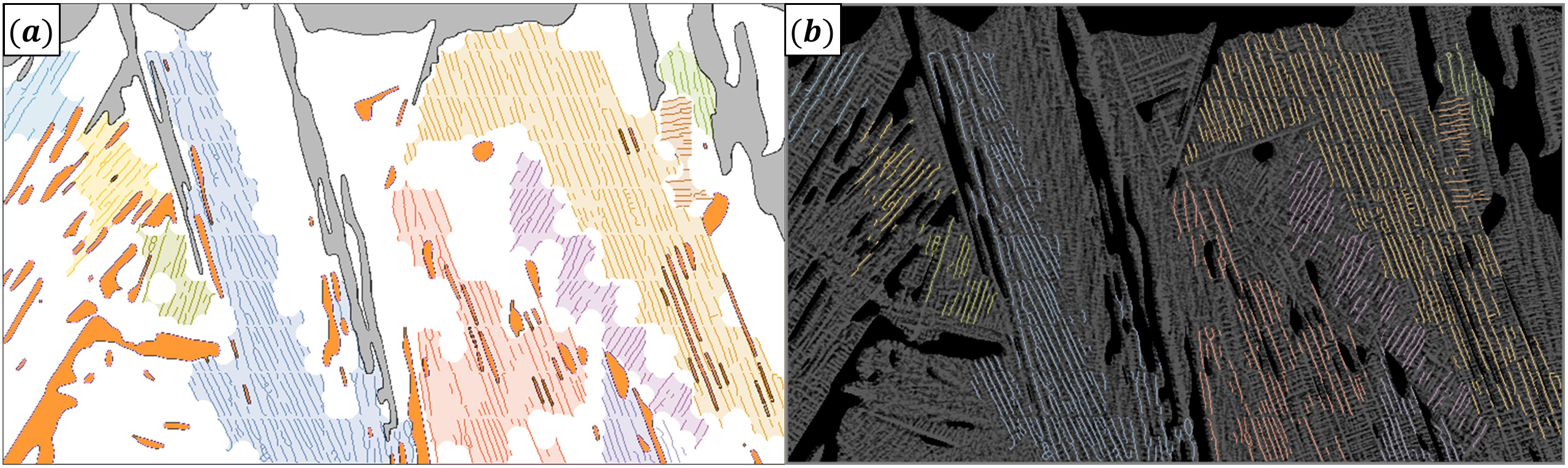}
\caption{Results of DGD for the case corresponding to Figure \ref{fig:dgd-example-1}: (a) dendrite segments colorized by dendrite grain IDs with grain mask overlays, with orange areas representing liquid metal cavities and bulk liquid with channels highlighted as gray areas, and (b) grain dendrite skeletons overlays. The background in (b) is a post-processed raw image (Algorithm \ref{alg:pre-processing} followed by image inversion, reference-less FFC, CTM, re-scaling and sharpening) with the LZ masked.}
\label{fig:dgd-results-1}
\end{figure}

\begin{figure}[H]
\centering
\includegraphics[width=1\textwidth]{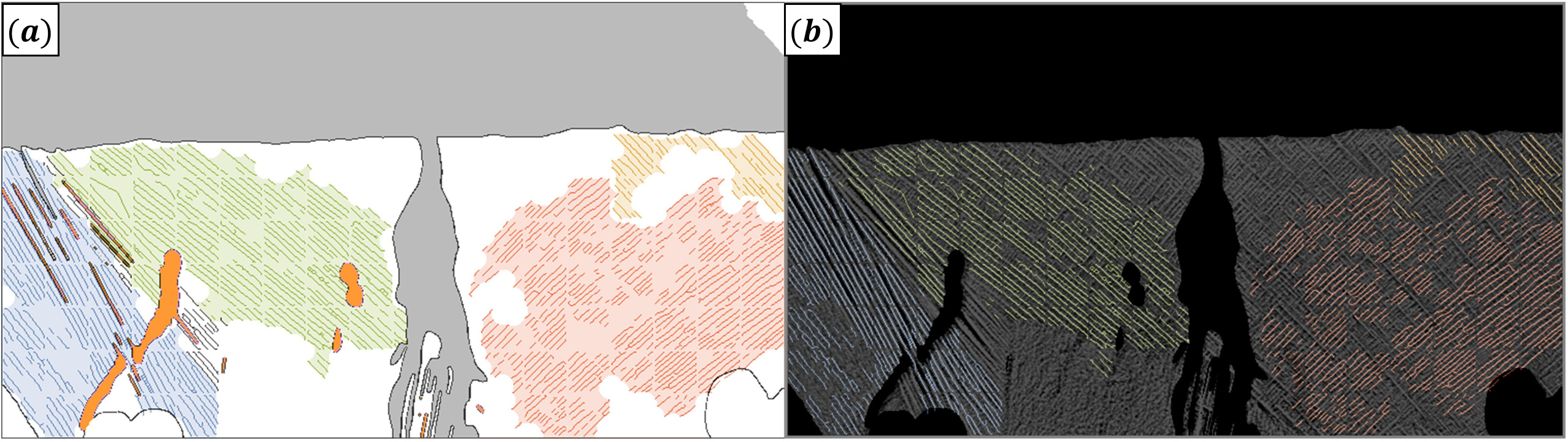}
\caption{Results of DGD for the case considered in Figure \ref{fig:dgd-example-3}. Note the transition zone (black-colored dendrite segments) in (a) between the two grains to the left.}
\label{fig:dgd-results-2}
\end{figure}

When the refined scans and grain cleanup (Section \ref{sec:refined-grain-scan} and Algorithm \ref{alg:resolve-grain-ambiguities}) are done, one can assemble the resulting grains (both their masks and dendrite segments) within the FOV, examples of which are shown in Figures \ref{fig:dgd-results-1} and \ref{fig:dgd-results-2}. It becomes visible in Figure \ref{fig:dgd-results-1}, and especially (b) that the detected grains indeed constitute the major dendrite clusters with coherent (sufficiently similar) $\varphi$ sets. Note also that in (a) the grains separated by the liquid cavities (e.g. upper-left corner and the lower-right part of the FOV) are correctly separated, even the ones with sufficiently similar $\varphi$. This is because liquid cavities are accounted for in the grain cleanup process. Observe that the detected dendrite skeletons match the landscape seen in the background of (b).

\clearpage

The same holds for Figure \ref{fig:dgd-results-2}, where in (a) one can see a black-colored cluster of skeletons designated as undetermined, i.e. they do not belong to either of the two adjacent grains with certainty and are a transition zone. This is because both grains are very close to this area and the $\varphi$ of dendrites within this zone is somewhere between the mean $\varphi$ for the two grains in question. It might seem that the lower part of one of the grains in (a), highlighted with light blue, should be treated as a separate smaller grain because the long diagonal cavities seem to split it in two. However, one can clearly see in (b) that they are actually connected via one of the dendrites that bridges a narrow gap between the two liquid metal cavities. A similar situation holds for the grain highlighted with light red in Figure \ref{fig:dgd-results-1}a. If one does observe some small clusters of grains with similar $\varphi$ in Figures \ref{fig:dgd-results-1}b and \ref{fig:dgd-results-2}b that were discarded by the DGD process, then these grains are most likely below the size threshold. One can, of course, always adjust settings accordingly if capturing smaller grains is necessary.

\clearpage

\begin{figure}[H]
\centering
\includegraphics[width=0.7\textwidth]{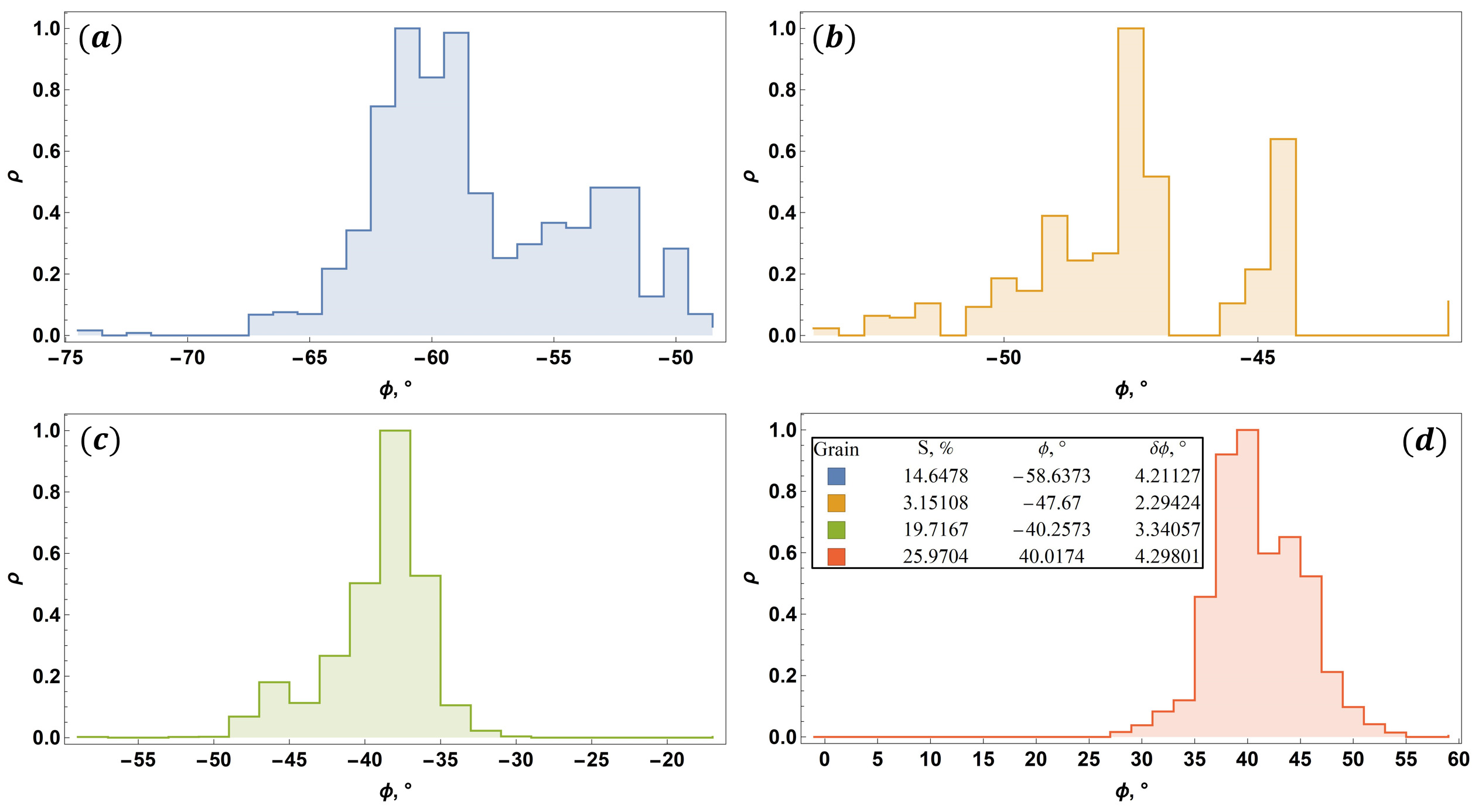}
\caption{Dendrite orientation $\varphi$ spectra for the grains identified via the DGD process as shown in Figure \ref{fig:dgd-results-2}. Colors for plots (a-b) correspond to the grain colors in Figure \ref{fig:dgd-results-2}. The legend in (d) shows the grain area fraction $S$ (with respect to the SZ mask area), mean dendrite orientation $\varphi$ and its standard deviation $\delta \varphi$.}
\label{fig:grain-spectra}
\end{figure}

Once the dominant grains are detected, their relative areas (with respect to the SZ area) and $\varphi$ statistics can be determined, which is demonstrated in Figure \ref{fig:grain-spectra} for the case seen in Figure \ref{fig:dgd-results-2}. Again, note that this can be done for all or selected frames in an image sequence to observe the dynamics of grain formation, fragmentation and how their statistics change.

\subsection{Solidification front dynamics}

\begin{figure}[H]
\centering
\includegraphics[width=0.675\textwidth]{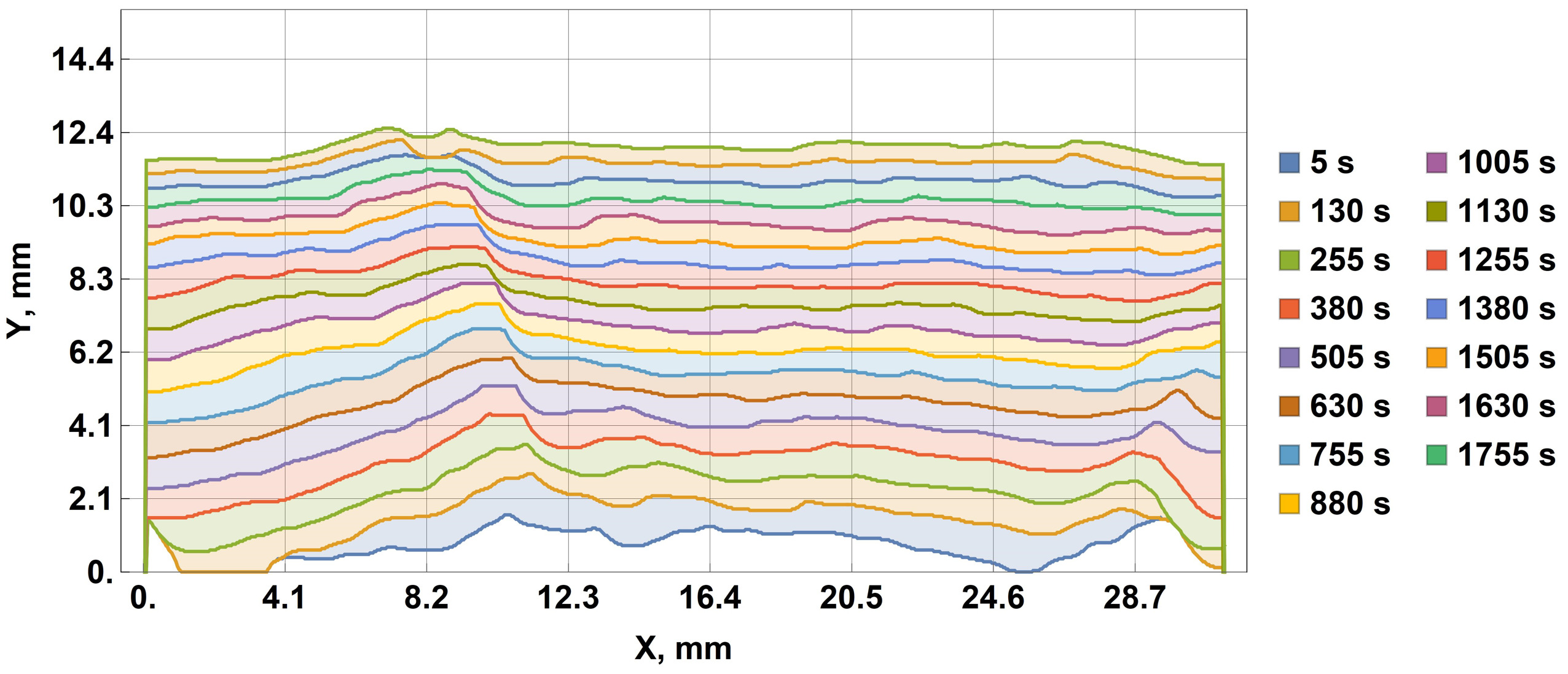}
\caption{SF height $Y$ over cell width $X$ at different time stamps (legend to the right) for the case shown in Figure \ref{fig:fov-segmentation-1}.}
\label{fig:front-height-instances}
\end{figure}

\begin{figure}[H]
\centering
\includegraphics[width=0.65\textwidth]{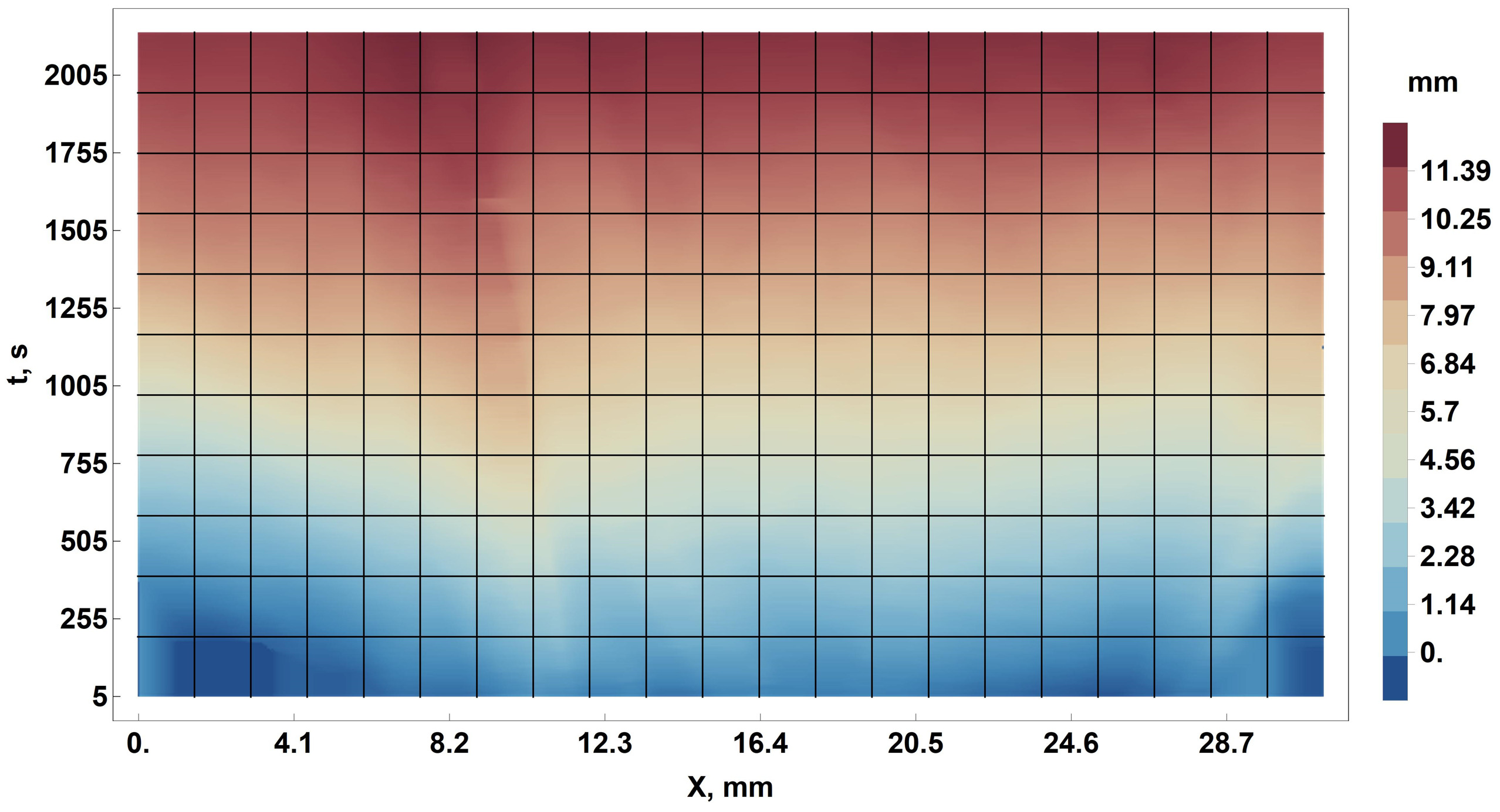}
\caption{SF height (color-coded, with color legend to the right) over cell width $X$ and time $t$ for the case shown in Figure \ref{fig:fov-segmentation-1}.}
\label{fig:front-height-matrix}
\end{figure}

\begin{figure}[H]
\centering
\includegraphics[width=0.65\textwidth]{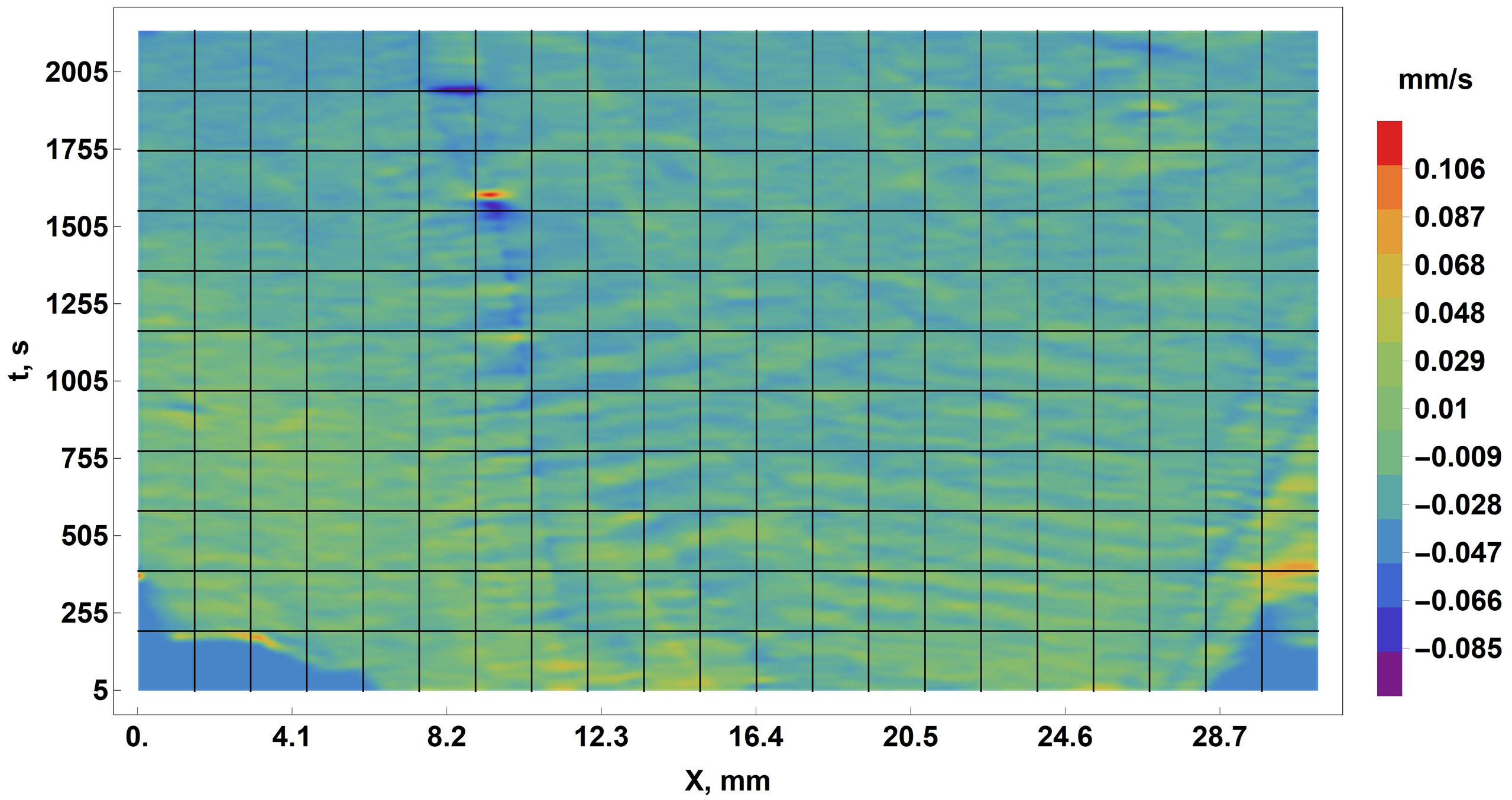}
\caption{SF vertical velocity (color-coded, with color legend to the right) over cell width $X$ and time $t$ for the case shown in Figure \ref{fig:fov-segmentation-1}.}
\label{fig:front-local-velocity}
\end{figure}

\begin{figure}[H]
\centering
\includegraphics[width=0.875\textwidth]{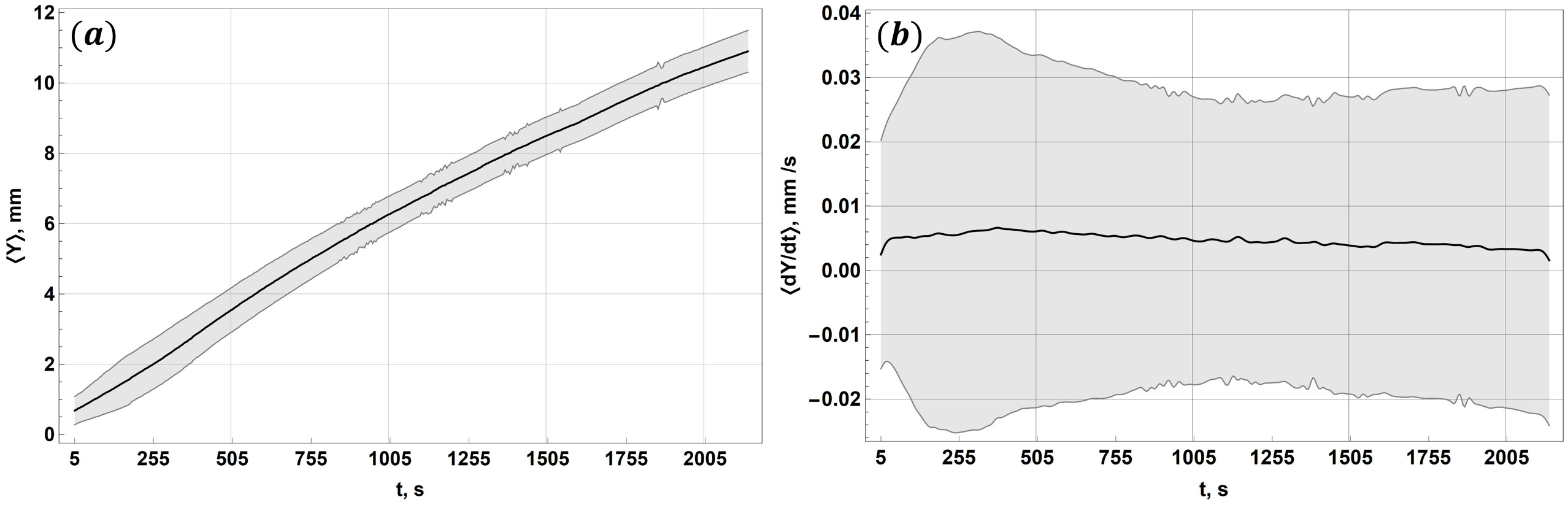}
\caption{Mean SF dynamics over cell width: (a) height $\left< Y \right>$ and (b) vertical velocity $\left< dY/dt \right>$ over time $t$ for the case shown in Figure \ref{fig:fov-segmentation-1}.}
\label{fig:front-mean-height-velocity}
\end{figure}

In addition to the above, having derived the SF, one can now look at its dynamics. Figure \ref{fig:front-height-instances} shows how the SF height over the cell width changes over time for the case shown in Figure \ref{fig:fov-segmentation-1}, as the SF advances vertically upwards and the SFF increases. This also contains the shape information to some degree. If the latter is not of interest, or if a more continuous dynamics visualization is needed, one can generate an image as in Figure \ref{fig:front-height-matrix} where the color encodes the front height and variations over time and cell width can be clearly seen for an entire image sequence. Figure \ref{fig:front-height-matrix} is obtained by median-filtering the front height matrix (elements indexed by time and width coordinates) with a 2-pixel (element) kernel radius. The resulting matrix can then be used to calculate the matrix of instantaneous SF propagation velocity, which is shown in Figure \ref{fig:front-local-velocity}. In the case of velocity, outlier removal (as in Algorithm \ref{alg:pre-processing}) and bilateral filtering ($\mu_\text{b} = 2$ pixel value range factor and Gaussian kernel scale $\sigma_\text{b} = 21$) are performed. If mean dynamics are of interest, they can be readily derived from the above matrices. The corresponding results can be seen in Figure \ref{fig:front-mean-height-velocity}.

\clearpage

\begin{figure}[H]
\centering
\includegraphics[width=0.65\textwidth]{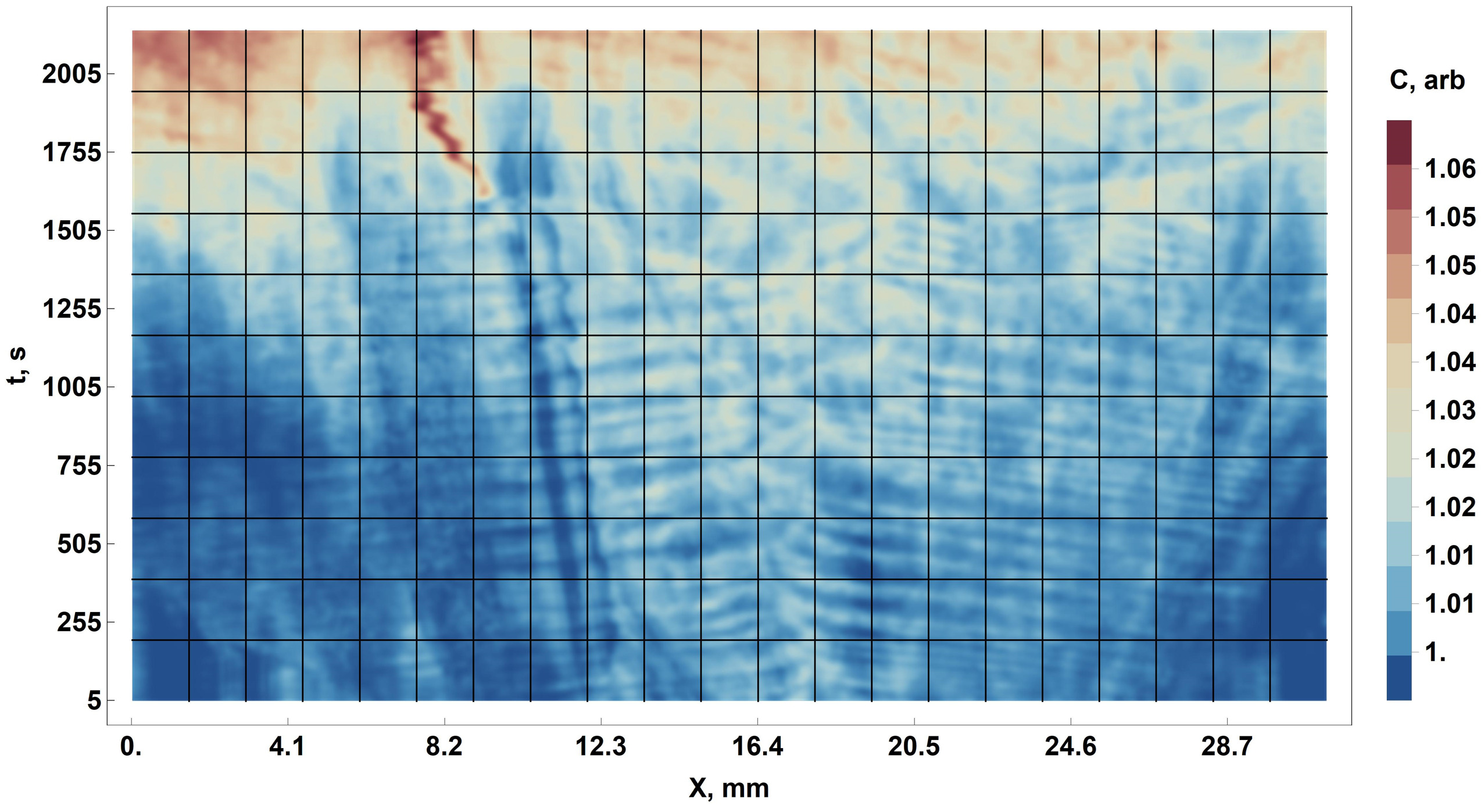}
\caption{Solute concentration dynamics above the SF and along the cell width $X$ over time $t$ for the case shown in Figure \ref{fig:fov-segmentation-1}. Concentration is expressed via image luminance (encoding $Ga$/$In$ concentration via the Beer-Lambert law) in relative units (with respect to the initial frame after Algorithm \ref{alg:concentration-measurement-above-sf} is applied) and is color-coded as shown in the color bar to the right.}
\label{fig:concentration-matrix-above-front}
\end{figure}

Solute concentration can be measured above the SF as explained in Algorithm \ref{alg:concentration-measurement-above-sf} and plotted for different locations along the cell width over time, which is showcased in Figure \ref{fig:concentration-matrix-above-front}.

Note the maximum spot within $X \in (8.2;12.3)~mm$ and $t \in (1505;1755)~s$, which corresponds to a rapid channel opening at the SF (Figure \ref{fig:fov-segmentation-1}). Figure \ref{alg:concentration-measurement-above-sf} reveals the sudden appearance of a highly concentrated "trace" in the same region as the velocity minimum/maximum in Figure \ref{fig:front-local-velocity}. Once a liquid metal cavity breaches the SF and a channel forms, much greater X-ray transmission is consistently measured, as it should be, implying increased $Ga$ concentration. The trail shift to the left is simply due to the change in the location of the channel outlet at the SF, which is caused by remelting. Observe also the banded structure for $t \lesssim 1500$ -- these are not artifacts due to data processing, but rather physical concentration oscillations above the SF, as well as the result of SF fluctuations.

\subsection{Convective plume segmentation}

\begin{figure}[H]
\centering
\includegraphics[width=1\textwidth]{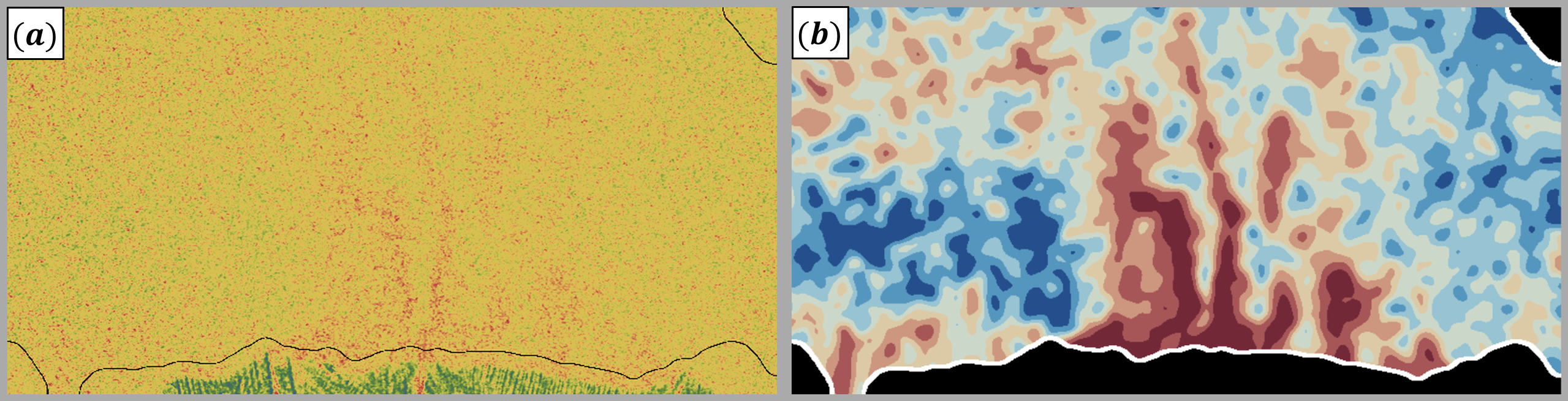}
\caption{Analyzing $Ga$-rich convective plumes in the LZ above the SF (Algorithm \ref{alg:convective-plume-segmentation}): (a) pre-processed (Algorithm \ref{alg:pre-processing}) image with the outlines of the buffer zone (black contours) extended from the SF and the artifacts (e.g., lower-left and upper-right corners), and (b) convective plumes segmented and highlighted outside of the buffer zone (black areas with white boundaries). The level sets representing the plumes are colorized by image luminance (representative of the solute concentration). The color scheme is as in Figure \ref{fig:concentration-matrix-above-front}. This case corresponds to the one shown in Figure \ref{fig:fov-segmentation-1} (close to the beginning of the image sequence where the plumes are the most intense).}
\label{fig:plume-segmentation-1}
\end{figure}

\begin{figure}[H]
\centering
\includegraphics[width=1\textwidth]{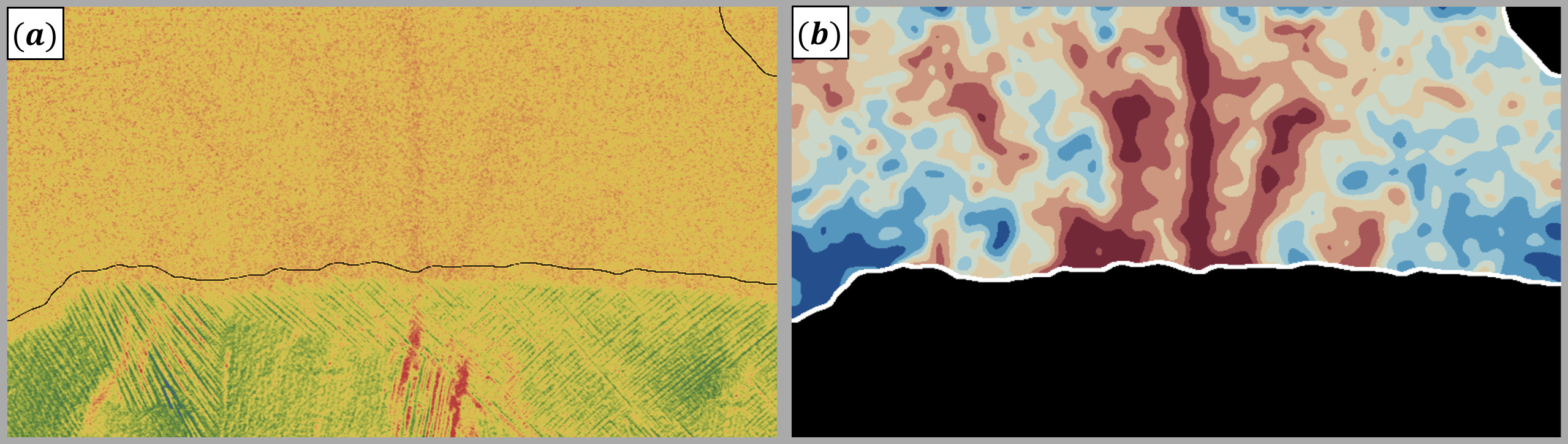}
\caption{An example of plume segmentation in the LZ. Here the case considered in Figure \ref{fig:dgd-results-2} (one of the earlier frames of the image sequence) is shown.}
\label{fig:plume-segmentation-2}
\end{figure}

\begin{figure}[H]
\centering
\includegraphics[width=1\textwidth]{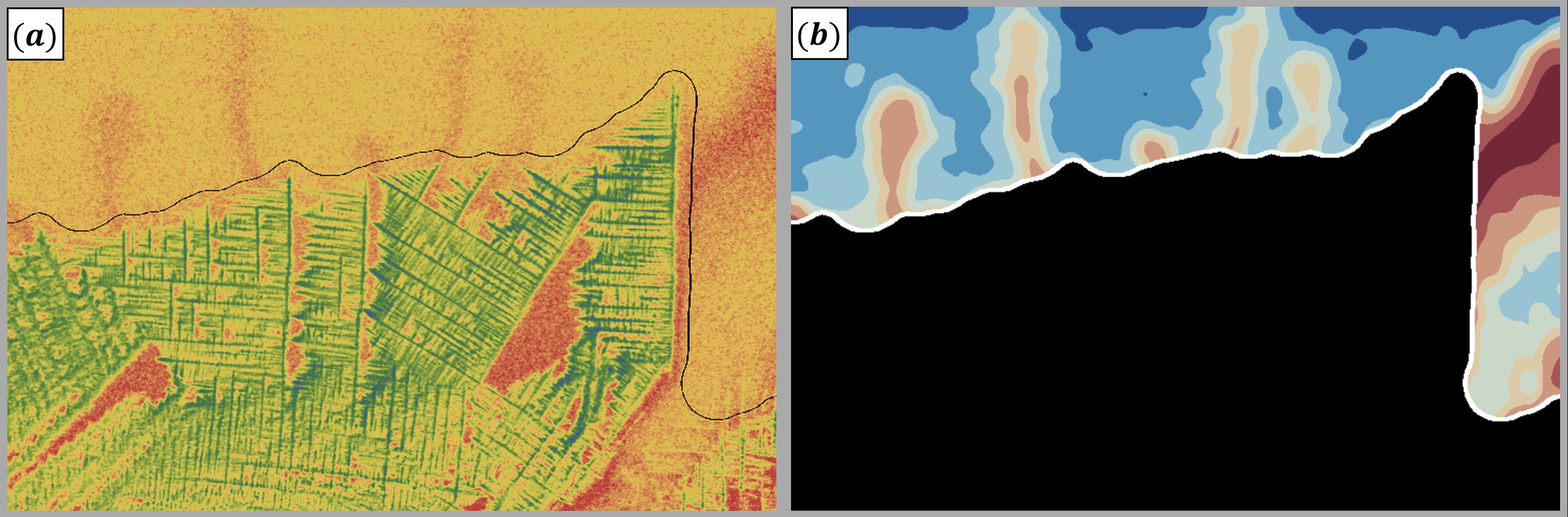}
\caption{Plume segmentation for the case seen in Figure \ref{fig:fov-segmentation-2}.}
\label{fig:plume-segmentation-3}
\end{figure}

\begin{figure}[htbp]
\centering
\includegraphics[width=1\textwidth]{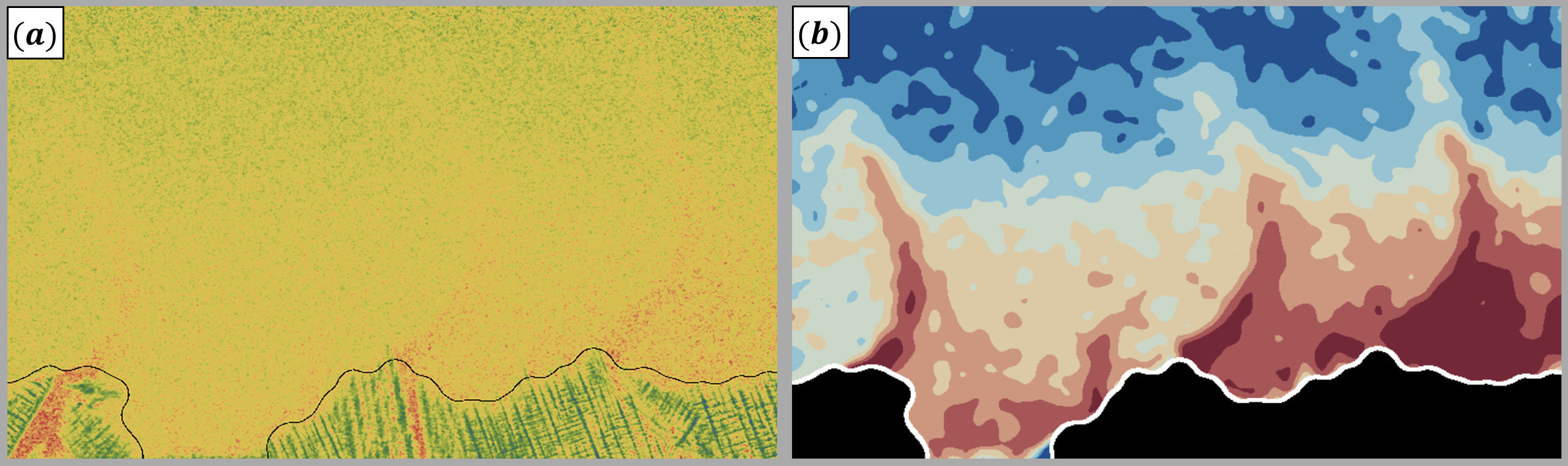}
\caption{Another example of convective plumes segmented in the LZ. Here an early frame from the image sequence corresponding to Figure \ref{fig:dgd-results-1} is shown.}
\label{fig:plume-segmentation-4}
\end{figure}

Finally, one can analyze what occurs above the SF in the bulk of the LZ further by examining the convection plumes using Algorithm \ref{alg:convective-plume-segmentation}. The results of its applications to example frames representing different cases are shown in Figures \ref{fig:plume-segmentation-1}-\ref{fig:plume-segmentation-4}. As with the concentration measurements just above the SF, the buffer zone mask is intended to prevent the interference from potentially sticking out dendrite tips that in general may occur above the derived SF for low SNR/CNR images. In addition, the artifact areas must be excluded, since any information contained there is otherwise meaningless. This makes obvious that the code successfully segments the plumes in the showcased examples and largely preserves their shapes despite the clearly visible large-grained noise.

\clearpage

\clearpage

This multiple level set representation, in conjunction with morphological analysis shown in Figures \ref{fig:segment-classification-1} and \ref{fig:segment-classification-2}, as well as the SF dynamics and SZ analysis, should provide a wealth of details, enabling in-depth analysis and physical interpretation of solidification processes studies using experimental setups that are similar in principle to those considered in this paper. It is worth noting that what can be seen in the above sub-figures (a) a purposely lowered SNR. Given the results seen in (b), higher SNR images should enable even better results. This, of course, also holds for the SZ analysis.

\section{Further improvements \& extensions}
\label{sec:improvements}

While many of the necessary features are already present in the code, there is most certainly room for functionality expansion and performance boosting. Specifically, we propose the following:

\begin{itemize}
    \item Transition to GPU implementations of BM3D and NM filters, such as the ones presented in \cite{bm3d-gpu-acceleration, bm3d-nl-means-gpu-acceleration} -- this should greatly reduce computational time per image sequence.
    \item Test a possibly better suited BM3D version, BM3D-SAPCA (shape-adaptive principal component analysis) \cite{bm3d-sapca}.
    \item Integrate an optical flow velocimetry code by Liu et. al. \cite{tianshu-optical-flow-code, tianshu-optical-flow-bubbles} directly into the presented image processing code via \href{http://matlink.org/documentation/}{\textit{MATLink}} for a more in-depth liquid flow analysis and seamless start-to-finish data processing. This will be invaluable, in particular when combined with the presented plume segmentation and the SF analysis routines.
    \item Improve upon the IW partitioning approach to eliminate the gaps between the skeletons from separate IWs -- this would require a robust method that combines the overlapping parts of the IWs. Successfully implementing this will further boost the quality of dendrite structure analysis.
    \item Derive solid phase thickness over the cell and Ga/In concentrations.
    \item Compute inter-dendrite primary spacing.
    \item Demonstrate that the showcased methodology readily translates to higher-resolution images, e.g. from synchrotron measurements and numerical simulation output.
\end{itemize}

\section{Conclusions}
\label{sec:conclusions}

To summarize, we have demonstrated a robust and noise-resilient image processing pipeline for analyzing directional metal alloy solidification processes in laboratory-scale experiments using Hele-Shaw liquid metal cells and dynamic X-ray imaging. The developed methodology at present allows one to segment liquid and solidified zones within the field of view, detect the skeletons of solidified structures in the solid zone, perform orientation analysis, detect dominant dendrite grains (if any), quantify the dynamics of the solidification front and solute concentration above it, detect and separate liquid metal channels and cavities, as well as segment and characterize convective plumes in the liquid zone. Even with artificially lowered SNR, the code performed reliably and the demonstrated performance is such that in-depth physical analysis of images is feasible. With the addition of extra features in the future, such as optical flow velocimetry for the liquid zone and convective plumes, as well as other improvements outlined above, the presented methods and code should be of great value to the relevant scientific community for further physics-focused research.

The code combines both existing and original methods, is open-source, and is available on \textit{GitHub}: \href{https://github.com/Mihails-Birjukovs/Meso-scale_Solidification_Analysis}{Mihails-Birjukovs/Meso-scale\_Solidification\_Analysis}.

\section*{Acknowledgements}

This research is supported by Hemlholtz-Zentrum Dresden-Rossendorf (HZDR) and a DAAD Short-Term Grant (2021, 57552336). The authors acknowledge the project ”Development of numerical modelling approaches to study complex multiphysical interactions in electromagnetic liquid metal technologies” (No. 1.1.1.1/18/A/108) wherein some of the utilized image processing methods were developed.

\printbibliography[title={References}]

\end{document}